\documentclass[a4paper]{aa}
\usepackage{ulem}
\usepackage{setspace}
\usepackage{geometry}
\usepackage{subfigure}
\usepackage{graphicx}
\usepackage{amsmath,bm}
\usepackage{multirow}
\usepackage{longtable}
\usepackage{subfigure}
\usepackage{color}
\usepackage{txfonts}
\begin{document} 

\title{ $\lambda$ = 2 mm spectroscopy observations toward the circumnuclear disk of NGC\,1068}

\author{Jianjie Qiu
          \inst{1},
          Jiangshui Zhang
          \inst{2}
             \thanks{e-mail: jszhang@gzhu.edu.cn},
          Yong Zhang
          \inst{1,3}
           \thanks{e-mail: zhangyong5@mail.sysu.edu.cn}, 
          Lanwei Jia
          \inst{2},
         Xindi Tang
          \inst{4,5,6}
          }

\institute{School of Physics and Astronomy, Sun Yat-sen University, Guangzhou 510275, PR China
               \and
             Center For Astrophysics, Guangzhou University, Guangzhou 510006, PR China
             \and
             Laboratory for Space Research, University of Hong Kong, Hong Kong, PR China
              \and
             Max-Planck-Institut f{\"u}r Radioastronomie, Auf dem H{\"u}gel 69, D-53121 Bonn, Germany
             \and 
             Xinjiang Astronomical Observatory, Chinese Academy of Sciences, 830011 Urumqi, PR China
             \and
             Key Laboratory of Radio Astronomy, Chinese Academy of Sciences, PR China
             }


 
\abstract
   {}
   {We investigate the physical and chemical conditions of molecular gas in the circumnuclear disk (CND) region of NGC\,1068.
   }
   { We carried out a spectral line survey with the IRAM 30m telescope toward the center of NGC\,1068 and mainly 
   focused on the 2 mm band with a frequency coverage of 160.7--168.6 GHz and 176.5--184.3 GHz. 
   }
   {Fifteen lines are detected in NGC\,1068, eight  of which are new detections for this galaxy. 
    We derive the rotation temperatures and column densities of fourteen molecular species.
      }
   {Based on the [HCO$^{+}$\,$(2-1)$]/[HOC$^{+}$\,$(2-1)$] ratio, we obtain a high ionization degree 
   in the CND of NGC\,1068. It is found that HC$_{3}$N is concentrated in the east knot, 
   while $^{13}$CCH, CH$_{3}$CN, SO, HOC$^{+}$, CS, CH$_{3}$CCH, and H$_{2}$CO are concentrated in the west knot. 
   Compared to the star-forming galaxies M\,82 and NGC\,253, the chemistry of NGC\,1068 might be less strongly affected by the UV radiation field, 
   and its kinetic temperature might be lower.
   }

\keywords{ISM: molecules --
        galaxies: ISM --
        galaxies: individual: NGC\,1068 --
        galaxies: nuclei --
        galaxies: active
        }
\titlerunning{ $\lambda$ = 2 mm spectroscopy observations toward the circumnuclear disk of NGC\,1068}
\authorrunning{Jianjie Qiu et al.}
\maketitle
%

\section{Introduction}

Molecular gas is the fuel of star formation.
So far, more than 60 molecular species have been detected in external galaxies \citep{McGuire18}. 
The relative abundances of different species vary with its surrounding astrophysical environment \citep{Omont07}.
The molecular rotational transition lines at millimeter wavelength are sensitive to and trace the physical and chemical properties of the interstellar medium.  
An unbiased line survey could detect molecular rotation transitions with different energy levels and/or different species and 
provide an unbiased view on the molecular environment.
Previous observations at millimeter band have shown that the properties of molecular gas are different 
between starbursts (SB) and active galactic nuclei (AGN) in external galaxies 
\citep{Kohno01, Krips08, Izumi13, Izumi16, Davis13, Aladro15, Nakajima18, Li19}.
The underlying causes of these variations, especially how the molecular composition might be effected by AGN, are still not fully clear.

NGC\,1068 is one of the nearest ($\sim$14.4~Mpc, 1$''$ = 72~pc, \citealt{Bland97}) and brightest ($L_{IR}$ = 3 $\times$ 10$^{11} L_{\odot}$) 
Seyfert II galaxies with a starburst.
It consists of a starburst ring ($\sim$15\arcsec from the central AGN) and a central nuclear disk  (CND) \citep{Schinnerer00}. 
It was established that
the physical conditions between the starburst ring and the CND are different \citep{Viti14}.
The molecular gas in the CND  is denser and hotter than the gas in the starburst ring.
Multi-gas-phase components exist in the CND region \citep{Viti14}.
The gas temperature in the CND region is higher than 150~K, 
and the gas density is above 10$^{5}$~cm$^{-3}$  \citep{Viti14}.
The CND region could be spatially resolved into an east and a west knot. These knots are dominated by a fast and a slow shock, respectively \citep{Kelly17}. 
The mass of the central torus was estimated to be 2 $\times$ 10$^{5} M_{\odot}$ \citep{Garcia14}.
With an AGN-driven outflow \citep{Garcia14} and a past inflow driven by a minor merger \citep{Furuya16}, 
the torus shows a complex dynamical behavior \citep{Garcia16}.

NGC\,1068 is one of the best extragalactic targets for continuum and line observations in all wavelengths from 
X-rays to the radio domain (e.g., radio, \citealt{Greenhill96}; millimeter, \citealt{Schinnerer00}; 
infrared, \citealt{Jaffe04}; optical, \citealt{Antonucci85}; UV, \citealt{Antonucci94}; and X-ray, \citealt{Kinkhabwala02}). 
These results show that the strong UV or X-ray field heavily 
influence the physical conditions, kinematics, and chemistry 
in the CND of NGC\,1068 \citep{Usero04,Perez09,Garcia10,Nakajima11,Nakajima18,Aladro11,Aladro15,Takano14,Viti14}.

To better understand the chemistry in the CND of NGC\,1068, we performed
an observation of molecular lines at 3~mm bands \citep{Qiu18}, and detected 
CH$_3$OCH$_3$ for the first time  in  external galaxies. As a supplement to the previous
report, this paper presents observations of molecular lines
at the 2~mm band toward NGC\,1068. 
The organization of this paper is as follows: 
we present the observations and data reduction in Sect.~\ref{sec2}, 
and the main results are provided in Sect.~\ref{sec3}. 
We discuss the physical and chemical properties of molecules in Sect.~\ref{sec4}, and give a brief summary in Sect.~\ref{sec5}.

\section{Observations and data reduction}
\label{sec2}
The observations toward the center of NGC\,1068 (RA: 02:42:40.70 DEC: -00:00:48.0 J2000) 
were carried out in January 2017 with the IRAM 30m single-dish 
telescope at Pico Veleta Observatory (Spain) \footnote{This publication is based on data acquired with the IRAM 30-m telescope. 
IRAM is supported by INSU/CNRS (France), MPG (Germany), and IGN (Spain).}.
We used the Eight Mixer Receiver (EMIR) with dual polarization and the Fourier Transform 
Spectrometers (FTS) backend tuned.
The frequency channel spacing was 195 KHz ,
and the instantaneous frequency coverage per sideband and polarization was 8~GHz.
We used the standard wobbler-switching mode with a $\pm$ 110\arcsec offset and 1.5 second per phase.
Each scan included eight subscans of 30 seconds.
We checked pointing and focus every two hours.

\begin{figure}
  \centering
    \begin{minipage}{17cm}
        \includegraphics[width=17cm]{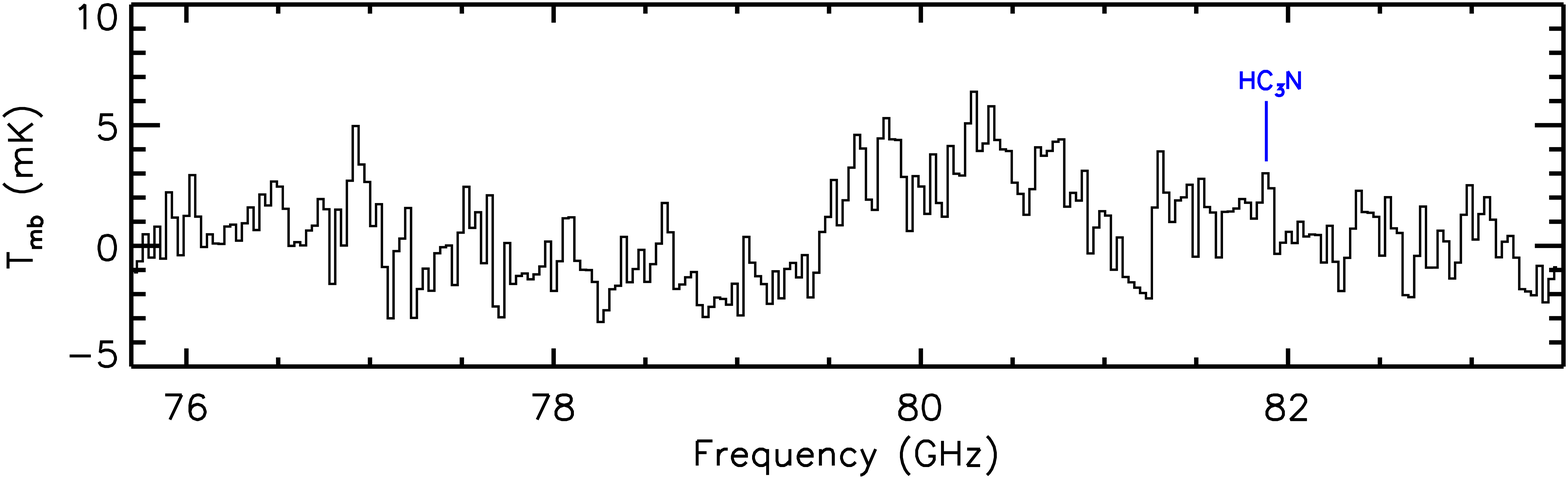}
    \end{minipage}
    
    \begin{minipage}{17cm}
        \includegraphics[width=17cm]{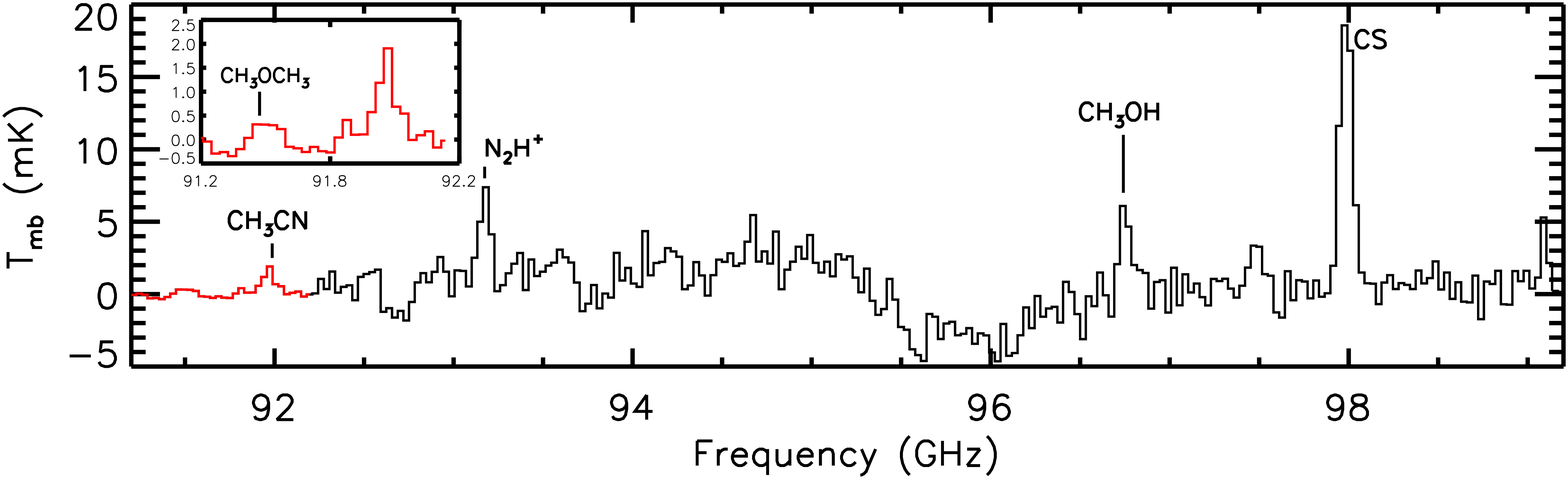}
    \end{minipage}
  
    \begin{minipage}{17cm}
        \includegraphics[width=17cm]{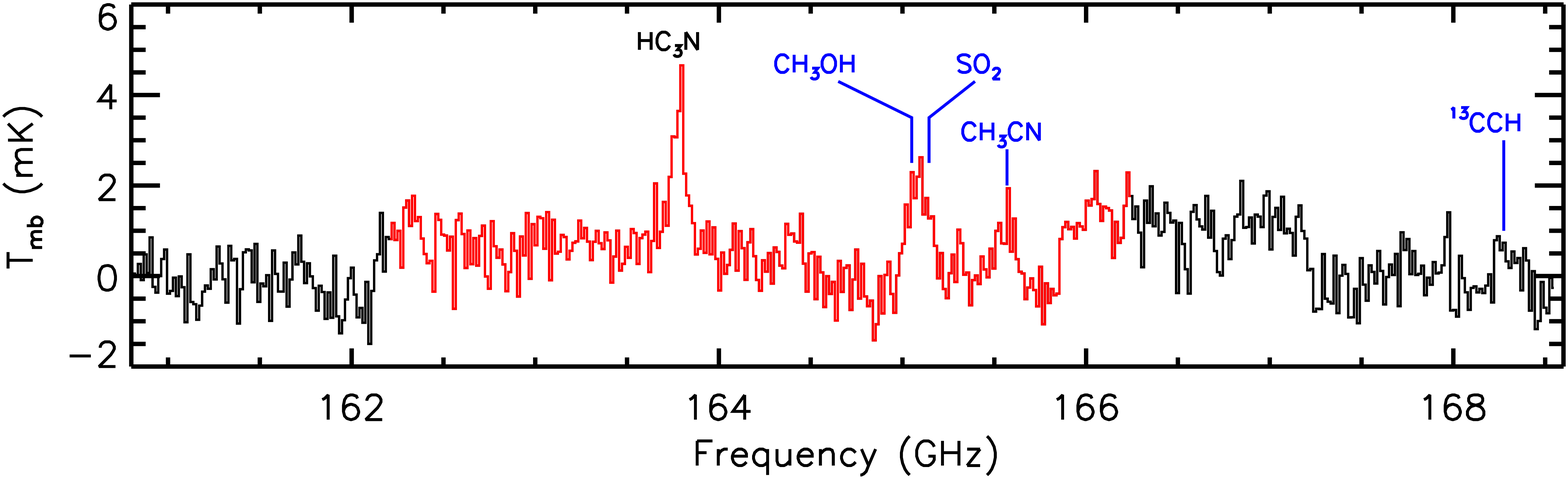}
    \end{minipage}
    
    \begin{minipage}{17cm}
        \includegraphics[width=17cm]{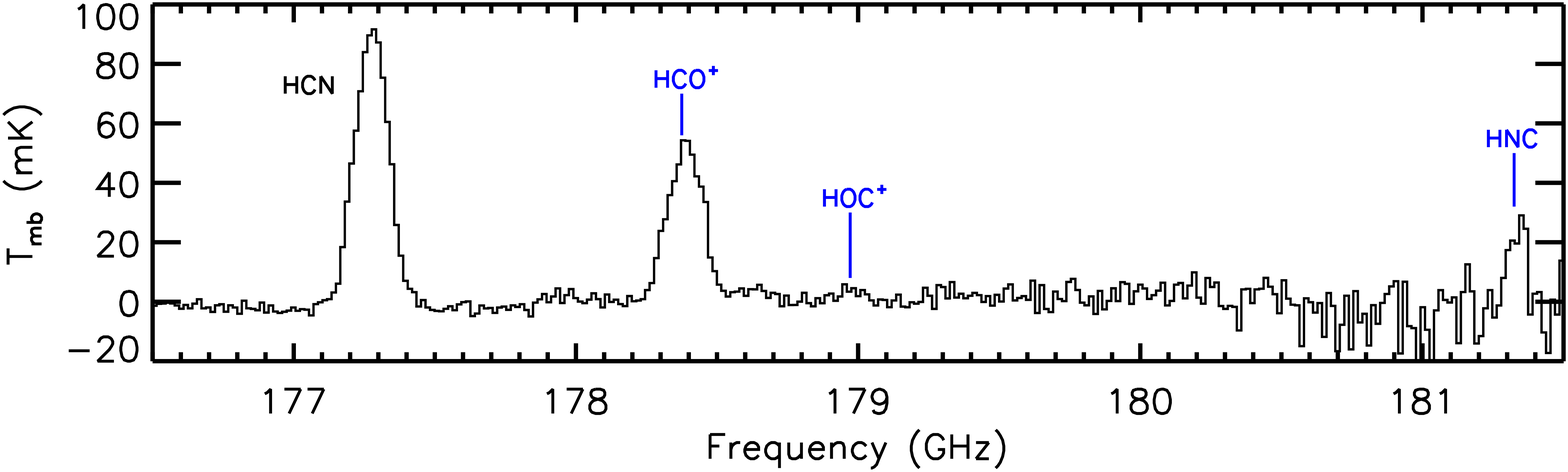}
    \end{minipage}

 \caption{Full spectra of NGC\,1068 obtained in the current observations. 
 Red represents a coaddition of the current spectrum and that in \cite{Qiu18}. 
 Blue represents the new detections in this galaxy.
}
\label{figure9}
\end{figure}

The observations were performed using two tunings in the 2 mm window 
(160.7--168.6 GHz and 176.5--184.3 GHz) and in the 3 mm window (75.7--83.5 GHz and 91.4--99.2 GHz). 
The 3 mm spectra were obtained with a short integration time to supplement the line survey 
in the 2 mm window (Table \ref{table_parameter}).
Figure \ref{figure9} presents the full spectra. 
The molecular line temperature scale was converted from the antenna temperature $T^{*}_{\rm A}$ 
into the main beam temperature $T_{\rm mb}$. The conversion relation between them is 
$T_{\rm mb} = (F_{\rm eff}/B_{\rm eff}) \times T^{*}_{\rm A}$, where $F_{\rm eff}$ and $B_{\rm eff}$ are 
the forward efficiency and main-beam efficiency of the telescope, respectively. The values assumed 
for $F_{\rm eff}$ and $B_{\rm eff}$ of each band are listed in Table \ref{table_parameter}.
The main observation parameters, including band range, half-power bandwidth (HPBW), 
and spatial resolution, are also listed in Table \ref{table_parameter}.

The data reduction procedures are similar to those described in \cite{Qiu18}.
Linear baseline subtraction and Gaussian profile fitting were made for all the detected lines.
The data reduction was performed using the CLASS software of the GILDAS 
package\footnote{CLASS http://www.iram.fr/IRAMFR/GILDAS}. 
 We identified each molecular transition by referring to the frequencies from 
the NIST database recommended 
rest frequencies for observed interstellar molecular microwave 
transitions\footnote{NIST http://pml.nist.gov/cgi-bin/micro/table5/start.pl} and 
from the splatalogue database for astronomical 
spectroscopy\footnote{http://www.cv.nrao.edu/php/splat/advanced.php}.
 
We performed a literature survey of previous observations made with single-dish telescopes toward NGC\,1068. The lines
detected by other research groups, together with those detected by us, are listed in Table 2.
This table may serve as a reference source for future studies of this object.
We assume a heliocentric systemic velocity $v_{sys}$ $\sim$ 1137~km~s$^{-1}$ 
(from NASA/IPAC Extragalactic Database (NED)\footnote{http://ned.ipac.caltech.edu}) for this source throughout.
 
\section{Results}
\label{sec3}

Overall, 15 lines belonging to 12 different molecular species are detected.
Figure \ref{f_list_1} shows the individual line profiles in velocity units, 
and Table \ref{totaltable} summarizes the parameters derived from Gaussian fitting with CLASS to the observed lines.   
The main results include the detection of the $J = 2 - 1$ transitions of $^{13}$CCH, HCO$^{+}$, HOC$^{+}$, and HNC, 
and emission lines of  
CH$_{3}$OH\,$(J_{1,k-1}-J_{0,k})$E, SO$_{2}$\,$(5_{2,4}-5_{1,5})$, and CH$_{3}$CN\,$(9_{k}-8_{k})$.
To our knowledge, these lines are detected  for the first time in NGC\,1068.
In addition, we also detect the lines of CH$_{3}$CN\,$(5_{k}-4_{k})$, 
N$_{2}$H$^{+}$\,$(1-0)$, CH$_{3}$OCH$_{3}$\,$(3_{2,2}-3_{1,3})$, 
CH$_{3}$OH\,$(2_{k}-1_{k})$, CS\,$(2-1)$, HC$_{3}$N\,$(18-17)$, and HCN\,$(2-1)$.

CH$_{3}$CN\,$(5_{k}-4_{k})$, HC$_{3}$N\,$(18-17)$, and CH$_{3}$OCH$_{3}$\,$(3_{2,2}-3_{1,3})$,  
have been detected or tentatively detected in previous observations \citep{Qiu18}. 
To enhance the signal-to-noise ratios (S/N), we combined the data of \cite{Qiu18} for these three transitions. 
CH$_{3}$OCH$_{3}$\,$(3_{2,2}-3_{1,3})$ was only marginally detected toward NGC\,1068 by \cite{Qiu18}. 
No stronger line is known in this wavelength.
For this line, we also combined the spectral data of \cite{Aladro13} and found that the S/N  increases from 1.6 to  2.3, 
 probably supporting the potential existence of CH$_{3}$OCH$_{3}$ in this galaxy. 
We checked the raw data of \cite{Qiu18}, and found that the CH$_{3}$CCH\,$(14_{0}-13_{0})$ line 
at 239~GHz is marginally detected, as shown in Fig. \ref{f_list_2}.

\subsection{Line ratios}
\subsubsection{HCN/HCO$^{+}$}
Table  \ref{Tab_line_ratio} lists the velocity-integrated intensity ratio of the $J = 2 - 1$ transitions of five species, 
including$^{13}$CCH, HCN, HCO$^{+}$, HOC$^{+}$, and HNC, detected in the CND of NGC\,1068.
The flux ratio of HCN to HCO$^{+}$ can be used to distinguish galaxies powered by starburst and AGN 
\citep{Kohno01, Krips08, Imanishi09, Izumi16, Aladro18, Li19}.
AGN galaxies usually exhibit higher [HCN/HCO$^{+}$] ratios than starburst galaxies because
high X-ray radiation in AGNs can enhance the abundance of HCN \citep{Kohno01, Privon17}.
Our observations show that the flux ratio of HCN-to-HCO$^{+}$ ($J = 2 - 1$) is 1.82 $\pm$ 0.03  in the CND of NGC\,1068. 
This is consistent with previous interferometer observations at other $J$-transitions.
For example, the flux ratio of HCN-to-HCO$^{+}$  is 1.6--2.0 at $J = 1 - 0$ \citep{Viti14}, 
1.6--3.3 at $J = 3 - 2$ \citep{Imanishi16}, and 2.1--2.8 at $J = 4 - 3$ \citep{Viti14} in the nucleus of NGC\,1068. 
Our single-dish and previous interferometer observational results confirm that the abundance of HCN is enhanced 
in the CND of NGC\,1068, which maybe related to the outflow of AGN \citep{Garcia14}.

\subsubsection{HNC/HCN}
HCN and its isomer HNC are tracers of dense gas, and commonly coexist in different environments \citep{Gao04, Aalto12}.
Some galaxies exhibit a high HNC/HCN ratio, 
which may be caused by low temperature, high ion density, high optical depth, and/or IR pumping.
Theoretical studies \citep{Meijerink05,Meijerink07}  suggest that the HNC/HCN ratio 
 increases in X-ray dominated regions (XDRs) to a higher degree than that in photon-dominated regions (PDRs) and quiescent-cloud regions.
 Observations of three galaxies (Arp\,220, NGC\,4418, and Mrk\,231), which have strong X-ray emission produced by embedded AGN,
indicate a high HNC-to-HCN ($J = 3 - 2$) flux ratio of 1.9 $\pm$ 0.3 in Arp\,220, 1.5 $\pm$ 0.2 in Mrk\,231, and 2.3 $\pm$ 0.3 in NGC\,4418 \citep{Aalto07}. 
Our observations show that the integrated line ratio of HNC to HCN ($J = 2 - 1$) toward the CND of NGC\,1068 is 0.16 $\pm$ 0.03.
This is in agreement with $J = 3 - 2$ and $J = 4 - 3$ line ratios (0.15 $\pm$ 0.03 and 0.19 $\pm$ 0.02, respectively; \citealt{Perez07,Perez09} ).
This means that we do not 
find enhanced HNC emission in the CND of NGC\,1068, and an XDR exists here.
Because the HNC line is narrower than that of HCN, we exclude the IR pumping scenario: the IR pumping dominated scenario will show a broader line width of HNC than HCN, as described in \citet{Aalto02}.
The ammonia (NH$_{3}$) observations show that some components of the molecular cloud have a higher temperature of 140 $\pm$ 30~K 
in the CND of NGC\,1068 \citep{Ao11}.
A reasonable explanation of the lower  HNC-to-HCN ratio  is that the warmer molecular cloud  leads to the HNC deficiency.

\subsection{Line profile}

The line profile can be taken as a diagnostic tool to investigate the properties of CND molecular gas. 
\cite{Usero04} found that all molecular line profiles in the CND of NGC\,1068 are distributed asymmetrically.
They therefore suggested that the east and west knots of the CND are chemically differentiated.

Previous interferometer observations of CO\,$(1-0)$ showed that NGC\,1068 has a circumnuclear molecular ring with 
diameter ranging from $\sim$20$''$ to 40$''$ \citep{Schinnerer00}.  
The beam size of the IRAM 30m telescope at the 3 mm band is larger than the inner diameter of the two spiral arms of NGC\,1068. 
In contrast, the beam sizes at the 2 mm and 1 mm bands are smaller than the inner diameter of the the two spiral arms. 
The HPBW of HOC$^{+}$\,$(1-0)$ and $(2-1)$, for instance, is about 27.5$''$ at 89.5~GHz and 13.7$''$ at 179.0~GHz, as shown in Fig.~\ref{f_region}.
High-resolution images show that the most prominent emission stems from the CND \citep{Takano14,Viti14}.
Therefore, the molecular lines lying within the 2 mm and 1 mm bands can preclude most of the contamination from the two spiral arms. 
Using a similar method as \cite{Usero04}, 
we divided each line into its blue and red velocity components  (see Fig. \ref{CNDspectra}), 
which dominantly arise from the east and west knots, respectively. 
The line profiles in Fig. \ref{CNDspectra} show that the local systemic velocity 
of these lines ranges from  $v - v_{sys}$ = $-$200  km s$^{-1}$ to  200 km s$^{-1}$. 
We defined the blue and red components as those lying
within an interval of $-200 < v - v_{sys} < 0$ km s$^{-1}$ and $0 < v - v_{sys} < 200$ km s$^{-1}$, respectively.
The velocity-integrated intensities of the two components are listed in Table \ref{tab_ratio}. 
As shown in Fig. \ref{CNDspectra}, there are noticeable differences between the line shapes of the CND spectra. 
Some molecular lines, including HC$_{3}$N\,$(18-17)$, SiO\,$(5_5-4_4)$, SO$_{2}$\,$(14_{3,11}-14_{2,12})$, HNC\,$(2-1)$, 
HCN\,$(2-1)$, HCO$^{+}$\,$(2-1)$, HNCO\,$(11-10)$, $^{13}$CCH\,$(2-1)$, and CH$_{3}$CN\,$(9-8)$, 
show a stronger blue component, while the others, 
including SO\,$(5_5-4_4)$, HOC$^{+}$\,$(2-1)$, CS\,$(5-4)$, CH$_{3}$CCH\,$(14_{k}-13_{k})$, and H$_{2}$CO\,$(3_{1,2}-2_{1,1}),$ 
show a stronger red component.

Interferometer observations showed that the SiO\,$(3-2)$ transition is  dominantly distributed in the east knot and the HNCO\,$(6-5)$ emission in the west knot \citep{Kelly17}.
Therefore, we hypothesize that 
the molecular species whose blue-to-red flux ratios are higher than that of SiO are mainly located in the east knot, 
while those with blue-to-red flux ratios lower than that of HNCO are mainly located in the west knot.
Our single-dish observations of SiO\,$(5_5-4_4)$ and HNCO\,$(11-10)$
suggest a blue-to-red flux ratio (R$_{\rm E/W}$) of 2.1 and 1.3, respectively.
The blue-to-red flux ratio of HC$_{3}$N is higher than that of SiO, and thus 
is mainly located in the east knot.
The blue-to-red flux ratios of the  $^{13}$CCH\,$(2-1)$, CH$_{3}$CN\,$(9-8)$, SO\,$(5_{5}-4_{4})$, HOC$^{+}$\,$(2-1)$, CS\,$(5-4)$, 
CH$_{3}$CCH\,$(14_k-13_k)$, and H$_{2}$CO\,$(3_{1,2}-2_{1,1})$ transitions are lower than that of the HNCO\,$(11-10)$ lines, and thus mainly
arise from the west knot. The
$^{13}$CCH\,$(2-1)$, CH$_{3}$CN\,$(9-8)$, and CH$_{3}$CCH\,$(14-13)$ transitions include several multiple components, 
which may affect the determination of their blue-to-red flux ratios.
However, as shown in Fig. \ref{CNDspectra}, the strongest components have small frequency splits, suggesting that this effect can be neglected.

In Fig. \ref{figure_three} we compare the line profiles of HCN, HNC, HCO$^{+}$, and HOC$^{+}$  and those of the $^{13}$C isomers.
We find that the ($J = 2 - 1$)-to-($J = 1 - 0$) flux ratios of HCN, HNC, and HCO$^{+}$ are 
higher for the blue components than those for the red components, while there is no 
significant difference for HOC$^{+}$. Under the assumption of
$^{12}$C/$^{13}$C, \cite{Wang14} suggested that the redshifted parts of HCN\,$(1-0)$ and HCO$^{+}$\,$(1-0)$ 
have lower optical depths and more likely come from the spiral arms than from the nuclear region.  
Because the beam size of $J = 2 - 1$  is much smaller than that of  $J = 1 - 0$, 
the different line profiles between the $J = 2 - 1$ and the $J = 1 - 0$ transitions may be attributed to
the difference of their emission regions.
The similar line profiles between the $J = 2 - 1$ lines of HCN and HCO$^{+}$ and their isotopic lines 
(H$^{13}$CN\,$(1-0)$ and H$^{13}$CO$^{+}$\,$(1-0)$) 
suggest that  they may come from the same location.  
The $J = 2 - 1$ and $J = 1 - 0$ transitions of  HOC$^{+}$ show similar profiles,
suggesting that the molecule might  be distributed in the CND region.

HCN and HCO$^{+}$ are tracers of dense gas. 
Their 
 $J = 2 - 1$ transitions show similar line profiles and comparable velocities (see the lower panels of Fig. \ref{figure_three}). 
Compared to the HCN\,$(2-1)$ line, the HNC\,$(2-1)$  line 
has a  similar line center velocity, but a noticeably narrower line width.
The HOC$^{+}$\,$(2-1)$-to-HCO$^{+}$\,$(2-1)$ flux ratio in the blue component is higher than that in the red component.  
However, this is not the case for HNC and its isomer HCN.
A possible explanation is that the $J = 2 - 1$ transitions of HCN, HNC, and HCO$^{+}$ are associated with each other 
in the CND of NGC\,1068, while the HOC$^{+}$\,$(2-1)$ transition does not peak at the same position.
This is consistent with the latest interferometer observation toward the center of 
 the Seyfert II galaxy Mrk\,273 \citep{Aladro18}, which shows  that the peak of the HOC$^{+}$\,$(3-2)$ 
emission is not at the same position as the dense gas of HCN and HCO$^{+}$. 
Further observations with higher spatial resolution of HOC$^{+}$ in the CND of NGC\,1068 are required to investigate the origin of this offset.

\subsection{Rotation diagram}

We applied the rotation diagram to derive excitation temperatures ($T_{\rm ex}$) and column densities ($N_{\rm tot}$) 
of the molecules we detected in our observations.
With the assumption of local thermodynamic equilibrium (LTE), optically thin conditions, and negligible background temperature,
we plot the populations of the upper levels ($N_{u}$) against the corresponding excitation energies ($E_{u}$) of the transitions 
(Fig. \ref{f_RT}), using the equation

\begin{equation}
{\rm ln} \frac{N_u}{g_u} = {\rm ln} \frac{8 \pi k \nu ^2 \int T_s d v }{h c^3 A_{ul} g_u} = {\rm ln} \frac{N_{\rm tot}}{Q(T_{\rm ex})} - \frac{E_u}{kT_{\rm ex}},
\end{equation}
where $k$ is the Boltzmann constant, 
$\nu$ is the rest frequency of the transition, 
$Q(T_{\rm ex})$ is the partition function,
$h$ is the Planck constant, 
$c$ is the light speed,
$A_{ul}$ is the spontaneous emission coefficient,
$g_{u}$ is the total degeneracy of upper energy level,
and $E_{u}/k$ is the upper level energy.
The values of $Q(T_{\rm ex})$, $A_{ul}$, $g_{u}$, and $E_{u}/k$ are taken from 
the Cologne Database for Molecular Spectroscopy (CDMS) catalog\footnote{https://cdms.astro.uni-koeln.de}
and the splatagogue astronomical spectroscopy database\footnote{http://www.cv.nrao.edu/php/splat/}.
$\int T_{s} \, d \nu$ is the detected transition-integrated intensity. 
$T_{s}$ is the source-averaged brightness temperature, corrected for beam dilution by 
$T_{s} = T_{\rm mb}(\theta ^{2}_{b} + \theta ^{2}_{s})/ \theta ^{2}_{s}$, where $\theta_{b}$ is the antenna HPBW and $\theta_{s}$ is the source size.
The antenna HPBW is estimated by HPBW($''$) = 2460/$\nu$(GHz). 
 Following the assumption made by other researchers (e.g., \citealt{Usero04,Bayet09,Krips08,Aladro13}), 
 we took a  $\theta_{s}$ value of 4$''$ that was obtained based on the interferometric observations of $^{12}$CO, HCN, and $^{13}$CO \citep{Helfer95,Schinnerer00}.  

The rotation diagrams of the 14 molecular species detected in our observations are shown in Fig. \ref{f_RT}. 
We combined previously published data for this plot. 
The derived values of $T_{\rm ex}$ and $N_{\rm tot}$ are listed in Table \ref{table_RT}. 
Two components were considered to fit the rotation diagrams of
8 molecules, including CS, HCN, HCO$^{+}$, HNC, SiO, CN, and CH$_{3}$OH.   
Alternatively, the nonlinear data distribution in the rotation diagrams can be caused by   
finite optical depths \citep{Goldsmith99}. 
In this scenario, Eq. 1 is modified to 
\begin{equation}
{\rm ln} \frac{N_u}{g_u} =   {\rm ln} \frac{N_{\rm tot}}{Q(T_{\rm ex})} - \frac{E_u}{kT_{\rm ex}} - {\rm ln\,C_\tau},
\end{equation}
where the optical depth correction factor $\rm ln\,C_\tau$ is expressed with $\rm ln\,C_\tau = ln\,[\tau/(1-e^{-\tau})]$.
However, using the method developed by \cite{Goldsmith99}, we found that the C$_\tau$ value ($< 0.005$) is too low to fit the
rotation diagrams.

The rotation temperatures of CS are consistent with the results derived by
 \citet{Aladro13} within the errors, whereas the cold components of HC$_{3}$N
and HCN have slightly higher rotation temperatures than the values obtained 
by \citet{Aladro13}. For HCO$^{+}$, HNC, and SiO, the rotation temperatures derived
by \citet{Aladro13} lie between those of the cold and warm components.
Because the SO\,$(5_{6}-4_{5})$ and CH$_{3}$CN\,$(12-11)$ transitions are strongly blended with the wings of the strong 
$^{13}$CO\,$(2-1)$ and C$^{18}$O\,$(2-1)$ transitions, 
their fluxes represent upper limits; see \cite{Qiu18}.
These two transitions were excluded in fitting the rotational diagrams.
The rotation temperature of SO is
lower than that derived by \citet{Aladro13} ($\sim$22.8 $\pm$ 18.6 K).
Because they deviate from the fitting lines, 
we did not use the transitions of 
SO$\,(5_4-4_4)$, HNCO$\,(6_{1,6}-5_{1,5})$, $(5_{1,4}-4_{1,3})$, and $(5_{1,5}-4_{1,4})$  
in the rotation diagrams.

\subsection{Molecular transitions detected with different single-dish telescopes}
\label{sec3.4}

Some of the molecular transitions have been detected by single-dish telescopes with different beam sizes.
NGC\,1068 consists of a starburst ring and a CND. 
As shown in Fig. 1 of \citet{Takano19}, 
the beam of the IRAM 30m telescope covers the CND and starburst ring at the 3 mm band, 
while the beam of the NRO 45m telescope only covers the CND. 
In the 1 and 2 mm windows, the JCMT beam covers the CND and starburst ring, while 
the beam of the IRAM 30m telescope only covers the CND.  By comparing the main-beam temperatures
of a certain transition detected by different telescopes, we could roughly estimate the size of
the regions from which this molecule arises. This approach has been used by \citet{Takano19}
to study the molecular emission regions in NGC\,1068 and NGC\,253. The authors concluded that
the distributions of molecules between these two galaxies are significantly different.
Combining these results with the data in the literature, we calculated the ratios  between
the observed intensities by different telescopes for 28 transitions (see Table \ref{flux_ratio}).

When the beams cover the same emission regions (i.e., the CND), the intensities measured by the telescope with
the wider beam should be lower because of the beam-dilution effect, and thus the ratios shown in
Fig. \ref{figure10} should be higher than unity. However, we find that the ratios for
$^{13}$CO\,$(1-0)$, CH$_{3}$OH\,$(2_k-1_k)$, HNC\,$(1-0)$, $^{13}$CO\,$(2-1)$, HCN\,$(3-2)$, C$^{18}$O\,$(1-0)$, 
CO\,$(1-0)$, SiO\,$(2-1)$, HCO$^{+}$\,$(1-0)$, and CS\,$(2-1)$ are lower than unity, 
suggesting that the surrounding starburst ring may significantly contribute to these transitions.
Theoretically, the line intensity ratio decreases with  increasing  percentage 
of the line emission from the starburst ring. 
The lines with intensity ratios much higher than unity may be mainly distributed in the CND.
This is supported by  interferometer observations. 
For instance, $^{13}$CN\,$(1_{1/2}-0_{1/2})$,  HC$_3$N\,$(11-10)$,  
HNCO\,$(5_{0,5}-4_{0,4})$, and CH$_3$CN\,$(6_k-5_k)$ have high intensity ratios.
The ALMA observations of \citet{Takano14} showed that they are concentrated in the CND.
Significant contributions from the ring have been revealed by ALMA for the $^{13}$CO\,$(1-0)$ emission,
which has a low intensity ratio  (Fig. \ref{figure10}).
The intensity ratio of the CS\,$(2-1)$ line is close to unity, again in agreement with the ALMA observations
that reveal low contributions from the ring.

Assuming that the telescopes with different beam sizes detect the same molecular regions,  we can
derive the source size by the expression
 \begin{equation}
\theta_s = \sqrt{ \frac{I_2\, \times\, \theta_{b2}^2\, -\, I_1\, \times\, \theta_{b1}^2}{ I_1\, -\, I_2} },\end{equation} 
where $I_1$ and $I_2$ are the integrated intensities detected with two different telescopes,
and $\theta_{b1}$ and $\theta_{b2}$ are the beam sizes of the corresponding telescopes.
As shown in Fig. \ref{figure10} and Table \ref{flux_ratio}, the estimated source sizes
are generally larger than 4$''$. However, these calculations
sensitively depend on the intensity ratio and on the assumption of a Gaussian brightness
distribution. Low pollution from outside of the CND may substantially increase
the estimated $\theta_s$ value.  Nevertheless, the assumption of $\theta_s=4$$''$ does not significantly
affect the results we obtain from the rotation diagrams. With $\theta_s$ increasing by a factor of two, 
the derived column densities would decrease by about 10\%, while the rotation temperatures
would be maintained at the same level.

\section{Discussion}
\label{sec4}

\subsection{HOC$^{+}$ in XDR}
\label{HOCp}
Based on  the observations of HCN and its isomer HNC in interstellar clouds, 
\cite{Herbst76} hypothesized that if HNC is present in interstellar clouds, then HOC$^{+}$, 
the energetically disfavored HCO$^{+}$ isomer, is also present.
Subsequently, the HOC$^{+}$\,$(1-0)$ transition was detected toward Sgr B2 by \cite{Woods83} and \cite{Ziurys95}, 
based on the experimental work of \cite{Gudeman82}.
So far, HOC$^{+}$ has been detected in diverse environments in our Galaxy and external galaxies, 
including diffuse clouds \citep{Liszt04}, an ultracompact HII region \citep{Rizzo03}, PDRs \citep{Apponi99,Fuente03}, 
dense molecular clouds \citep{Apponi97}, 
 starburst galaxies 
 \citep{Fuente05, Martin09a, Aladro15}, and Seyfert galaxies \citep{Usero04,Aalto15,Aladro18}.  
 These observations suggest that HOC$^{+}$ is widespread in the interstellar medium, with 
 the [HOC$^{+}$]-to-[HCO$^{+}$] ratio varying from tens to some thousand.
The formation routes of HOC$^{+}$ and its isomer HCO$^{+}$ were described by \citet{Aladro18} and references therein.
Interferometer observations toward Mrk 273, a nearby Seyfert II galaxy, show a
 global HCO$^{+}$/HOC$^{+}$ ($J = 3 - 2$) brightness temperature ratio of
 9 $\pm$ 4 and nuclear ratio of 5 $\pm$ 3 \citep{Aladro18}.
HOC$^{+}$ was tentatively detected toward the nucleus of Mrk 231, which hosts a powerful AGN, 
with the HCO$^{+}$/HOC$^{+}$ ($J = 3 - 2$) brightness temperature ratio ranging from 10 to 20 \citep{Aalto15}. 
These values are generally 
lower than those in other sources, suggesting  that the AGN might be  associated
with enhanced HOC$^{+}$.

We detect the $J = 2 - 1$ transitions of both HCO$^{+}$ and HOC$^{+}$ in the CND of NGC\,1068, 
allowing us to further investigate the association between AGN and the HOC$^{+}$ enhancement.
 Figure \ref{f_region} presents the CO\,$(1-0)$ map overlaid by beam sizes
of the IRAM 30m telescope at the HOC$^{+}$\,$(1-0)$ and $(2-1)$ lines.
The HPBW at the HOC$^{+}$\,$(1-0)$ line is larger than the inner diameter of the starburst ring, 
while that of the HOC$^{+}$\,$(2-1)$ line is smaller.
This means that the  HOC$^{+}$\,$(2-1)$ emission dominantly arises from the CND region and not 
from the two spirals that are dominated by star formation.
The situations are the same with the HCO$^{+}$\,$(1-0)$ and $(2-1)$ transitions.
Figure \ref{f_hoc} shows the transitions of $J = 1 - 0$ and $J = 2 - 1$ of HCO$^{+}$ and HOC$^{+}$, 
as well as the HCO$^{+}$-to-HOC$^{+}$ temperature ratio profiles.
The HCO$^{+}$-to-HOC$^{+}$ temperature ratio of each velocity channel ranges from 32 to 109 and  from 7 to 27 
for the $J = 1 - 0$  and $J = 2 - 1$ transitions, respectively.
These low values are consistent with the results of
\citet{Usero04}, who suggested that the high electron density in the XDR may lead to a low
[HCO$^{+}$]/[HOC$^{+}$] ratio.  
HCO$^{+}$ and HOC$^{+}$ in the XDR are produced through three chemical paths
(CO + H$_3^+$, CO$^+$ + H$_2$, and H$_2$O + C$^+$) \citep{Maloney96,Sternberg95}.
The HCO$^{+}$ and HOC$^{+}$ fractions formed in these paths were illustrated
by \citet{Usero04} (see their Fig.~8). Our detections suggest that the integrated
intensity ratio of HCO$^{+}$/HOC$^{+}$\,($J = 2 - 1$)  is 
15.8 $\pm$ 3.0, which indicates
an electron abundance with n(e$^-$)/n(H$_2$) = 10$^{-4.6}$
according to the XDR model of \citet{Usero04}. The CO + H$_3^+$ and H$_2$O + C$^+$
routes dominate the production of HCO$^{+}$ and HOC$^{+}$.
The high ionization degree suggests strong X/UV irradiation powered by AGN.

As shown in Fig. \ref{f_hoc}, the  profiles of the HOC$^{+}$\,$(1-0)$
and $(2-1)$ lines are noticeably asymmetrical with respect to $v_{sys}$.
The asymmetry can be quantified by  the ratio between the integrated
fluxes in the blue and red parts relative to the central frequency.
In Table \ref{tab_ratio} we compare  the blue-to-red flux ratio of
HOC$^{+}$ line with those of other lines. We find that the ratio for
HOC$^{+}$\,$(2-1)$ is 0.93, differing from those of HNC\,$(2-1)$ ($\sim$1.70), 
HCN\,$(2-1)$ ($\sim$1.47), and HCO$^{+}$\,$(2-1)$ ($\sim$1.38). This might suggest that
HOC$^{+}$ 
has a different distribution than 
 HNC, HCN, and HCO$^{+}$.
We note that the spatial resolved observations toward Mrk 273 
also showed different peak positions of the HOC$^{+}$ ,
HNC, HCN, and HCO$^{+}$ lines \citep{Aladro18}. Therefore, we hypothesize
that the formation of HOC$^{+}$ is more sensitive to the environments of
the XDR than the formation of other molecules.
Future interferometer observations of HOC$^{+}$ in NGC\,1068 would allow a firmer conclusion.

\subsection{Comparison of the molecules in NGC\,1068,  M82, and NGC\,253}

To investigate the roles of AGN and starburst on molecular chemistry, we
compared the molecular lines detected in NGC\,1068 and in the two nearby typical starburst galaxies M\,82 and NGC\,253.
M\,82 is a prototype starburst galaxy with a high star formation rate ($\sim$9 M$_\odot$ yr$^{-1}$, \citealt{Strickland04}) 
at a distance of $\sim$3.6 Mpc \citep{Freedman94}.  
NGC\,253 is an almost edge-on barred spiral galaxy with an active nuclear starburst 
($\sim$3.6 M$_\odot$ yr$^{-1}$, \citealt{Strickland04} at a distance of $\sim$3 Mpc \citep{Mouhcine05}.  
The star formation in M\,82 and NGC\,253 is in the middle and later stages, respectively.  
The unbiased line surveys of NGC\,253 \citep{Martin06} and M\,82 \citep{Aladro11} 
at the millimeter  band have revealed that the two galaxies have clearly different chemical
characteristics.
In Figs. \ref{tvsname} and \ref{tvsn} we compare
the rotation temperatures, column densities, and fraction abundances of nine 
detected molecular species in three galaxies, including
SiO, CH$_{3}$CN, HC$_{3}$N, SO$_{2}$, SO, HNCO,  CS, CH$_{3}$OH, and CH$_{3}$CCH. 
 The molecular rotation temperatures in NGC\,1068  are generally lower than the temperatures
in M\,82 and NGC\,253. The only exception is the warm component of CS.
CS is a dense-gas tracer. Its high temperature probably suggests that  
the IR-photon heating, resulting from dust emission, is effective inside of the CS molecular cloud.  
Therefore, we conclude that 
the interior of the dense molecular cloud in 
the CND of NGC\,1068 is more severely obscured by dust than the interiors
in  M\,82 and NGC\,253. 
The heavy dust obscuration significantly shield molecules 
from a strong UV/X-ray radiation field, leading to the enhancement of complex molecules
such as CH$_{3}$OCH$_{3}$.

As shown in the upper panel of Fig. \ref{tvsn},
the column densities of molecular species in NGC\,1068 
are higher than those in M\,82 and NGC\,253 by one to two orders of magnitude,
 suggesting that NGC\,1068 is a rich reservoir of molecules.
The fraction abundance of SiO in NGC\,1068 is clearly larger than those in 
M\,82 and NGC\,253 (see the lower panel of Fig. \ref{tvsn}).
This is consistent with the interferometer observations of SiO\,$(3-2)$,
which suggest
that SiO is enhanced by a fast shock in the east knot of CND of NGC\,1068 
\citep{Kelly17}.

The [c-C$_{3}$H$_{2}$]/[HC$_{3}$N]  ratio can be used to trace the evolutionary stages 
of the star formation \citep{Fuente05}.
Taking the column density of c-C$_{3}$H$_{2}$ obtained by \cite{Nakajima11},
we derive [c-C$_{3}$H$_{2}$]/[HC$_{3}$N] = 0.07 for NGC\,1068, 
which is lower than that in M\,82 ($\sim2$) and NGC\,253 ($\sim0.2$). 
This suggests that the UV radiation field plays a less important  role in the chemistry 
of NGC\,1068 than that in M\,82 and NGC\,253, 
where HC$_{3}$N
is significantly dissociated and the cyclic molecule c-C$_{3}$H$_{2}$ is more resistant to UV photons.

CH$_{3}$CCH has a  small dipole moment (0.75 Debye), and thus its rotation 
temperature should be close to the kinetic temperature of the molecular cloud.
It follows that the kinetic temperature in NGC\,1068  is the lowest of the
three galaxies.
Through a large velocity gradient (LVG) analysis, \citet{Krips08} obtained 
that the kinetic temperature of M\,82 is 60--100 K. However, there are two solutions 
for the kinetic temperature of NGC\,1068 (20 K and 60--240 K).
The lower solution is close to the rotational temperature of CH$_{3}$CCH (28.0 $\pm$ 3.4 K).

The profile of the SO\,$(5_5-4_4)$ line suggests that SO is concentrated in the west knot of CND.  
Based on models of ice evaporation process, \citet{Viti04} suggested that 
SO is much more abundant than SO$_{2}$ at the beginning of the high-mass star formation, 
while the reverse is the case in the hot core. 
The  ALMA  1.1 mm continuum emission showed
that an ongoing star formation signature existed in the southwestern direction of the AGN with 
a distance of about 2 arcsecond \citep{Imanishi16}.
Therefore, it is likely that SO is enhanced by star formation
in the west knot of the CND in NGC\,1068.
In ocntrast,  the blue-to-red flux ratio of the SO$_{2}$\,$(5_5-4_4)$ 
line ($\sim$2.05) is close to that of the SiO\,$(55-44)$ line ($\sim$2.08). 
Therefore, we infer that SO$_{2}$ is more concentrated in the east knot.
Figure \ref{tvsn} shows that SO$_{2}$ is more abundant than SO in both NGC\,1068 and NGC\,253. 
This can be explained by the chemical model of  \citet{Viti04}, according to which SO$_{2}$ is more abundant than SO within the post-shock gas.
Previous observations have shown that 
the nuclear molecular clouds of NGC\,253 are dominated by large-scale low-velocity shocks \citep{Martin06} 
and that the east and west knot of NGC\,1068 are dominated by a fast and a slow shock, respectively \citep{Kelly17}.
Figure \ref{tvsn} also shows that the abundance ratio of SO/SO$_{2}$ in NGC\,1068 is lower than that in NGC\,253.
SO$_{2}$ can be photodissociated by interstellar UV radiation to form SO \citep{Willacy97}.
The lower [SO]/[SO$_{2}$] ratio in NGC\,1068 is consistent with our conclusion that 
the UV radiation field plays a less important role in the chemistry of NGC\,1068 than in that of NGC\,253.

\section{Summary}
\label{sec5}

We reported an observation toward the nuclear region of NGC\,1068 at the 2 mm bands.
The chemical properties of the CND in NGC\,1068 were investigated.
Our main conclusions are as follows.

1. Fifteen emission lines toward the center of NGC\,1068 are detected,
eight of which are  the first detection in this source.  
Based on the line profile diagnosis of molecular gas in CDN, we infer that 
HC$_{3}$N and SO$_2$ are mainly concentrated in the east knot of the CND, 
while $^{13}$CCH, CH$_{3}$CN, SO, HOC$^{+}$, CS, CH$_{3}$CCH, and H$_{2}$CO are in the west knot of the CND.

2. The CND of NGC\,1068 is highly ionized with an ionization degree of X(e$^{-}$) $\sim$ 10$^{-4.6}$. 
The high ionization degree is consistent with the spacial distribution of HOC$^{+}$, which is  enhanced in XDR and 
 is mainly distributed in the CND of NGC\,1068 and not in the surrounding starburst ring.

3. Based on the rotation-diagram, we derived the column densities and rotation temperatures of 14 molecular species in NGC\,1068.
With these results, we find that the physical conditions and chemical environments in NGC\,1068
may significantly differ from those in NGC\,253 and M\,82: the UV radiation filed in NGC\,1068 is lower than that in M\,82 and NGC\,253, and NGC\,1068 
has the lowest kinetic temperature.

It is clear that the CND of NGC\,1068 has complex chemical environments.
This paper demonstrates that the data obtained by single-dish
telescopes can provide significant supplements to interferometer observations
in investigating the physical and chemical environments of galaxies.

\begin{acknowledgements}
We thank the anonymous referee for the useful comments that improved the manuscript.
This work was supported by the Natural Science Foundation of China (NSFC) awarded to YZ (No. 11973099) and the China Postdoctoral Science Foundation funded project (No. 2019M653144) awarded to JJQ.
JSZ thanks the support of NSFC (No. 11590782). 
LWJ appreciates support by the Guangzhou Education Bureau (No. 1201410593). 
JJQ wishes to thank Dr. Jun-Zhi Wang for his useful suggestions and Mr. Deng-Rong Lu for his help with the data reduction. 
XDT acknowledges support by the Heaven Lake Hundred-Talent Program of Xinjiang Uygur Autonomous Region of China. 
We wish to express my gratitude to the staff at the IRAM 30m telescope for their kind help and support during our observations.
\end{acknowledgements}

%
%

\clearpage

\newgeometry{a4paper, left=0.5cm,right=0.5cm,top=2.5cm,bottom=2.5cm}

\begin{figure}
    \begin{minipage}{10cm}
        \includegraphics[width=10cm]{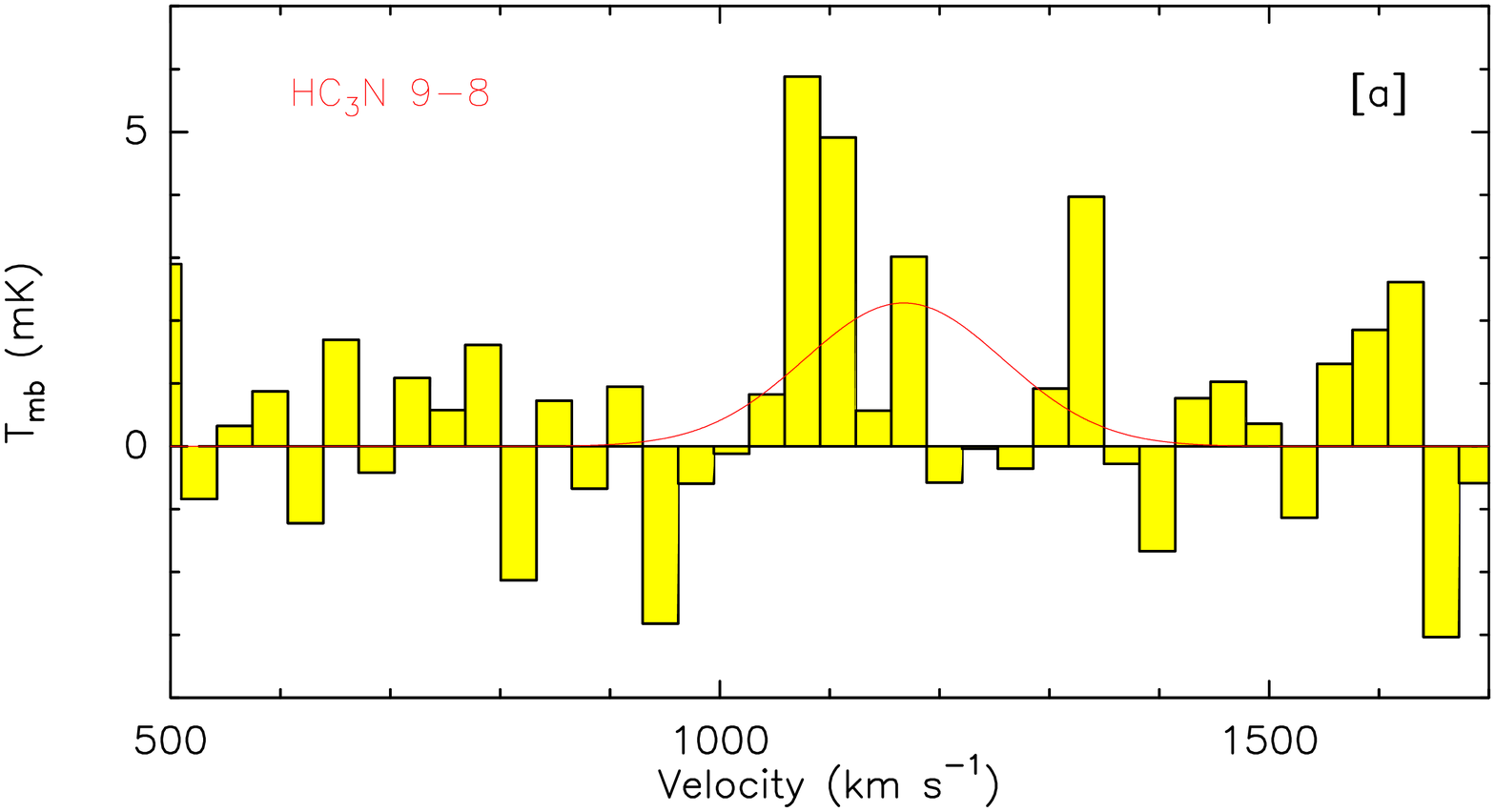}
    \end{minipage}
    \begin{minipage}{10cm}
        \includegraphics[width=10cm]{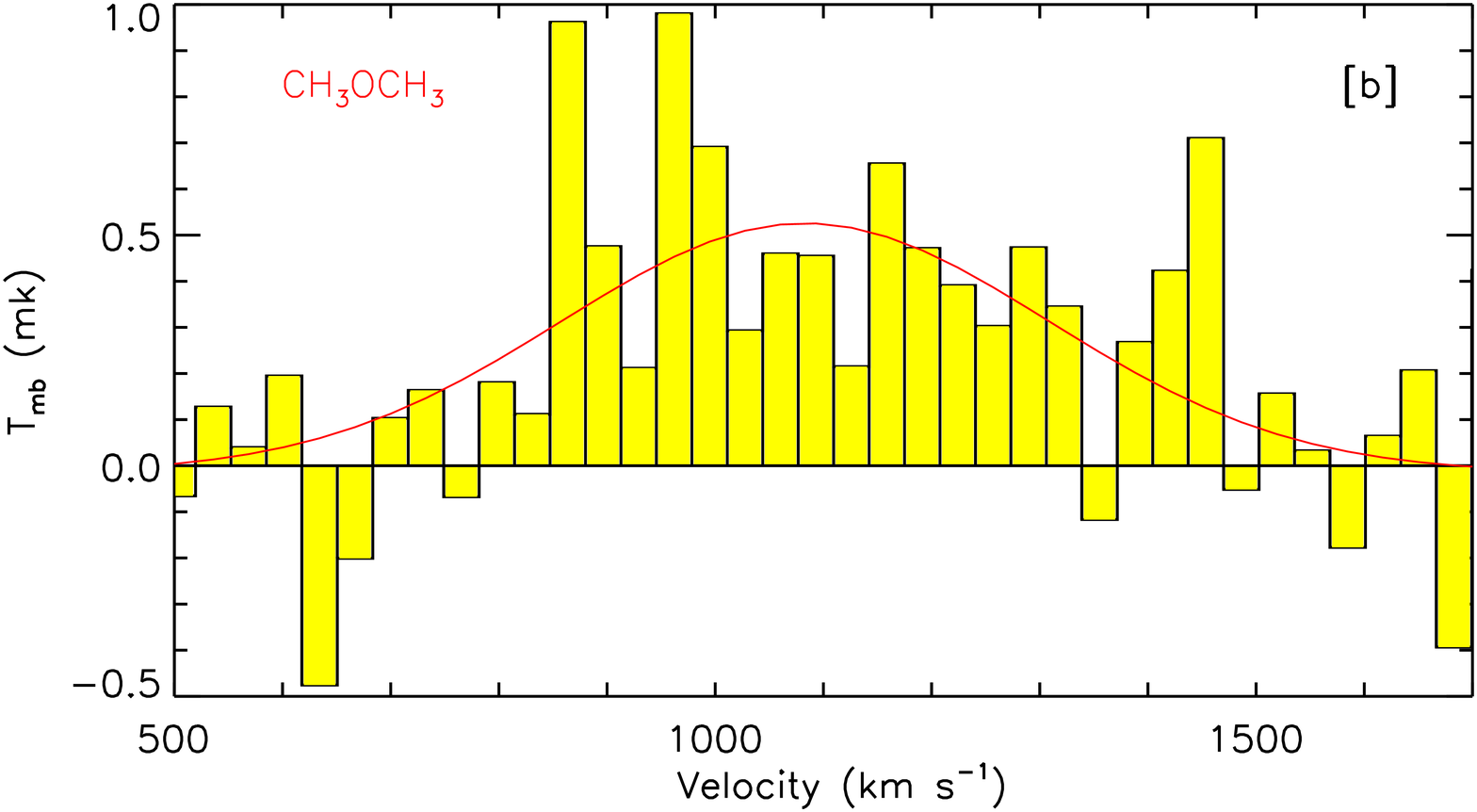}
    \end{minipage}
  
    \begin{minipage}{10cm}
        \includegraphics[width=10cm]{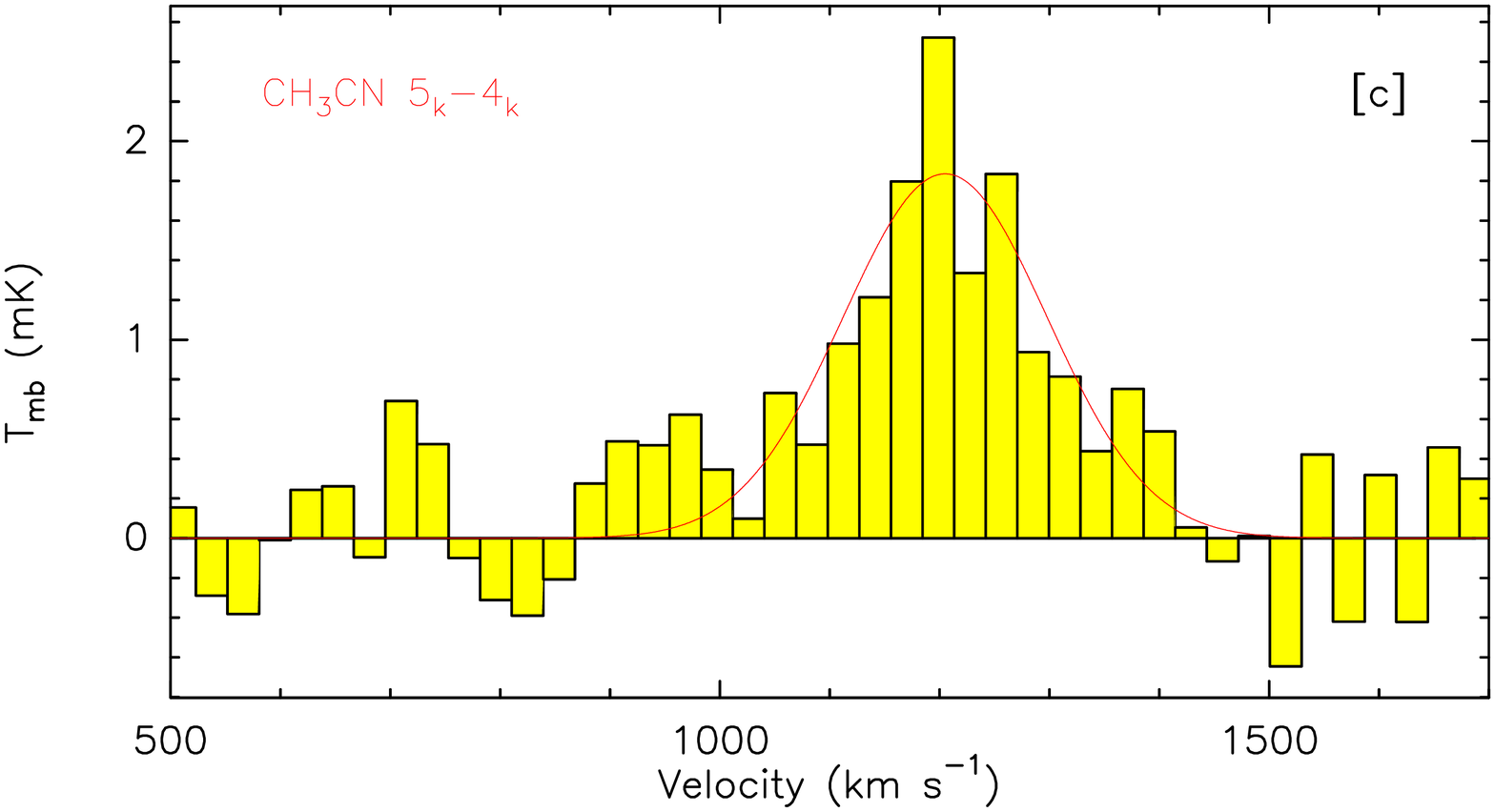}
    \end{minipage}
    \begin{minipage}{10cm}
        \includegraphics[width=10cm]{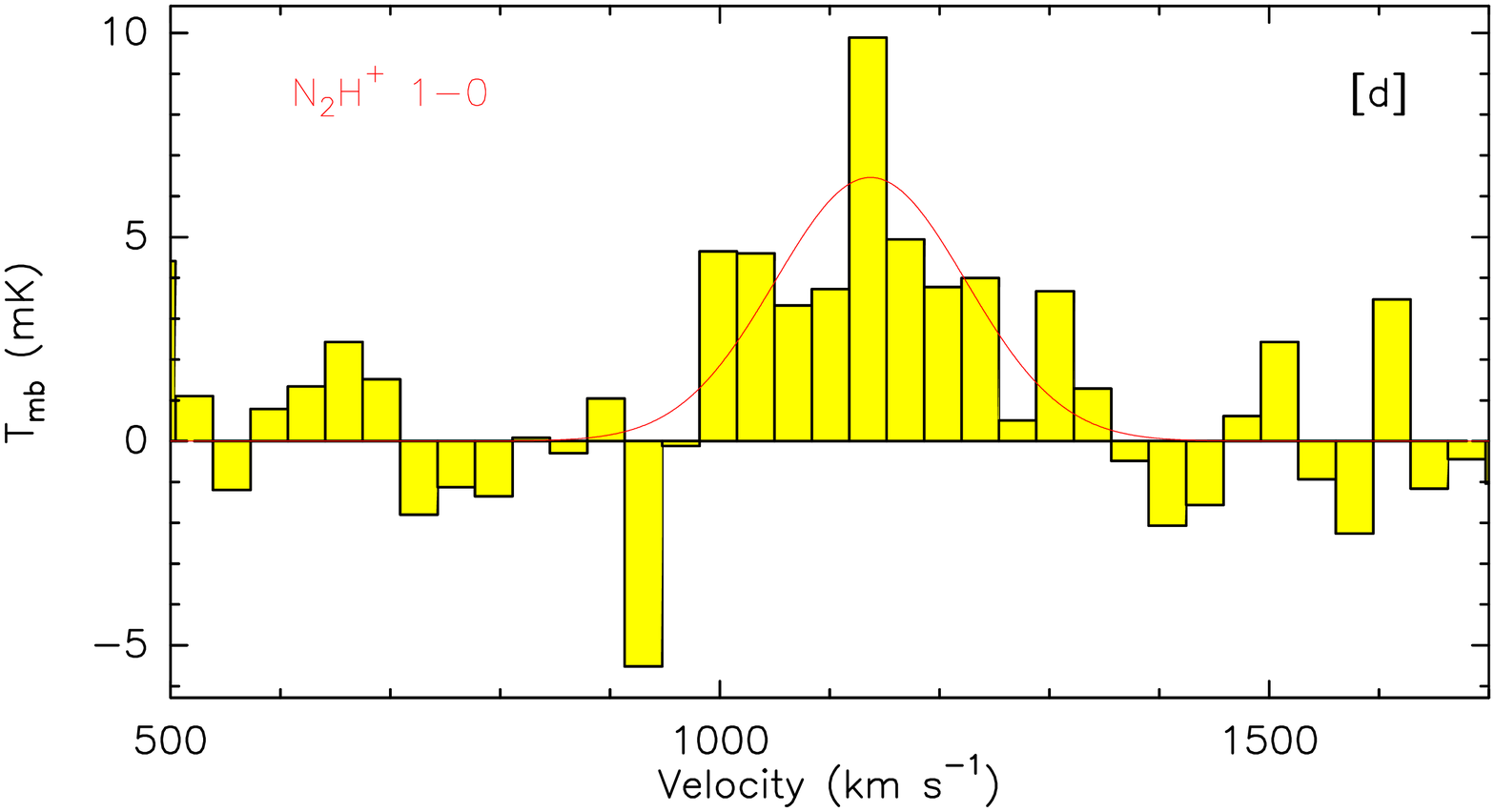}
    \end{minipage}

    \begin{minipage}{10cm}
        \includegraphics[width=10cm]{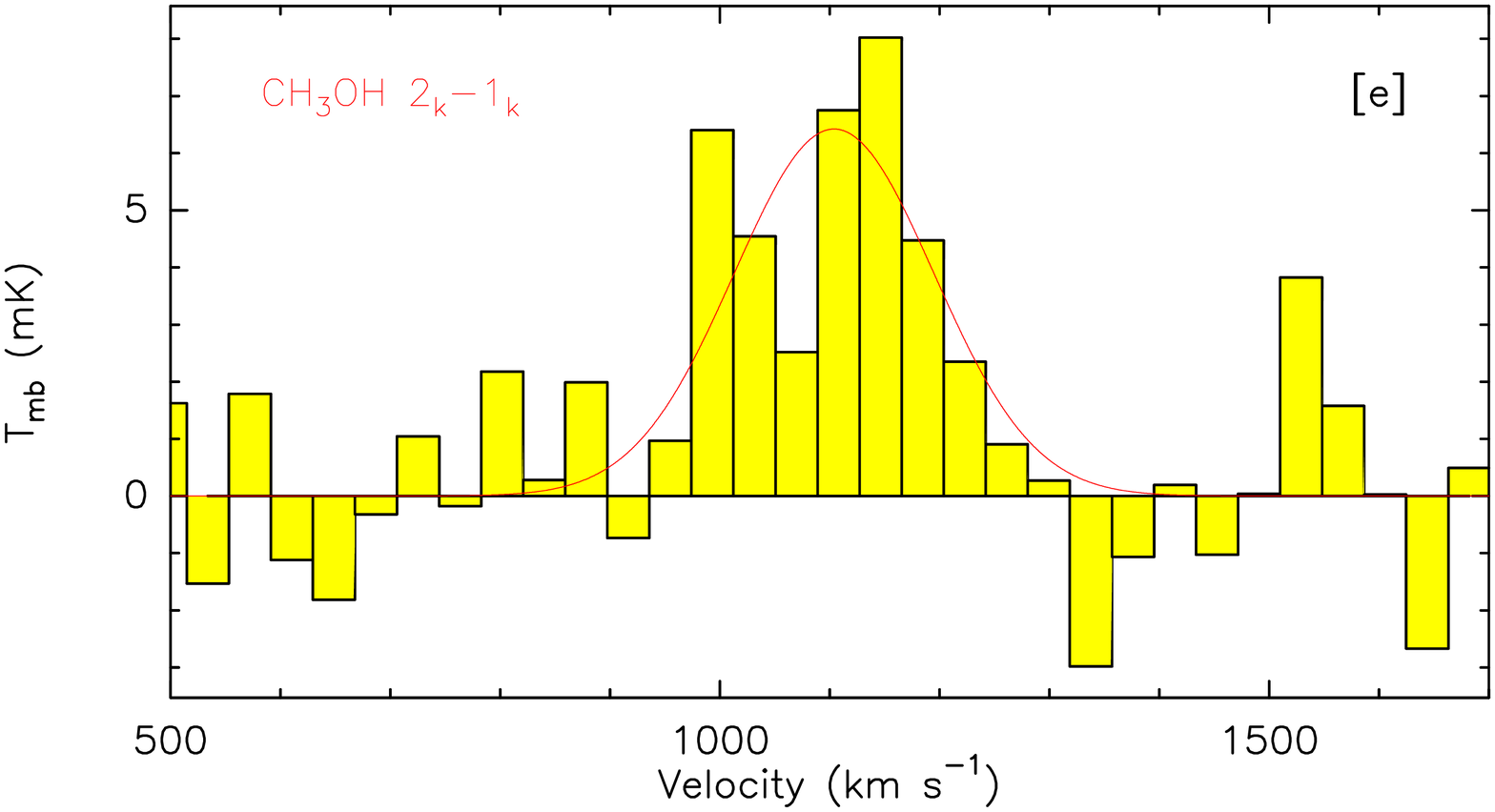}
    \end{minipage}
    \begin{minipage}{10cm}
        \includegraphics[width=10cm]{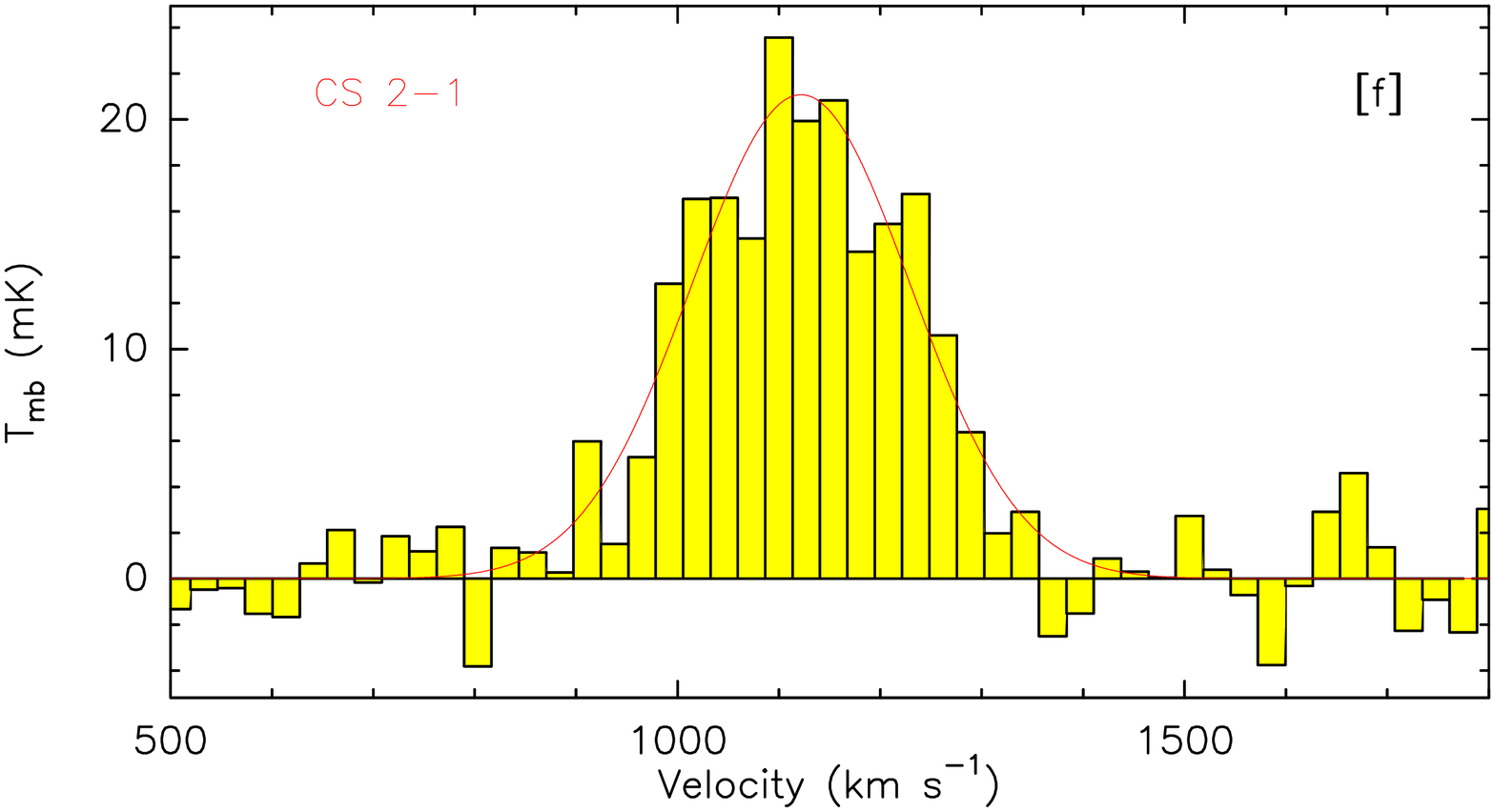}
    \end{minipage}
  
    \begin{minipage}{10cm}
        \includegraphics[width=10cm]{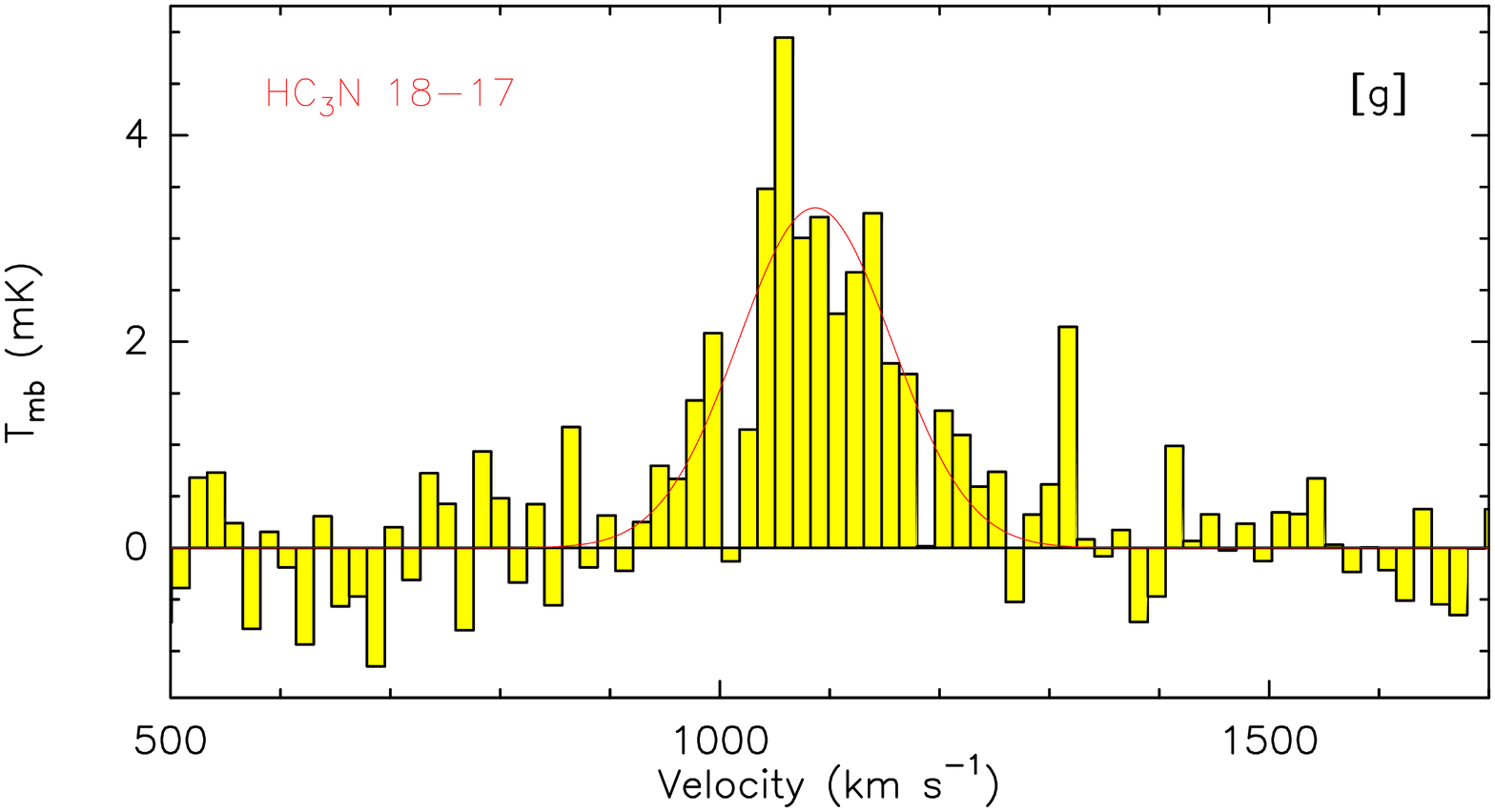}
    \end{minipage}
    \begin{minipage}{10cm}
        \includegraphics[width=10cm]{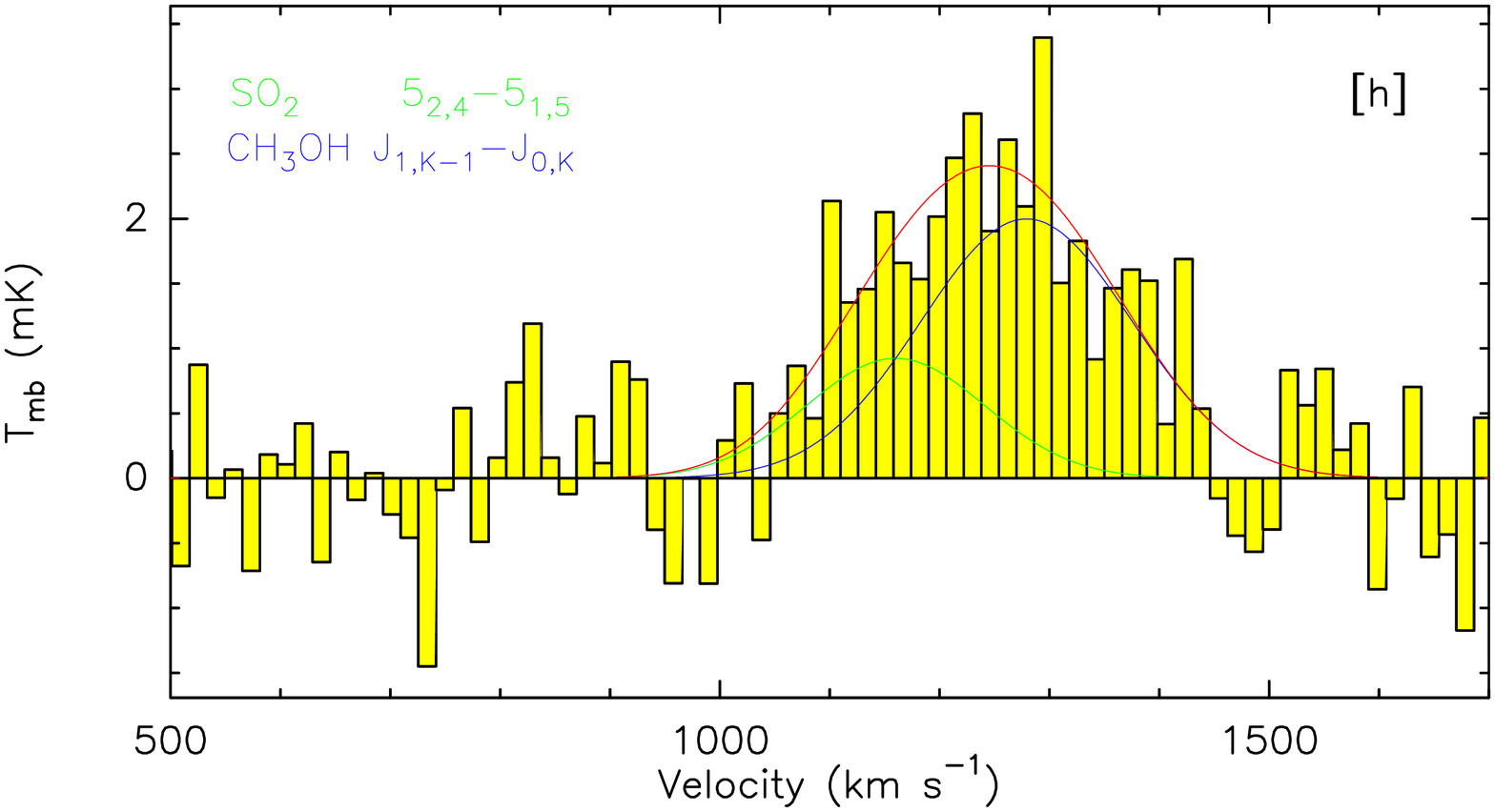}
    \end{minipage}
  
\caption{
      Molecular transitions toward NGC\,1068. The red curves
represent Gaussian fitting profiles.
}
\label{f_list_1}
\end{figure}
\addtocounter{figure}{-1}

\begin{figure}
    \begin{minipage}{10cm}
        \includegraphics[width=10cm]{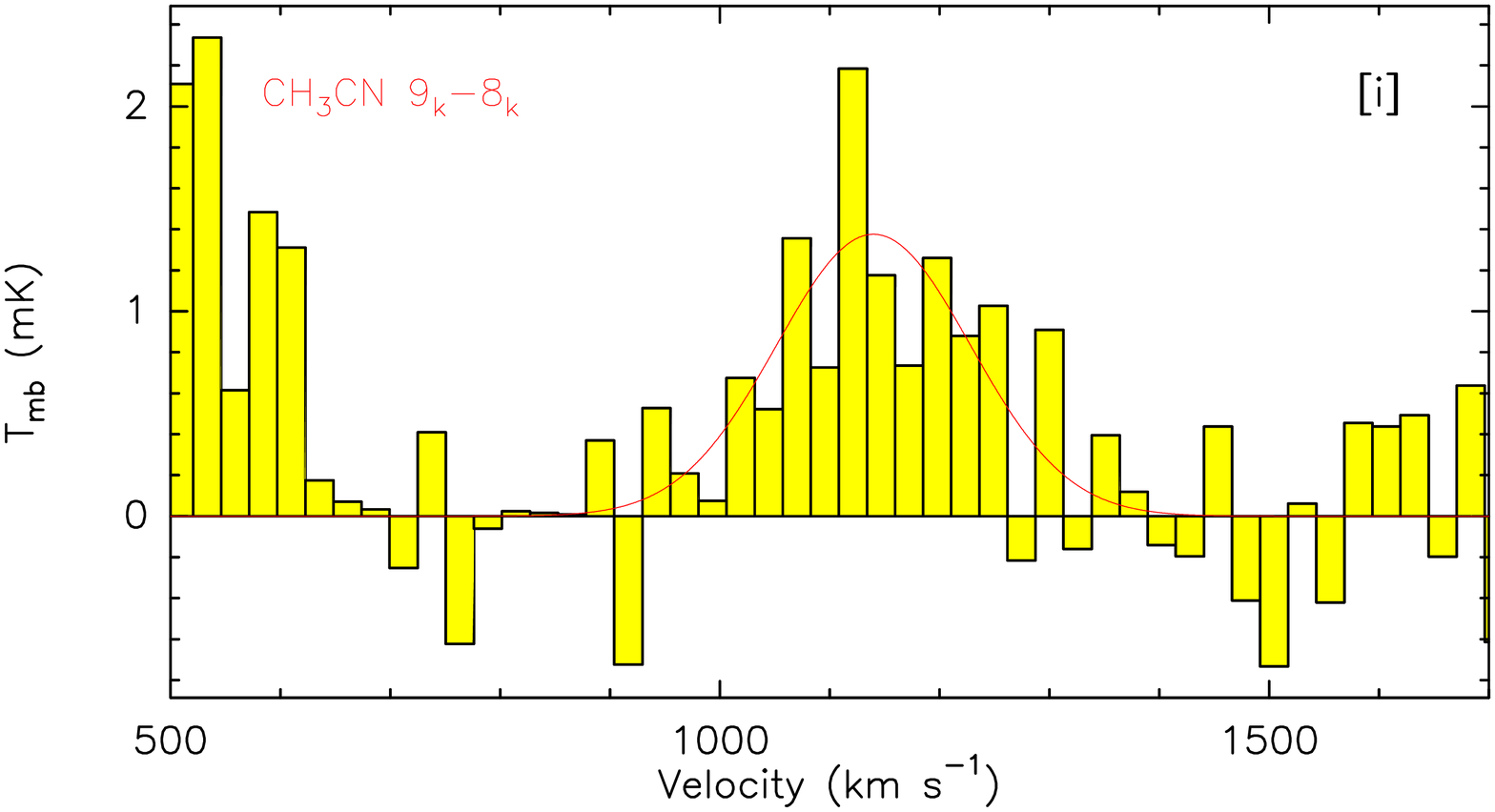}
    \end{minipage}
    \begin{minipage}{10cm}
        \includegraphics[width=10cm]{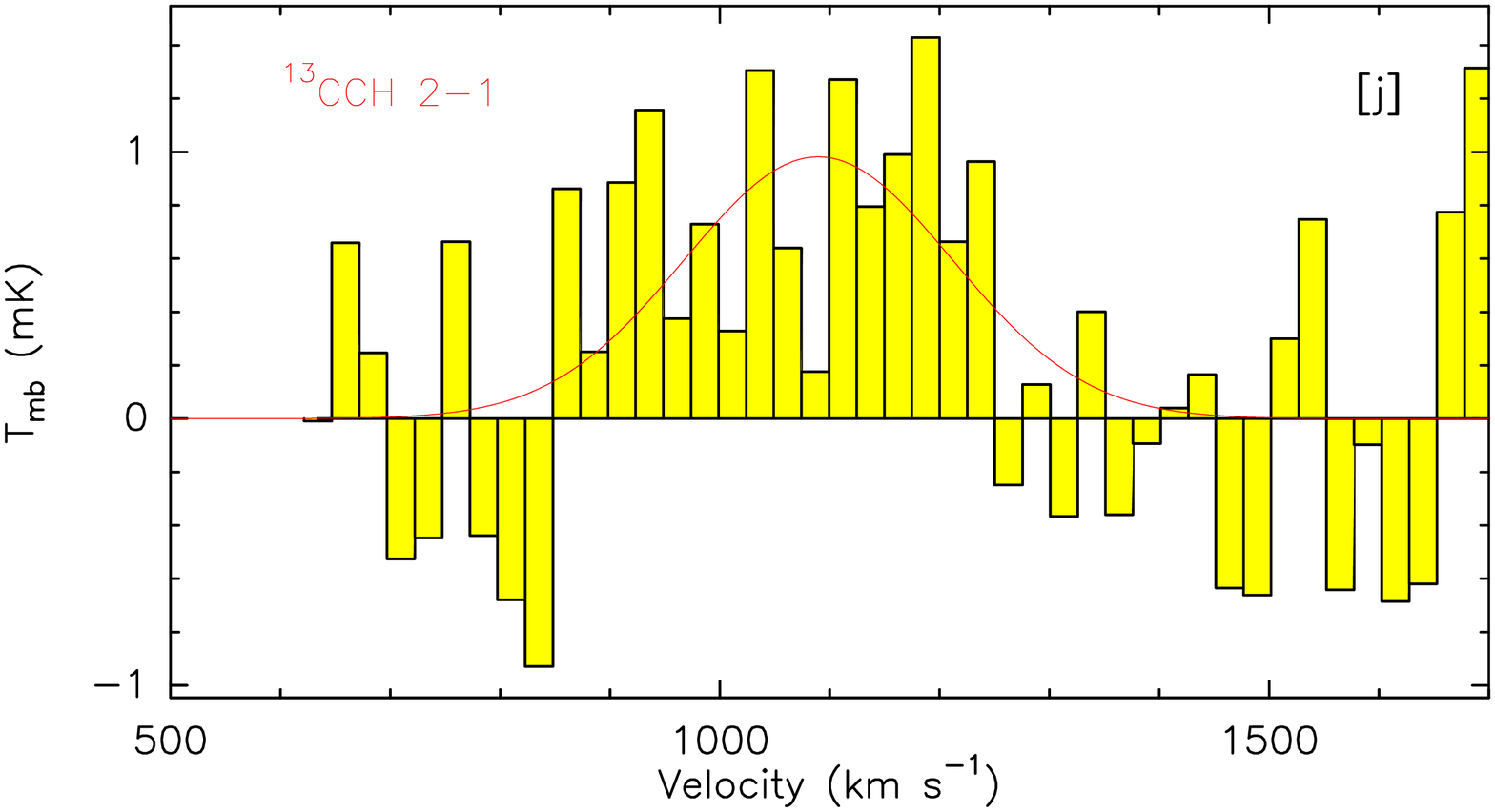}
    \end{minipage}
  
    \begin{minipage}{10cm}
        \includegraphics[width=10cm]{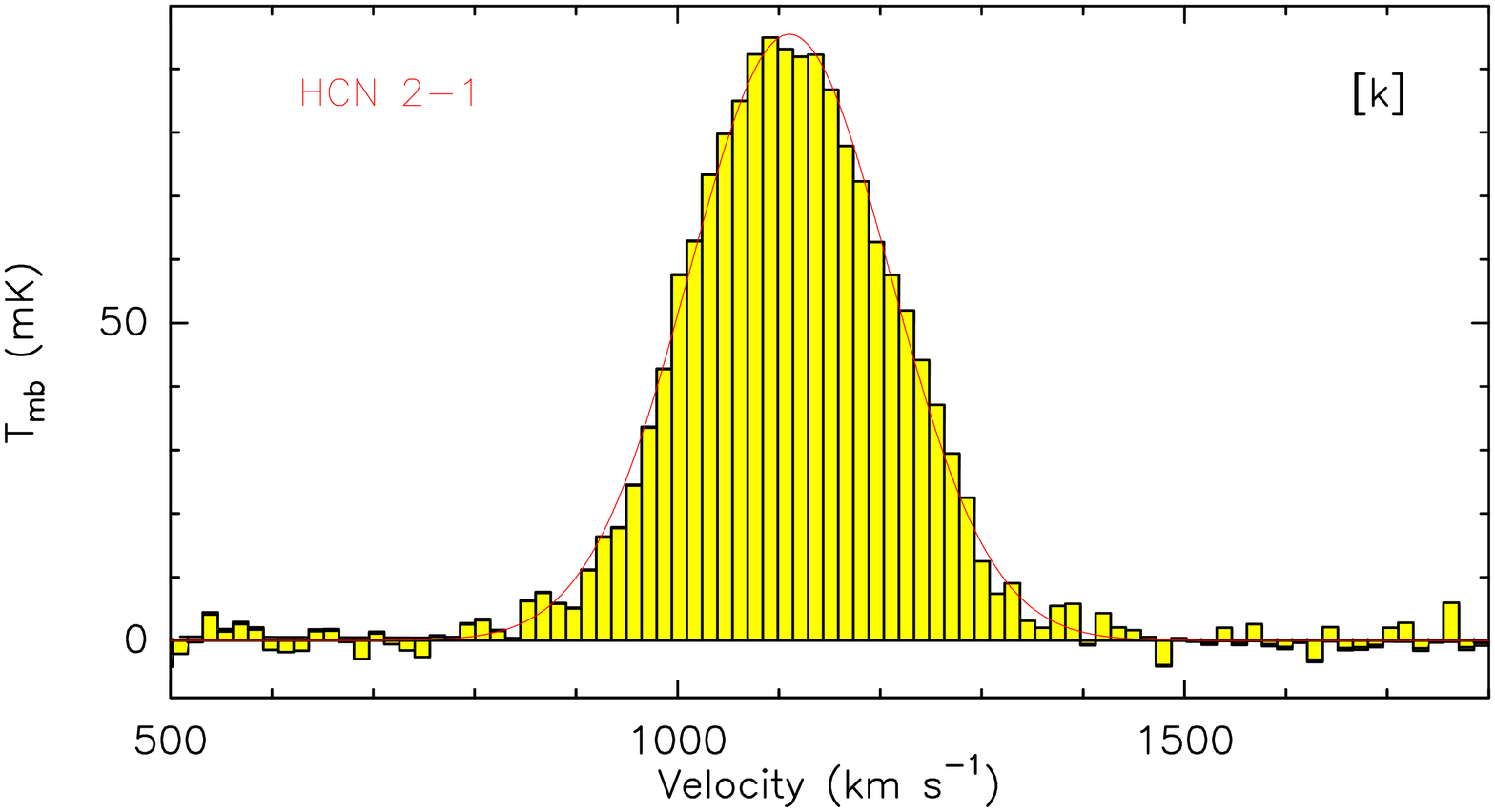}
    \end{minipage}
    \begin{minipage}{10cm}
        \includegraphics[width=10cm]{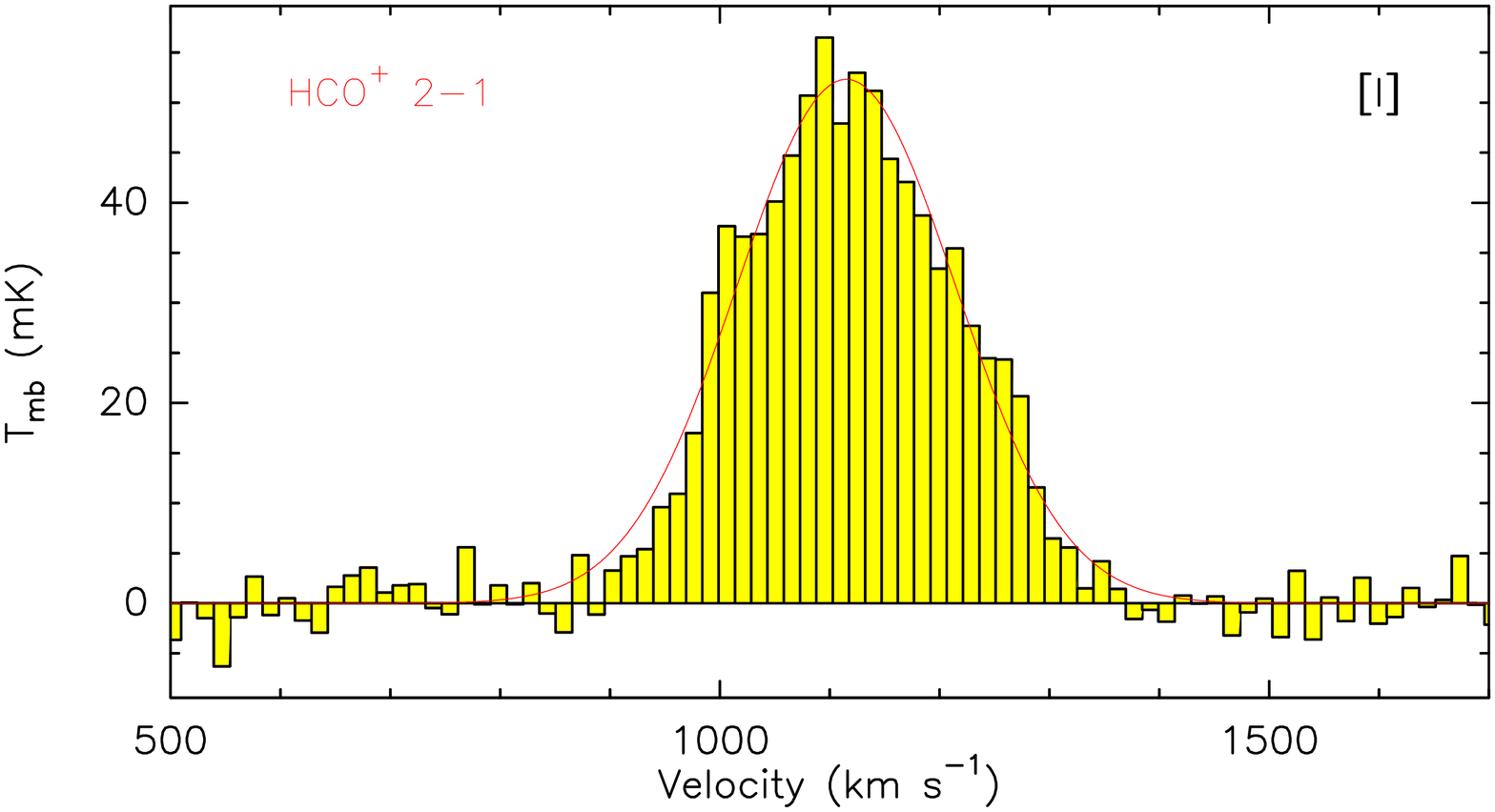}
    \end{minipage}
  
    \begin{minipage}{10cm}
        \includegraphics[width=10cm]{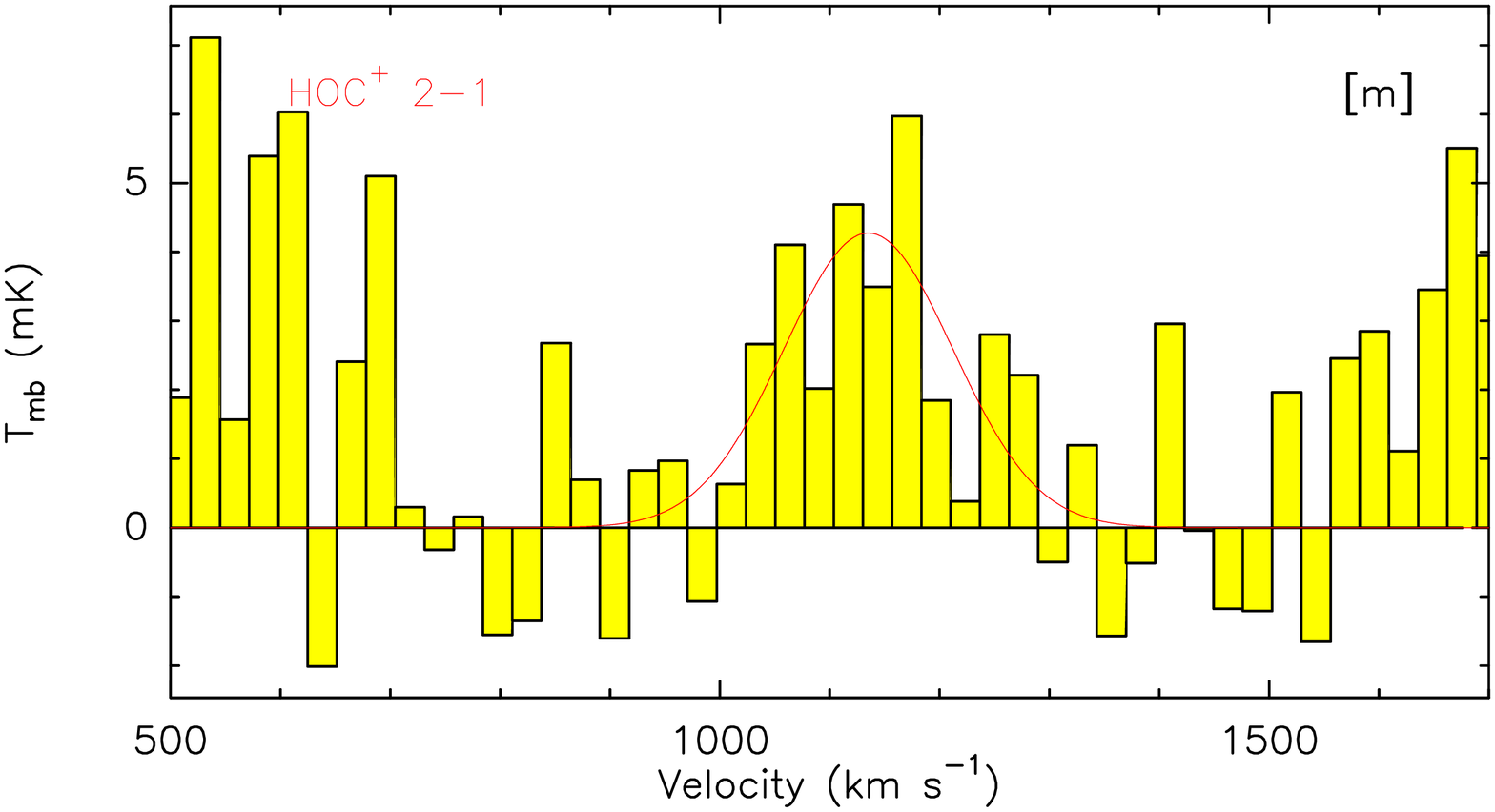}
    \end{minipage}
    \begin{minipage}{10cm}
        \includegraphics[width=10cm]{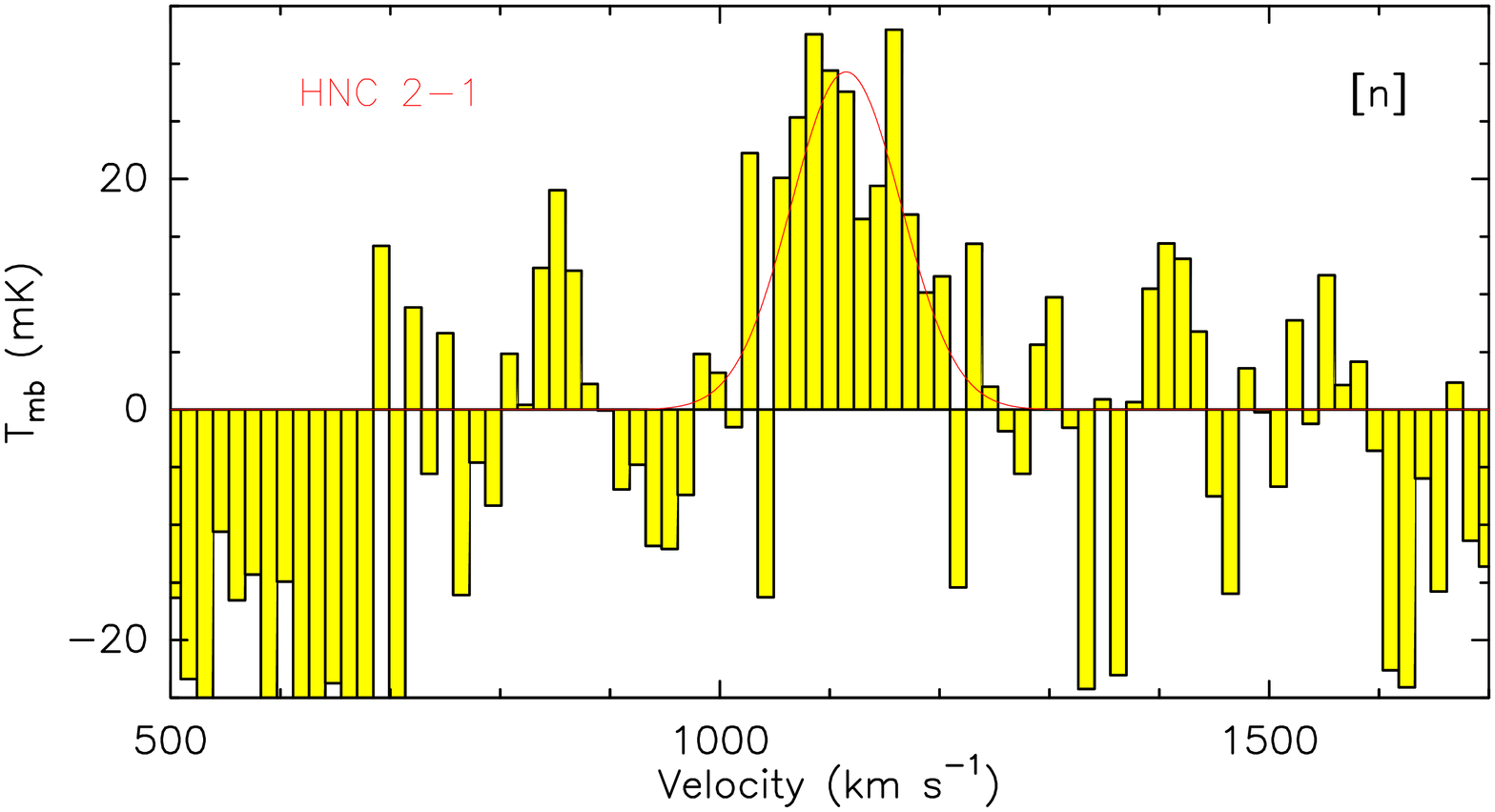}
    \end{minipage}
  
    \begin{minipage}{10cm}
        \includegraphics[width=10cm]{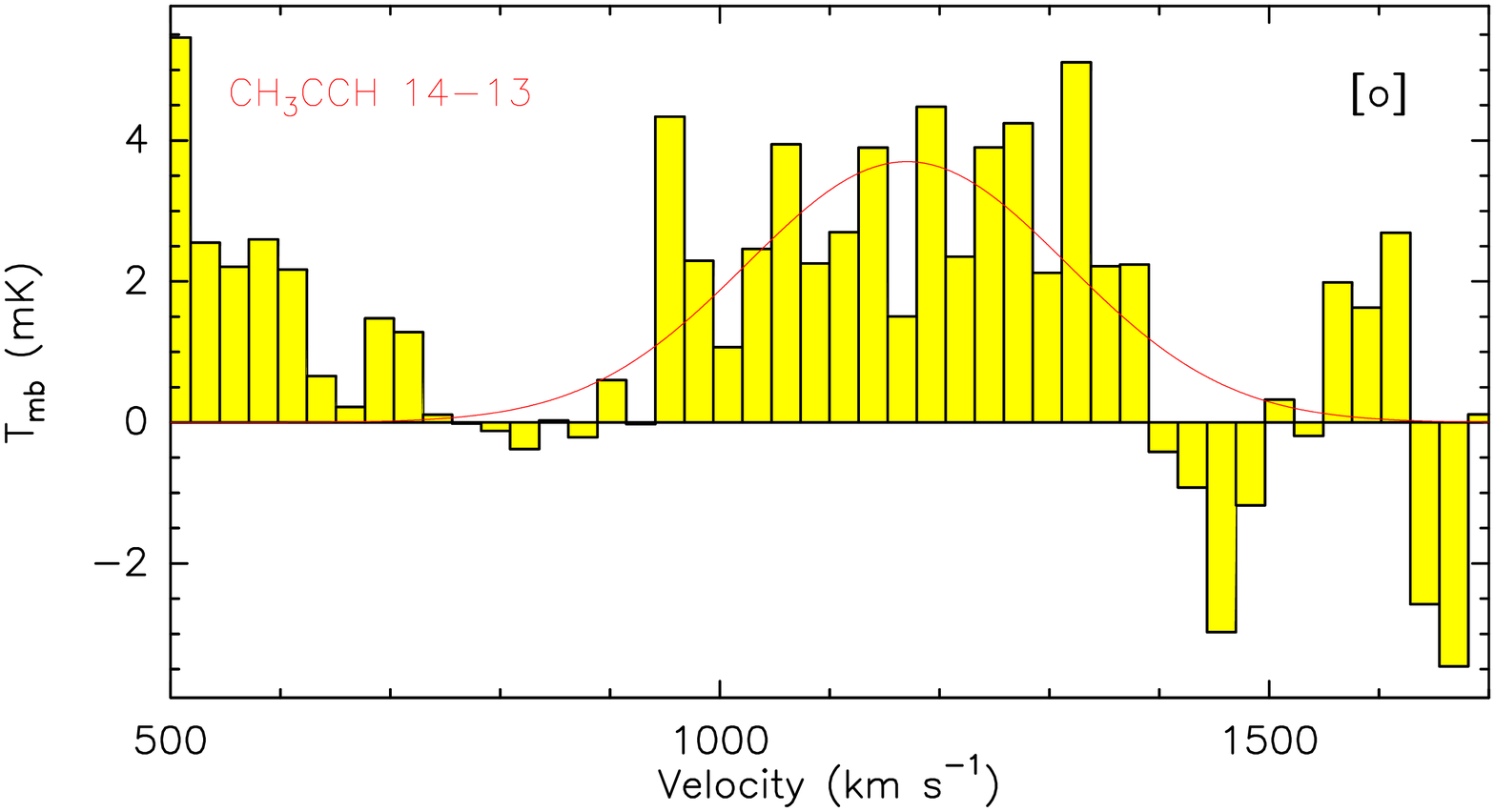}
    \end{minipage}

 \caption{continued.
}
\label{f_list_2}
\end{figure}

\begin{figure}
  \centering
        \includegraphics[width=18cm,bb= 60 0 900 500]{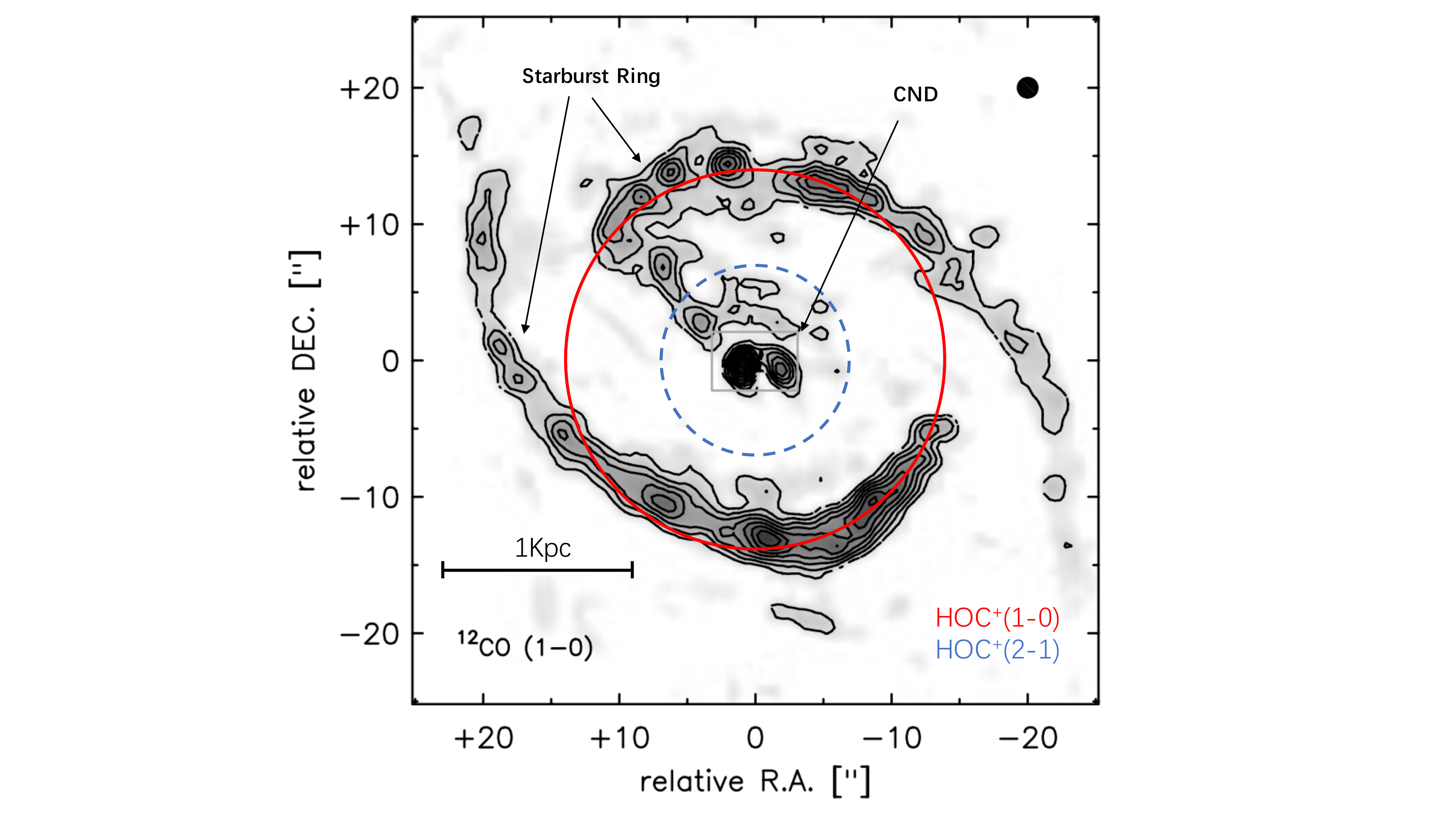}
\caption{
    CO\,$(1-0)$ emission of NGC\,1068 (gray scale and contours; taken from \cite{Schinnerer00}).
    The solid and dashed circles represent the beams of HOC$^{+}$\,$(1-0)$ and HOC$^{+}$\,$(2-1)$, respectively. 
    The central box marks the circumnuclear disk.
}
\label{f_region}
\end{figure}
\clearpage

\begin{figure}
  \centering
     \includegraphics[width=16cm]{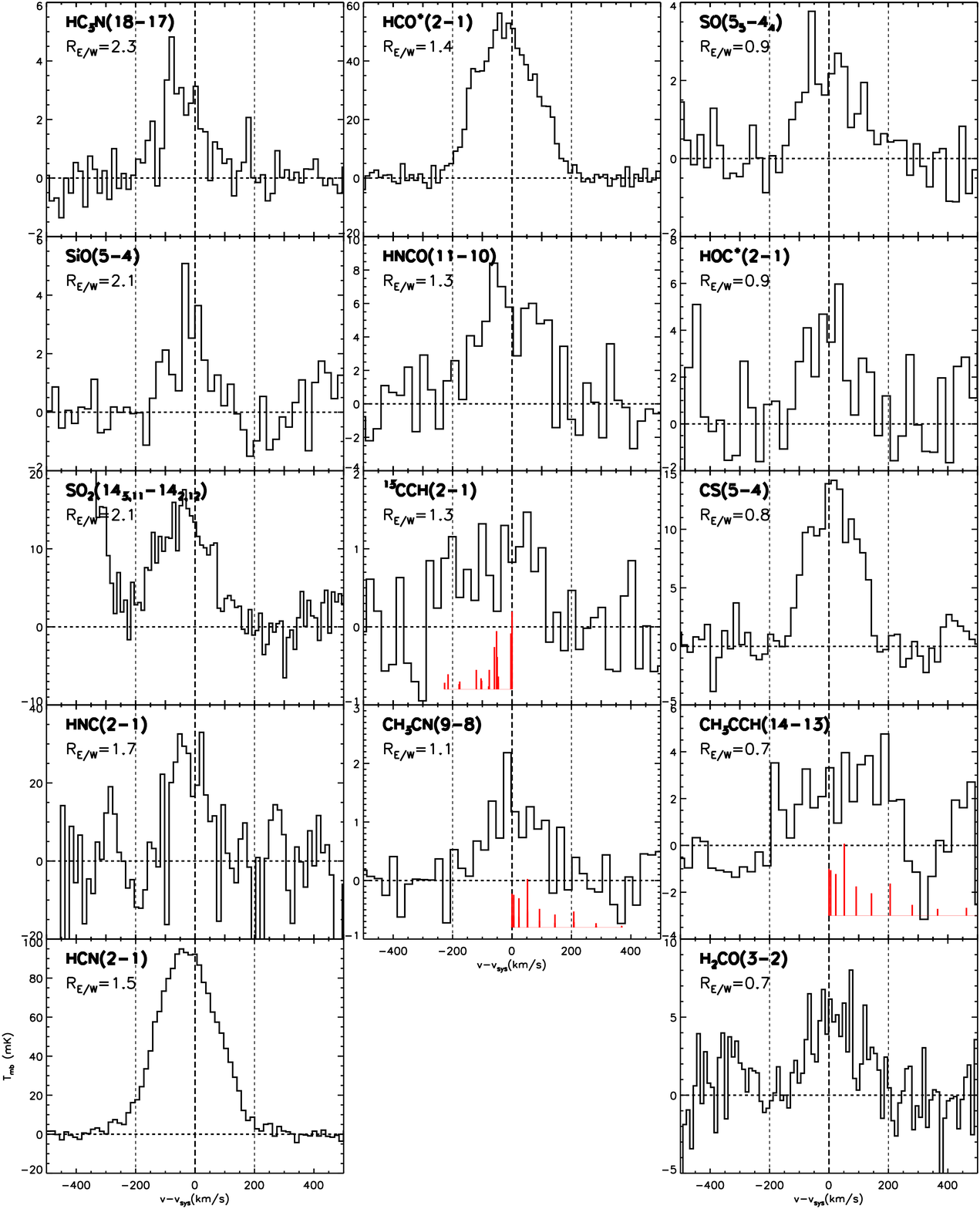}
\caption{
  Profiles of the molecular lines at 1 mm and 2 mm bands, as detected in \cite{Qiu18} and in the current observations. 
  The three vertical dotted lines at $v - v_{sys}$ = -200 km s$^{-1}$, $v - v_{sys}$ = 0 km s$^{-1}$, and $v - v_{sys}$ = +200 km s$^{-1}$  delimit the blue and red kinematical components. The wavelengths and relative intensities of the split components 
  of $^{13}$CCH\,$(2-1)$, CH$_{3}$CN\,$(9-8)$, and CH$_{3}$CCH\,$(14-13)$ are marked by red lines.
    }
\label{CNDspectra}
\end{figure}

\begin{figure}
    \begin{minipage}{10cm}
        \includegraphics[width=10cm]{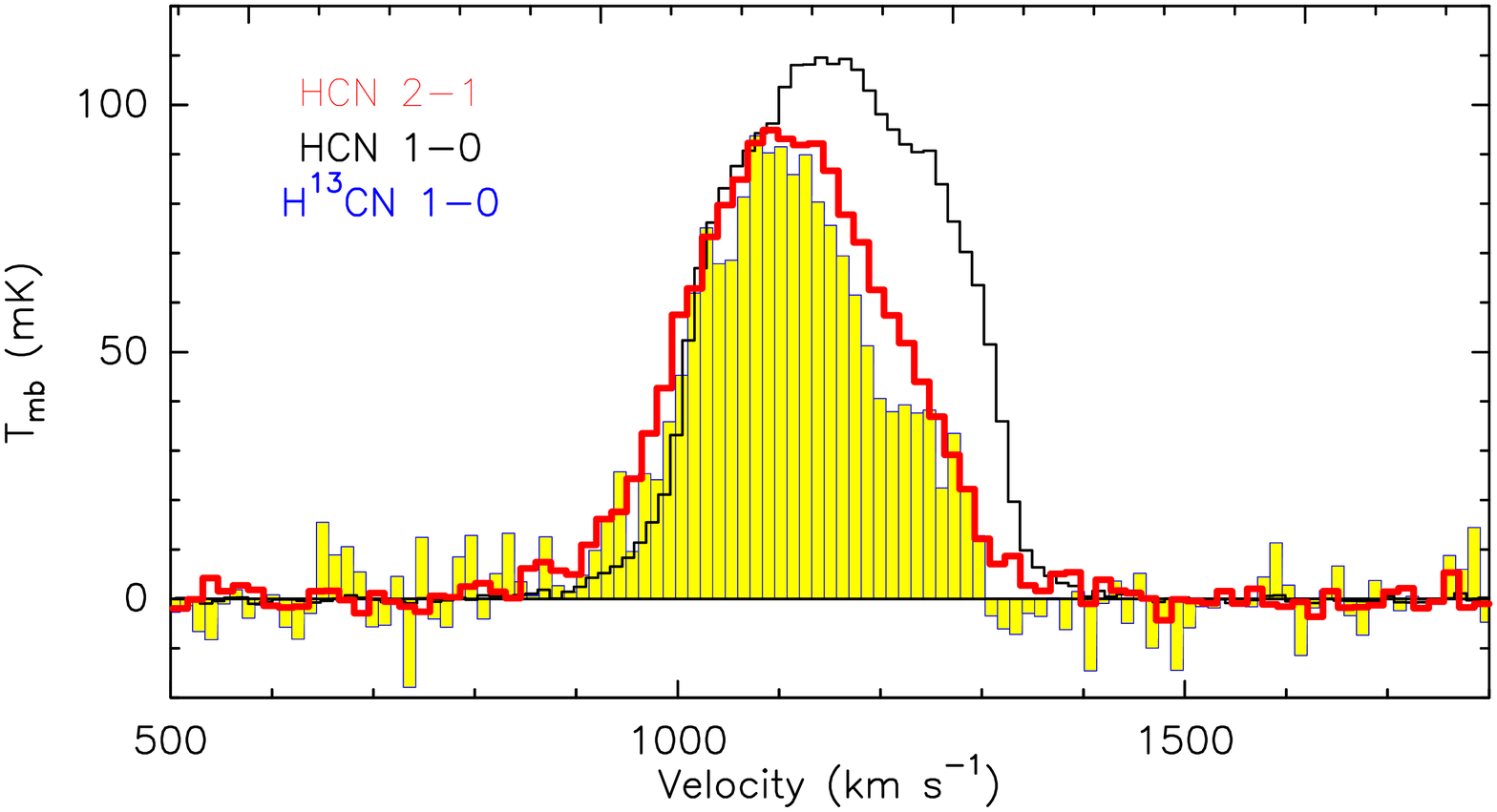}
    \end{minipage}
    \begin{minipage}{10cm}
        \includegraphics[width=10cm]{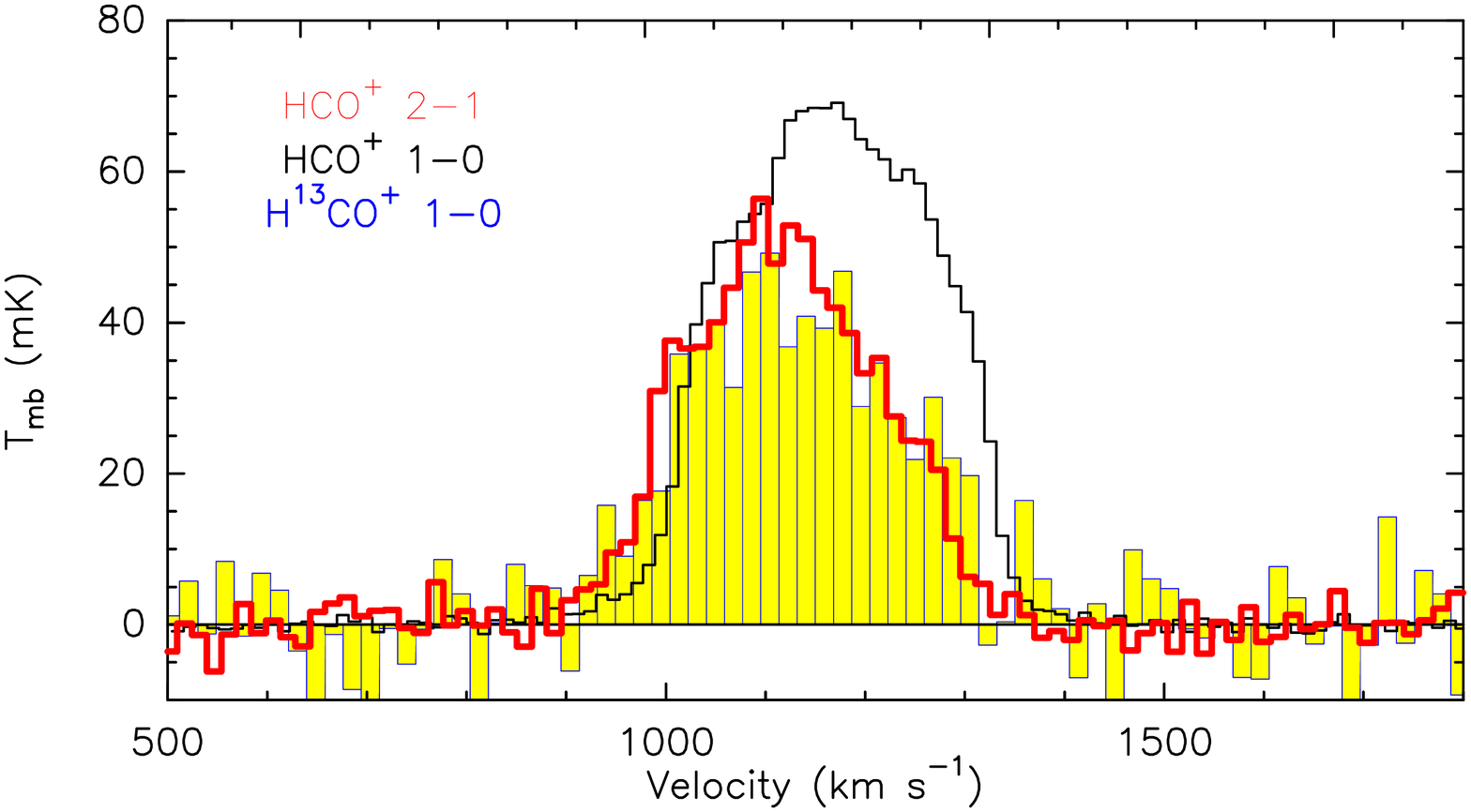}
    \end{minipage}

    \begin{minipage}{10cm}
        \includegraphics[width=10cm]{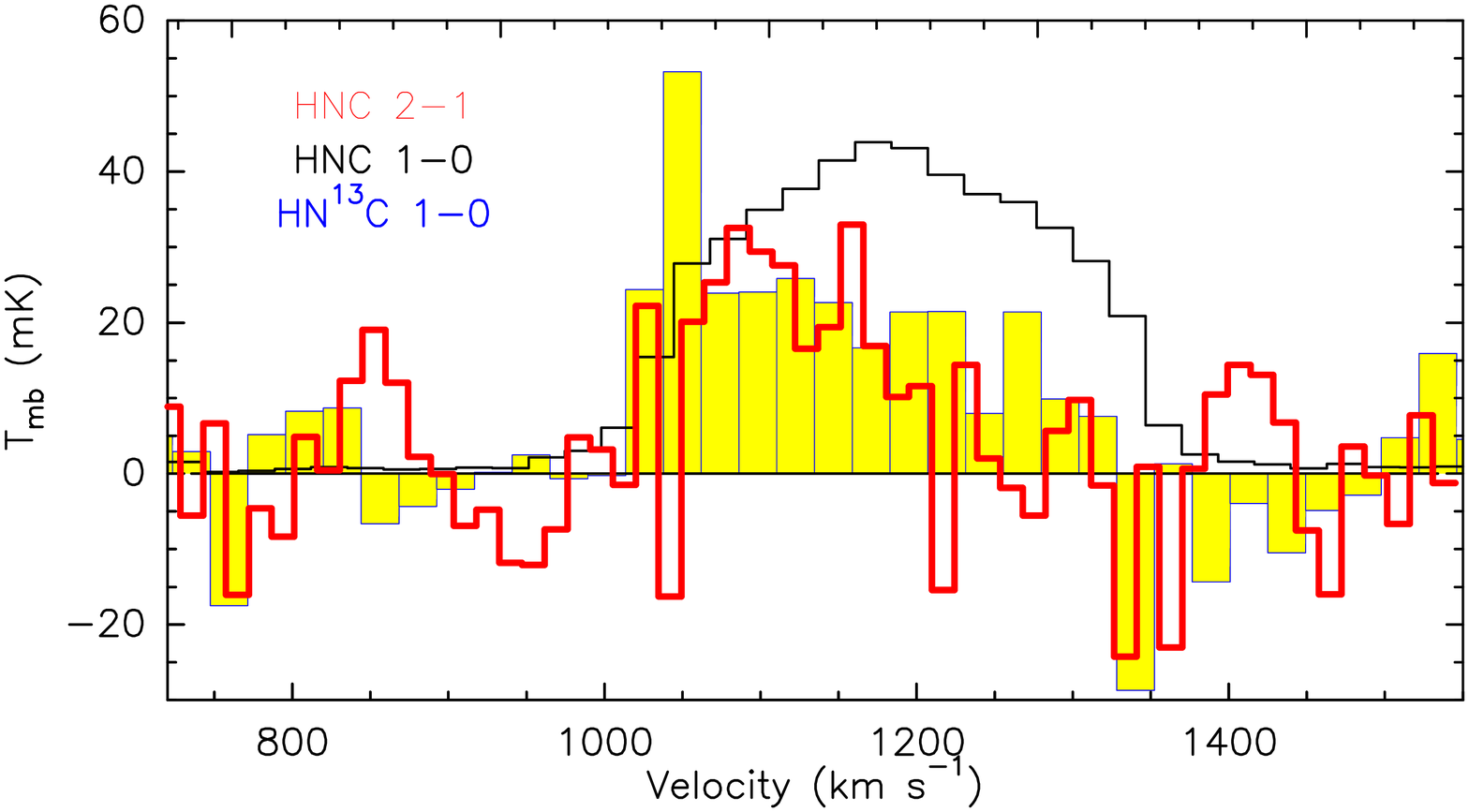}
    \end{minipage}
    \begin{minipage}{10cm}
        \includegraphics[width=10cm]{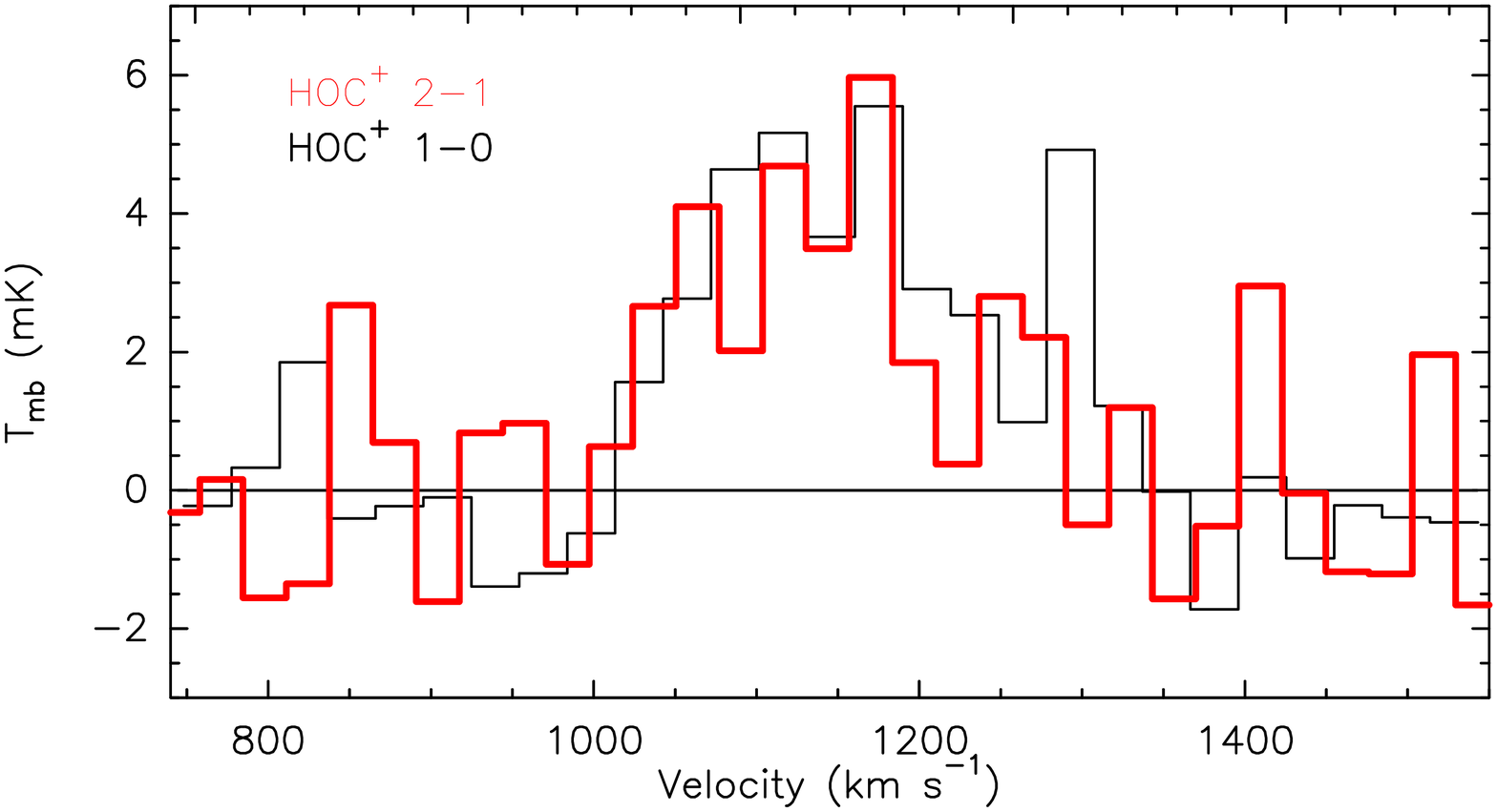}
    \end{minipage}

    \begin{minipage}{10cm}
        \includegraphics[width=10cm]{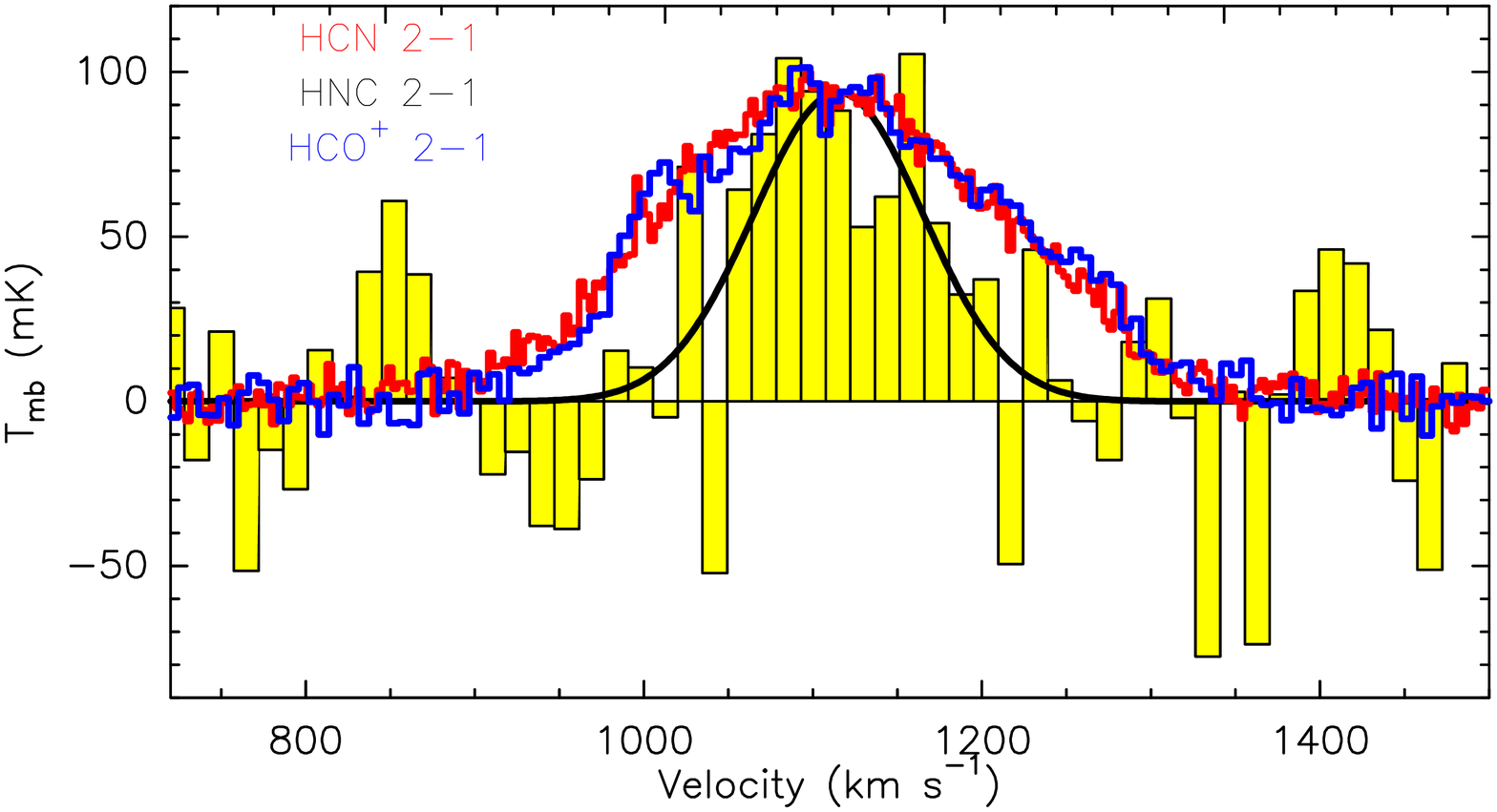}
    \end{minipage}
    \begin{minipage}{10cm}
        \includegraphics[width=10cm]{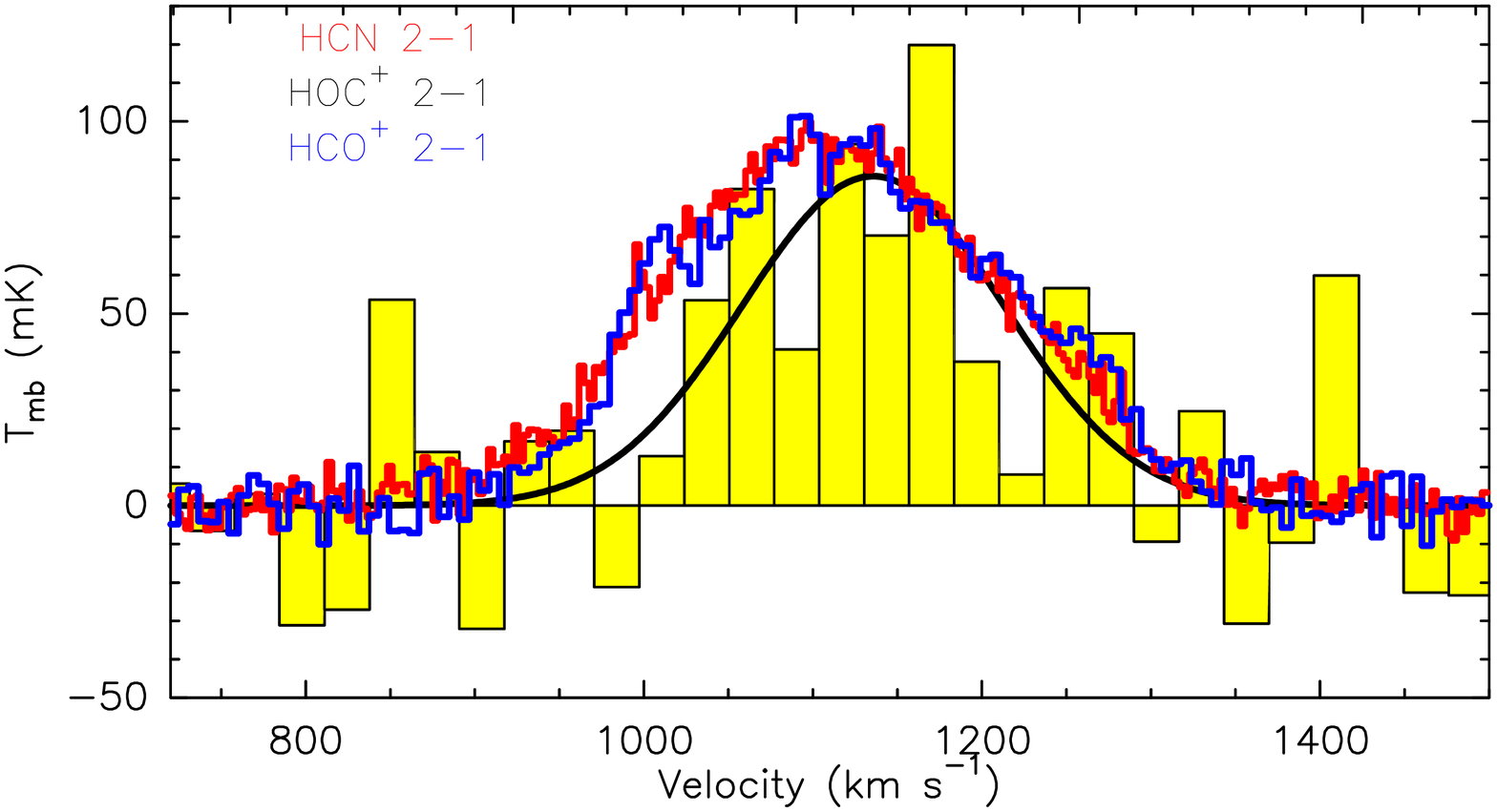}
    \end{minipage}

\caption{
    {\bf \emph{ Upper left panel:}} HCN\,$(2-1)$ (red line); H$^{13}$CN\,$(1-0)$ (blue line filled yellow, $\times$ 16); HCN\,$(1-0)$ (black line, $\times$ 1.5).
    {\bf \emph{Upper right panel:}} HCO$^{+}$\,$(2-1)$, H$^{13}$CO$^{+}$\,$(1-0)$ $\times$ 18, HCO$^{+}$\,$(1-0)$ $\times$ 1.3.
    {\bf \emph{Middle left panel:}} HNC\,$(2-1)$, HN$^{13}$C\,$(1-0)$ $\times$ 35, HNC\,$(1-0)$ $\times$ 1.5.
    {\bf \emph{Middle right panel:}} HOC$^{+}$\,$(2-1)$, HOC$^{+}$\,$(1-0)$ $\times$ 4.
    {\bf \emph{Lower left panel:}} HCN\,$(2-1)$ (red line, $\times$ 3.2); HNC\,$(2-1)$ (black line filled yellow, $\times$ 16); HCO$^{+}$\,$(2-1)$ (black line, $\times$ 1.8); the black line curve is the Gaussian fitting line of HNC\,$(2-1)$.
    {\bf \emph{Lower right panel:}} HCN\,$(2-1)$ (red line); HOC$^{+}$\,$(2-1)$ (black line filled yellow, $\times$ 20); HCO$^{+}$\,$(2-1)$ (black line, $\times$ 1.8); the black line curve is the Gaussian fitting line of HOC$^{+}$\,$(2-1)$.
    }
\label{figure_three}
\end{figure}

\begin{figure}
    \begin{minipage}{10cm}
        \includegraphics[width=10cm]{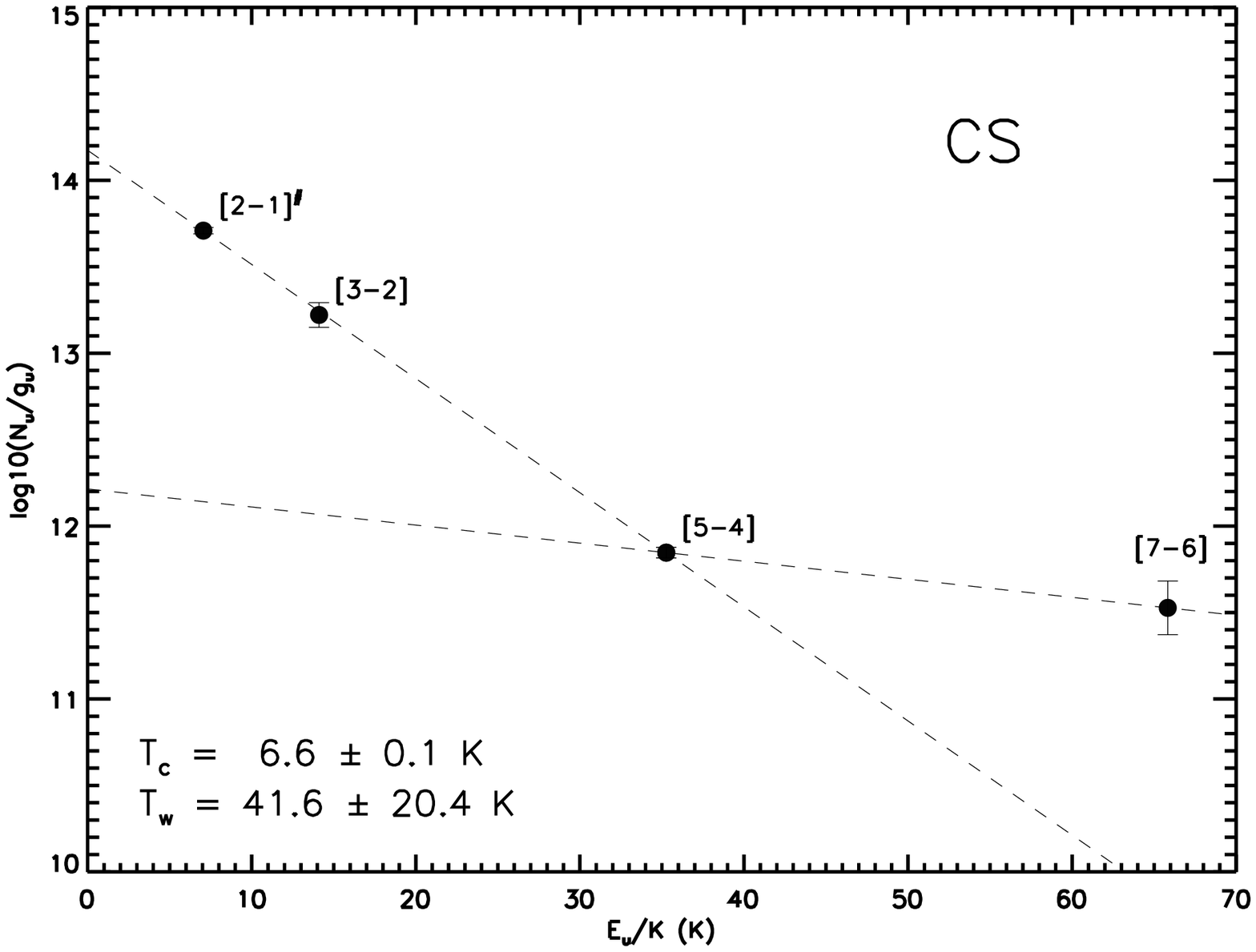}
    \end{minipage}
    \begin{minipage}{10cm}
        \includegraphics[width=10cm]{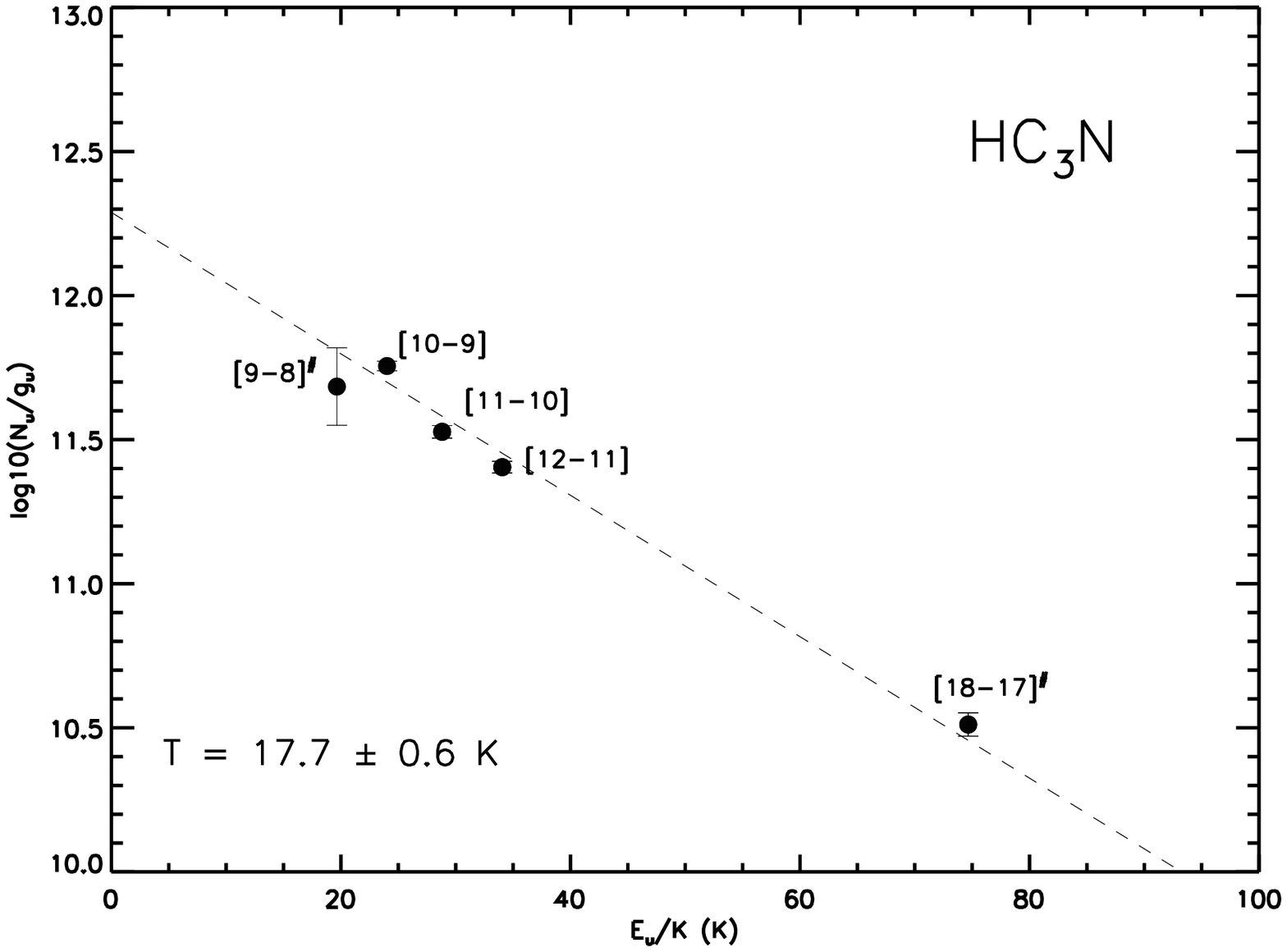}
    \end{minipage}

    \begin{minipage}{10cm}
        \includegraphics[width=10cm]{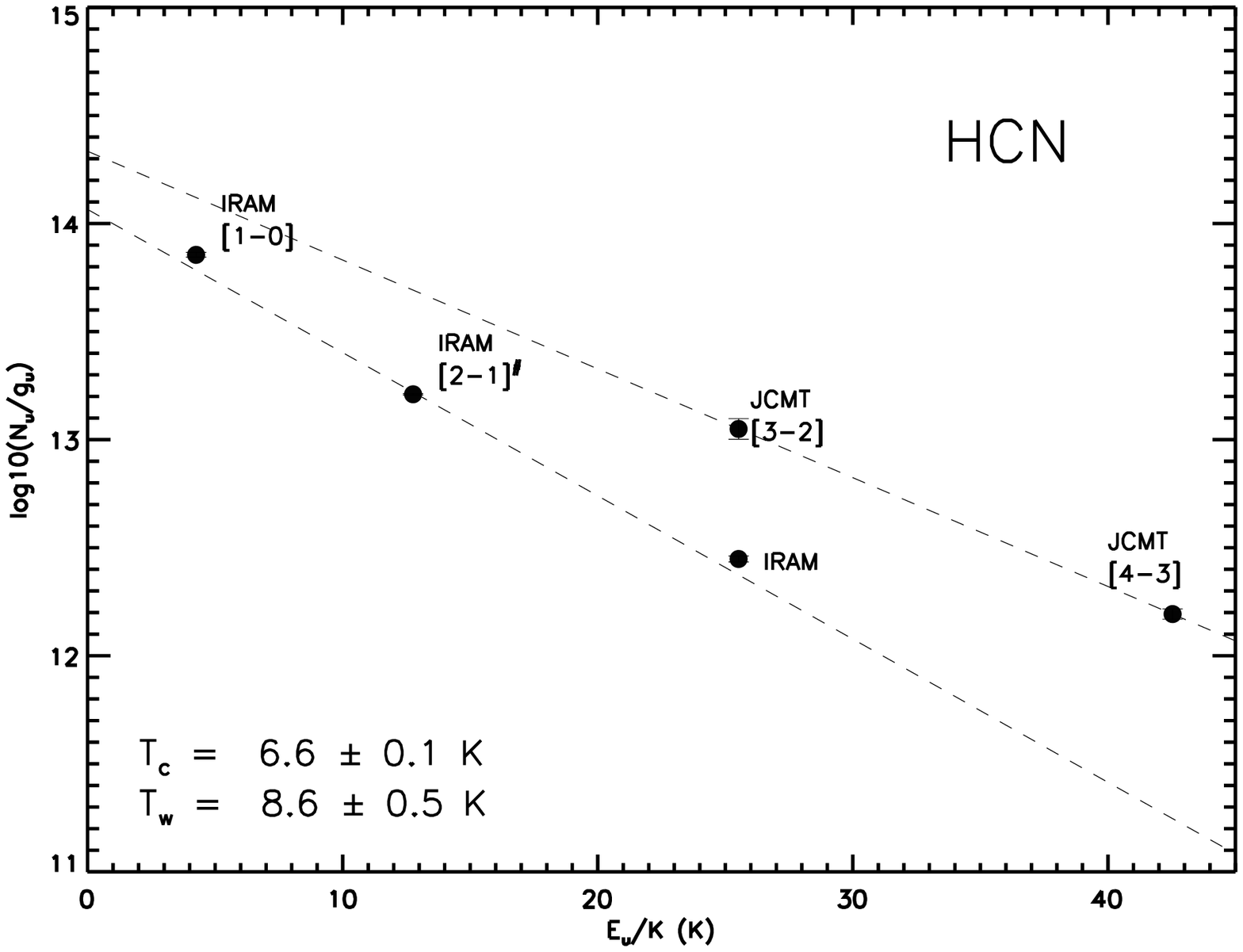}
    \end{minipage}
    \begin{minipage}{10cm}
        \includegraphics[width=10cm]{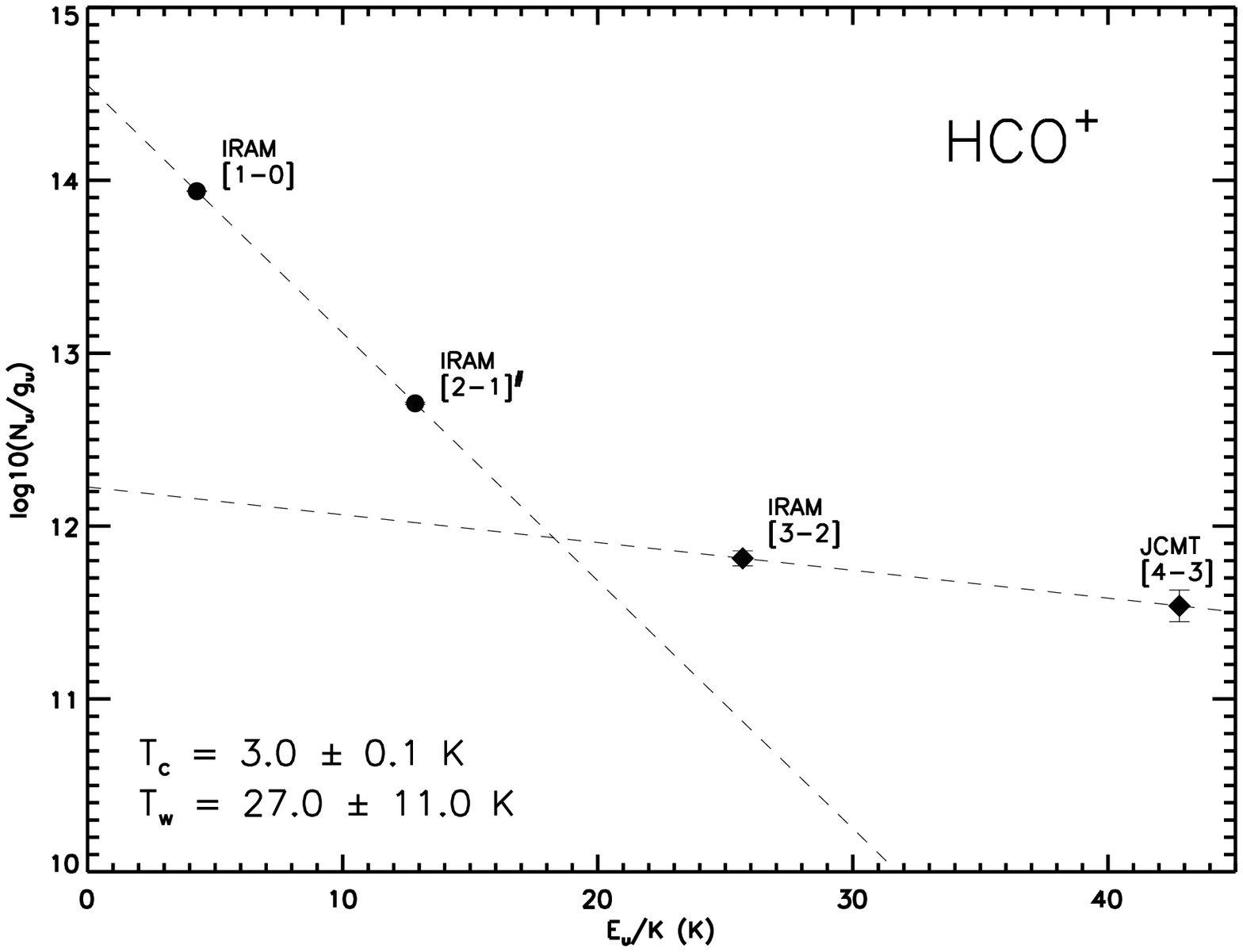}
    \end{minipage}

    \begin{minipage}{10cm}
        \includegraphics[width=10cm]{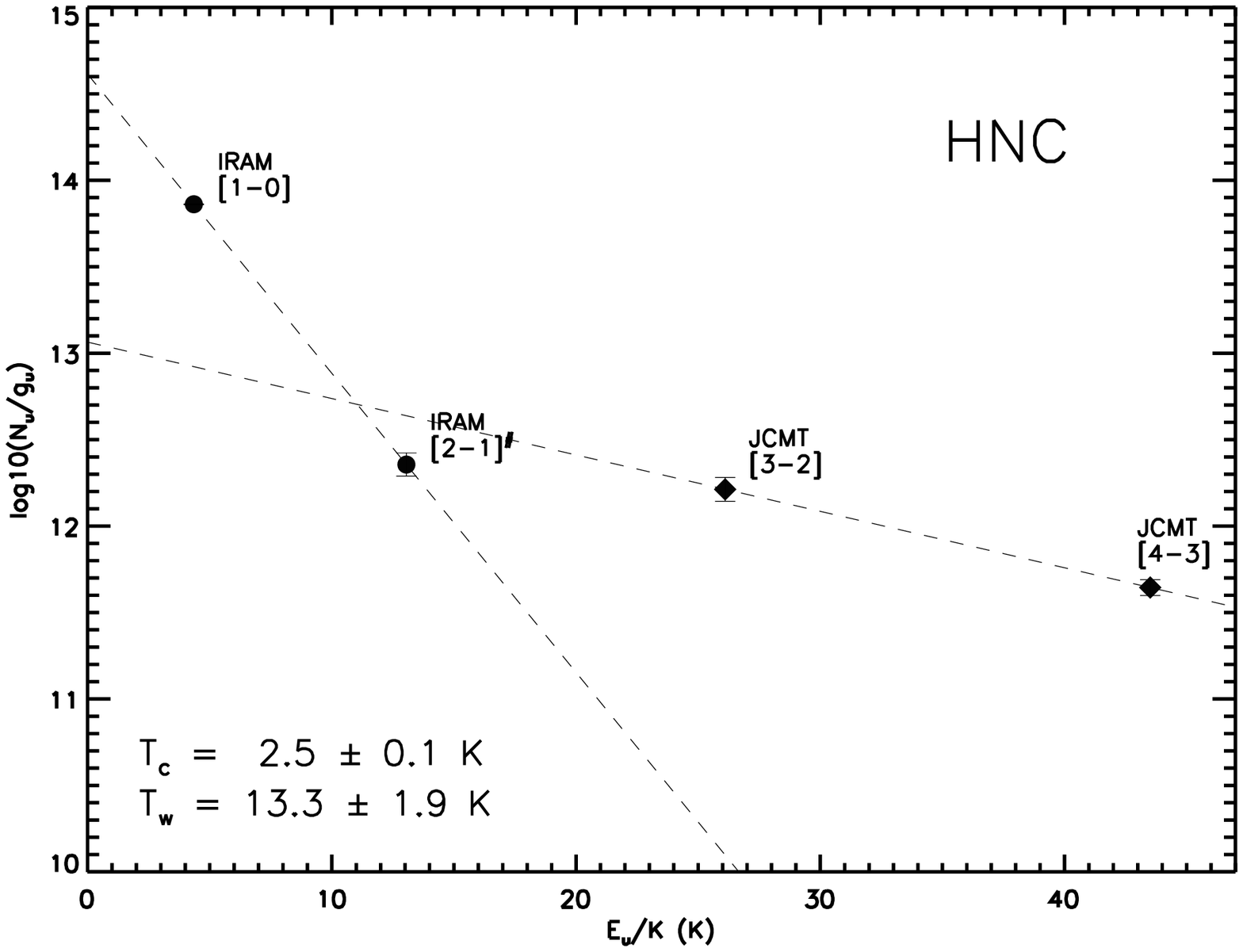}
    \end{minipage}
    \begin{minipage}{10cm}
        \includegraphics[width=10cm]{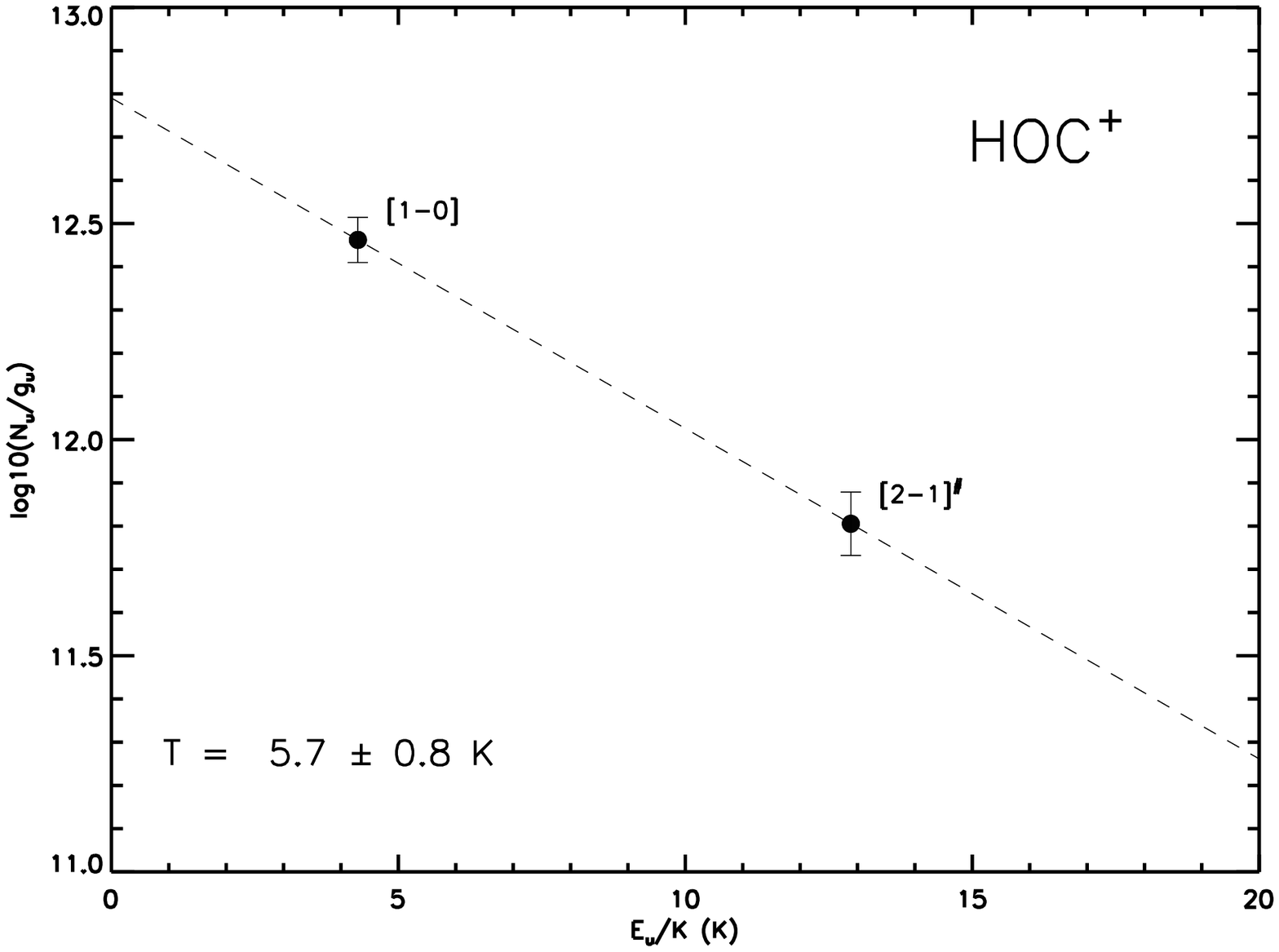}
    \end{minipage}
\caption{
Boltzmann diagrams. $T_{\rm c}$ is the rotation temperature of its cold component, and $T_{\rm w}$
is the rotation temperature of its warmer component. 
Transitions detected by the JCMT 15m telescope are also marked in the panels (filled diamond). 
The empty circles are discarded for the linear fitting.
Pound signs mark the lines that are detected for the first time in this work.
}
\label{f_RT}
\end{figure}

\addtocounter{figure}{-1}
\begin{figure}
    \begin{minipage}{10cm}
        \includegraphics[width=10cm]{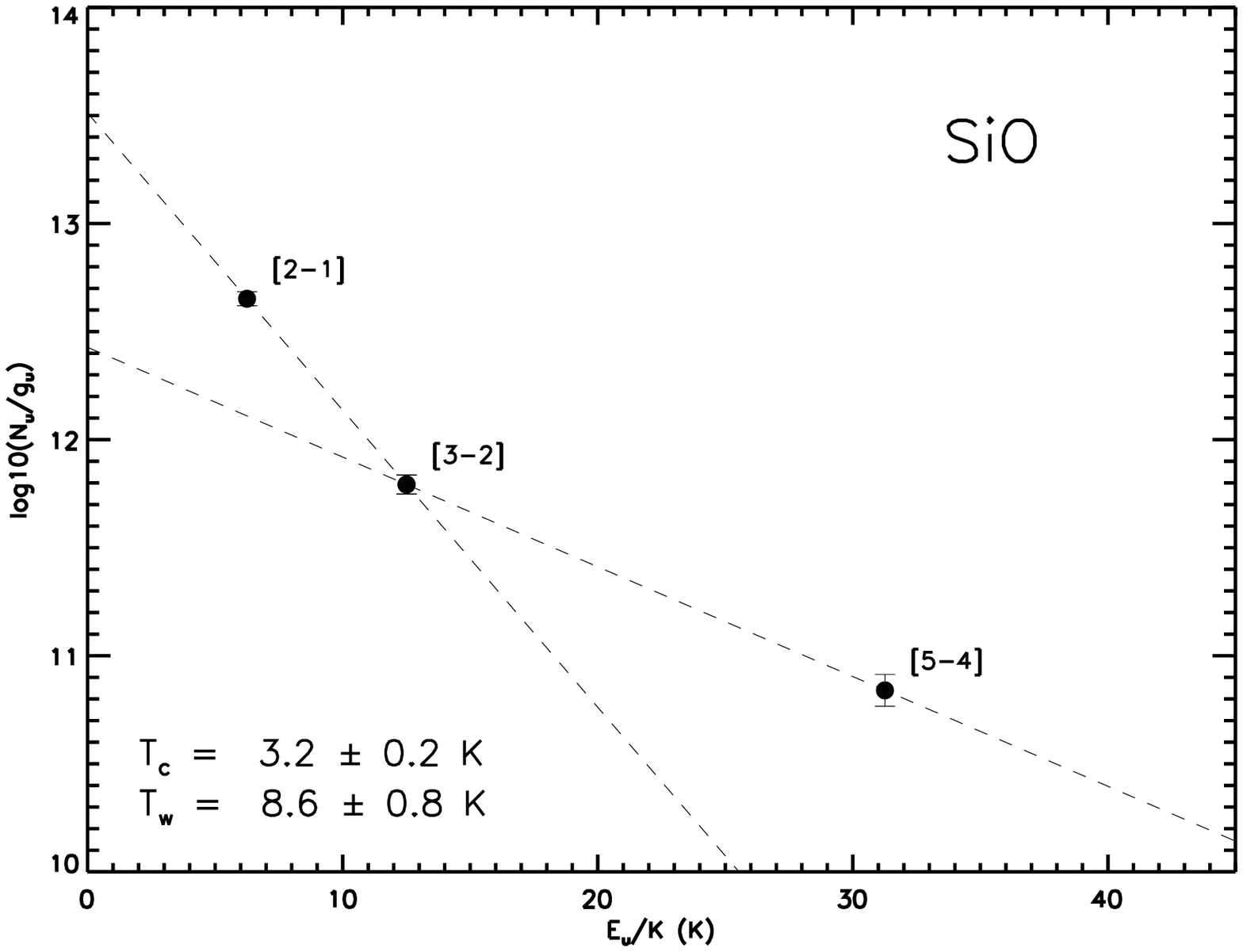}
    \end{minipage}
    \begin{minipage}{10cm}
        \includegraphics[width=10cm]{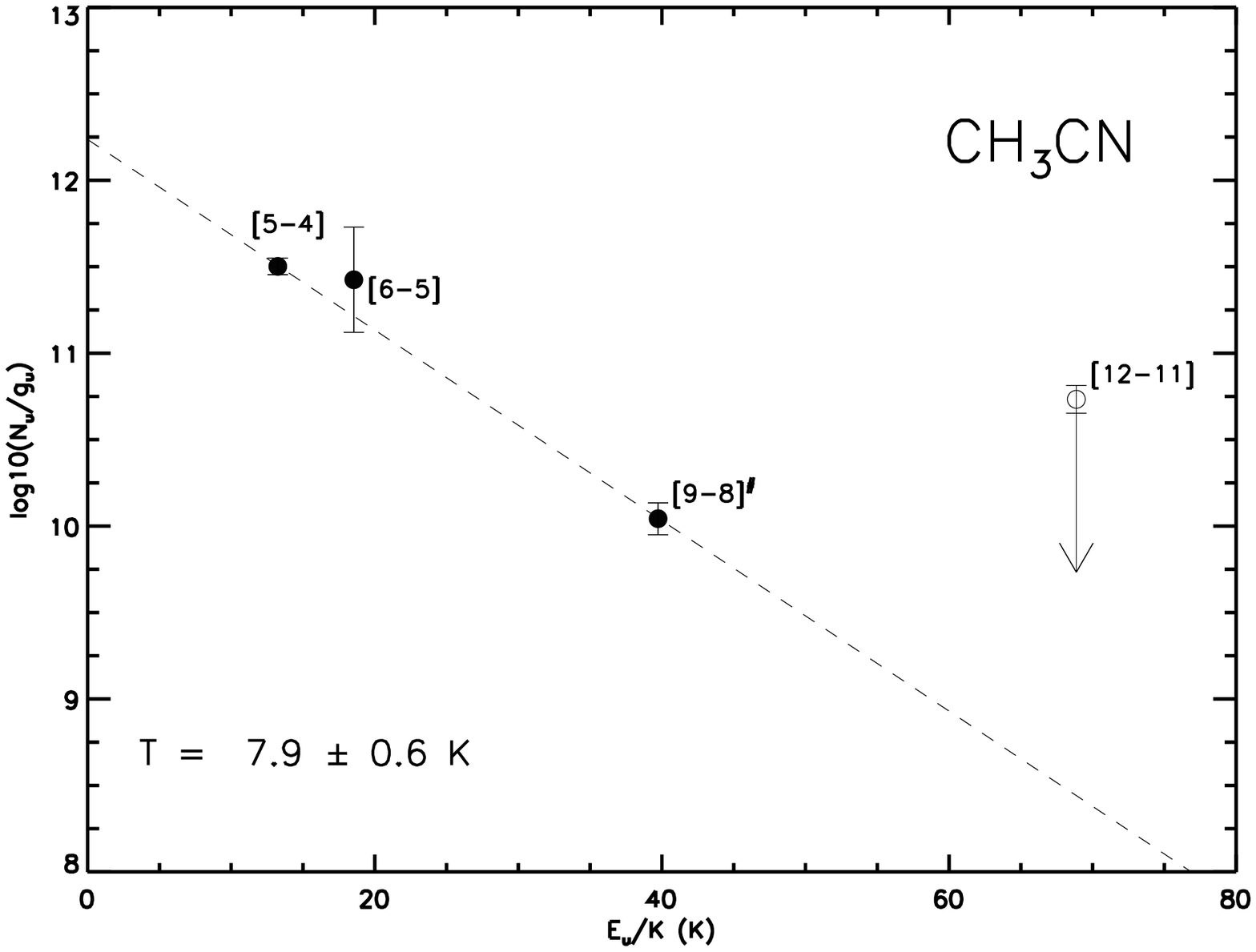}
    \end{minipage}

    \begin{minipage}{10cm}
        \includegraphics[width=10cm]{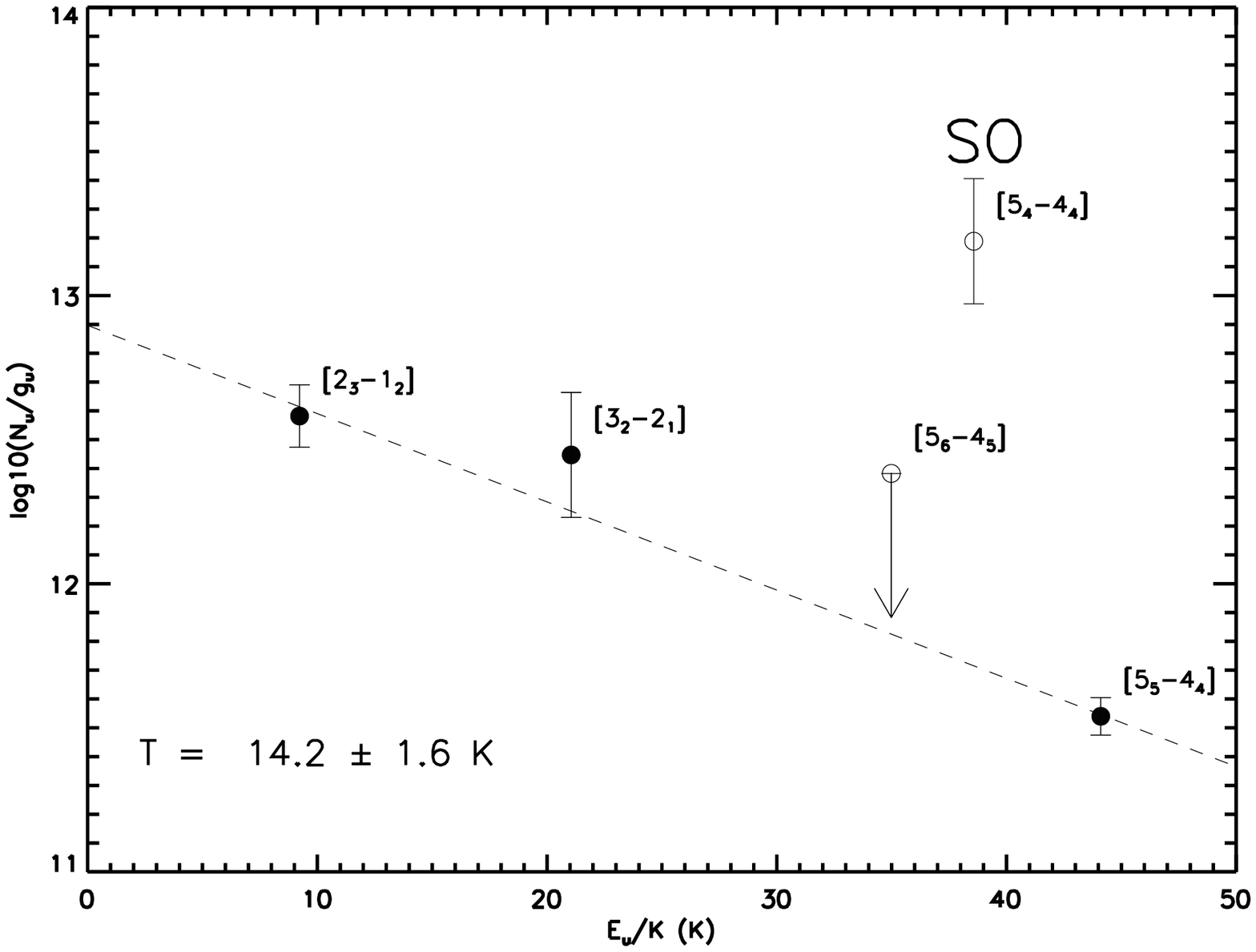}
    \end{minipage}
    \begin{minipage}{10cm}
        \includegraphics[width=10cm]{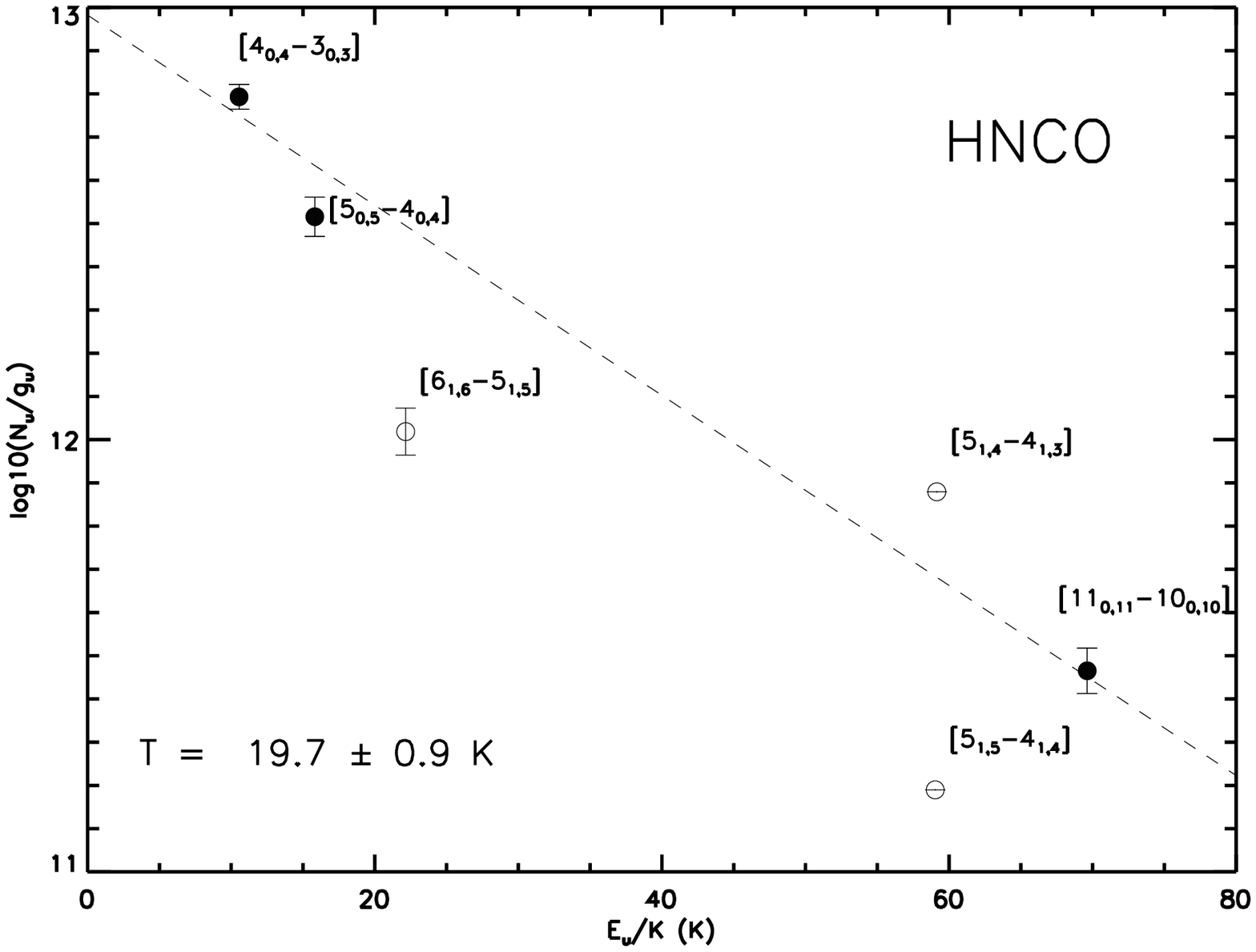}
    \end{minipage}

    \begin{minipage}{10cm}
        \includegraphics[width=10cm]{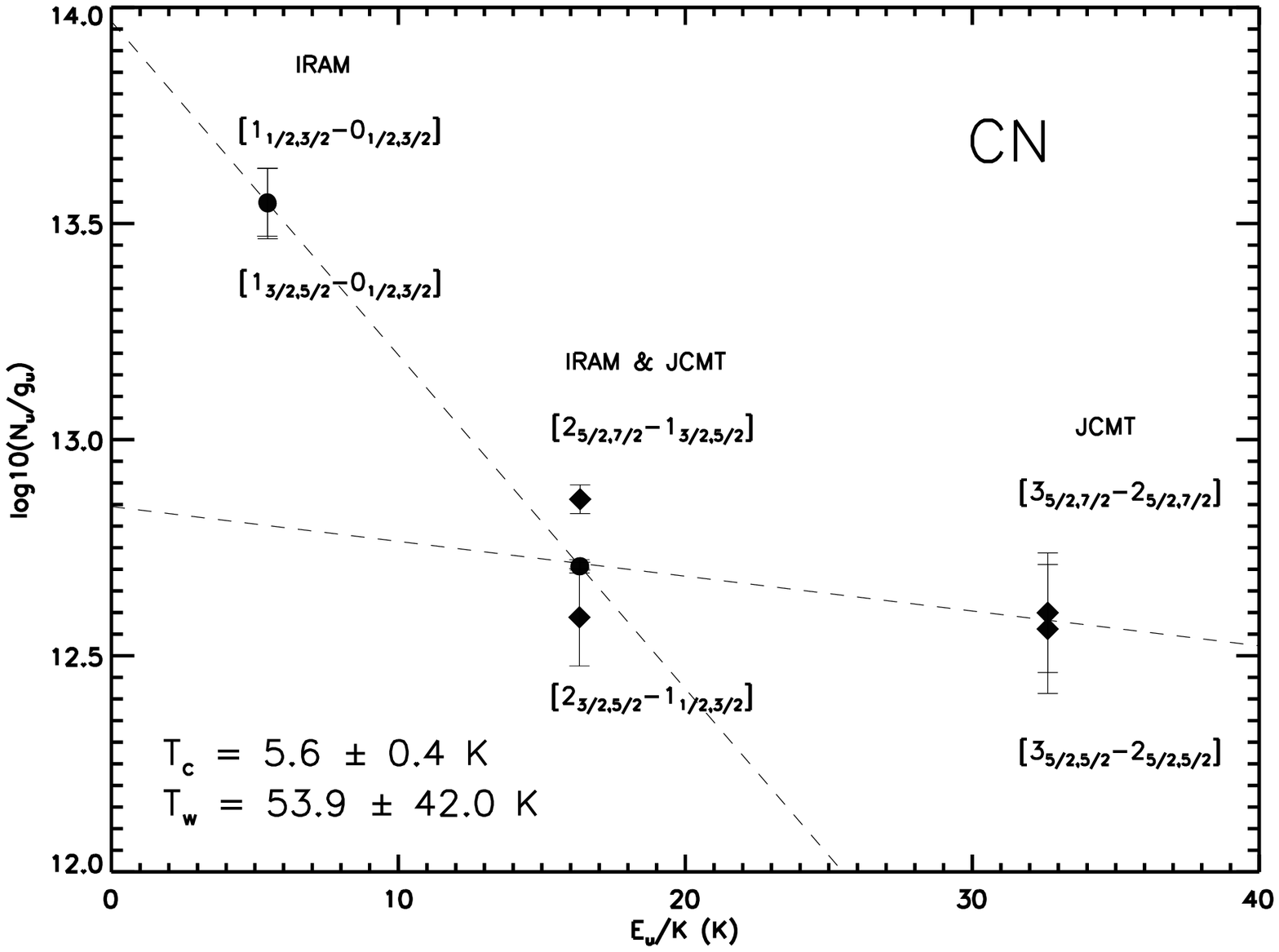}
    \end{minipage}
    \begin{minipage}{10cm}
        \includegraphics[width=10cm]{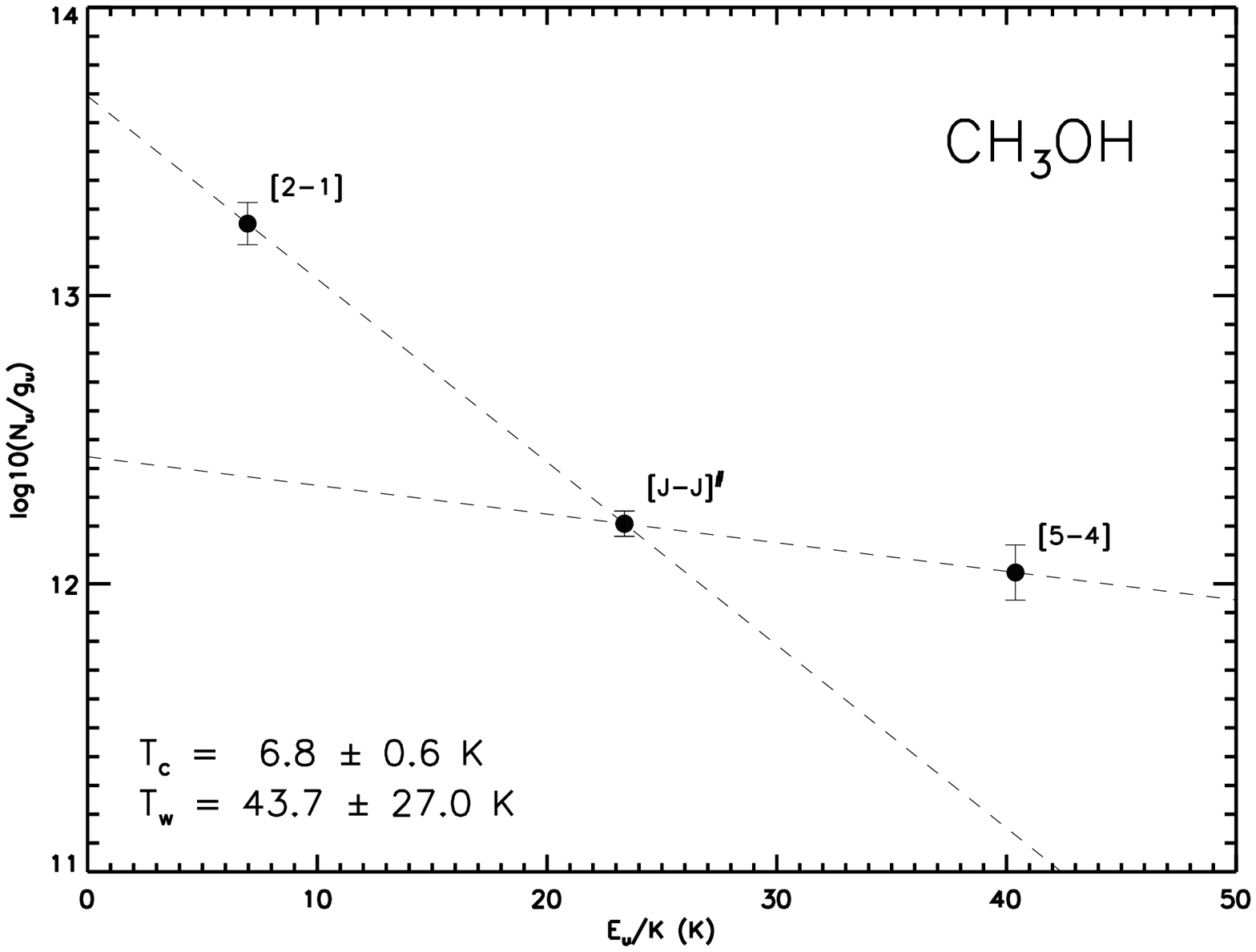}
    \end{minipage}

\caption{continued.
}
\label{f_RT}
\end{figure}

\addtocounter{figure}{-1}

\begin{figure}
    \begin{minipage}{10cm}
        \includegraphics[width=10cm]{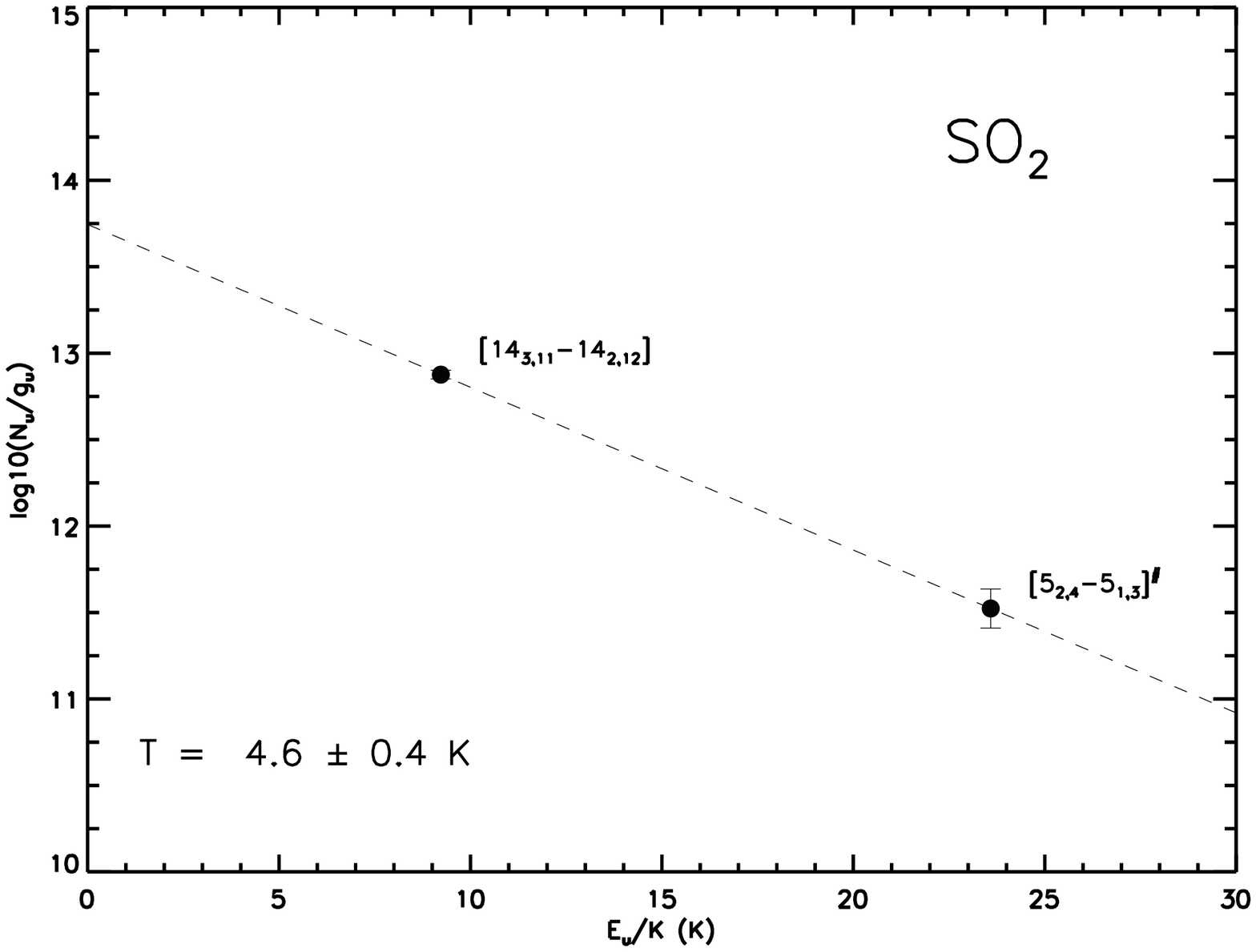}
    \end{minipage}
    \begin{minipage}{10cm}
        \includegraphics[width=10cm]{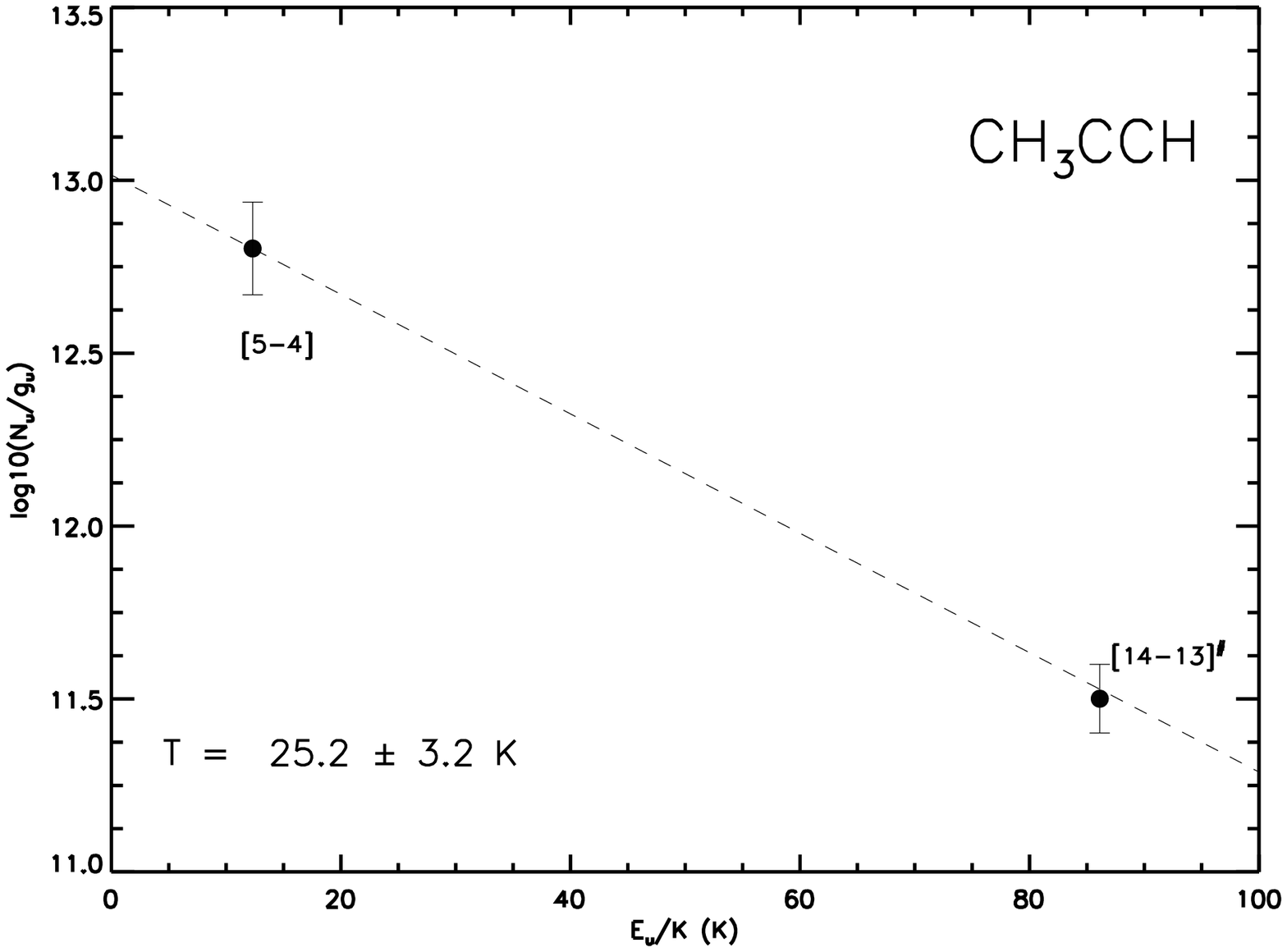}
    \end{minipage}

\caption{continued.
}
\label{f_RT}
\end{figure}

\clearpage

\begin{figure}
  \centering
    \begin{minipage}{17cm}
        \includegraphics[width=17cm]{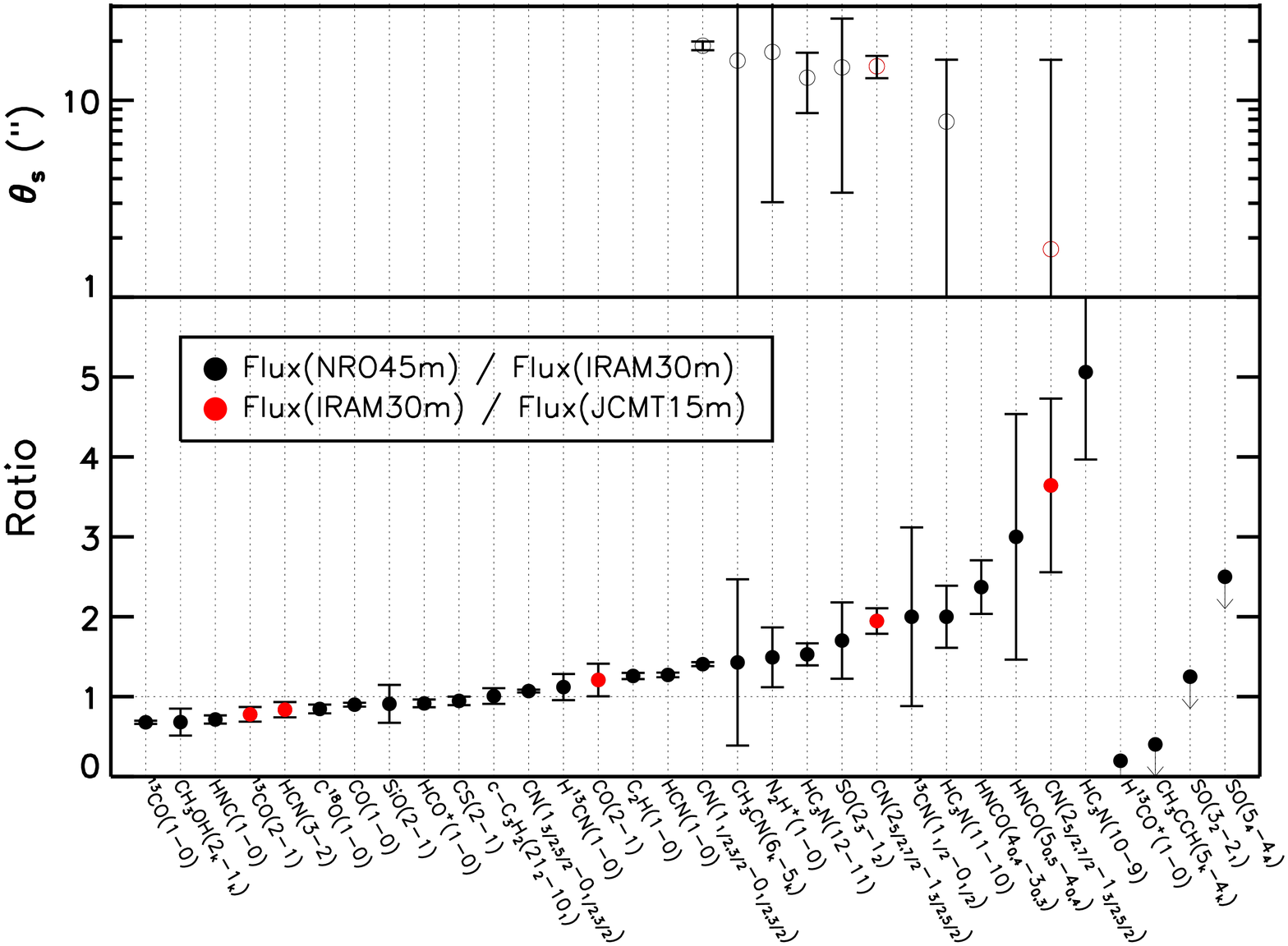}
    \end{minipage}

 \caption{ Integrated intensity ratios of the molecular lines detected with different single-dish telescopes
(lower panel). 
The calculated source sizes are shown in the upper panel.
}
\label{figure10}
\end{figure}
\clearpage


\begin{figure}
    \begin{minipage}{10cm}
        \includegraphics[width=10cm]{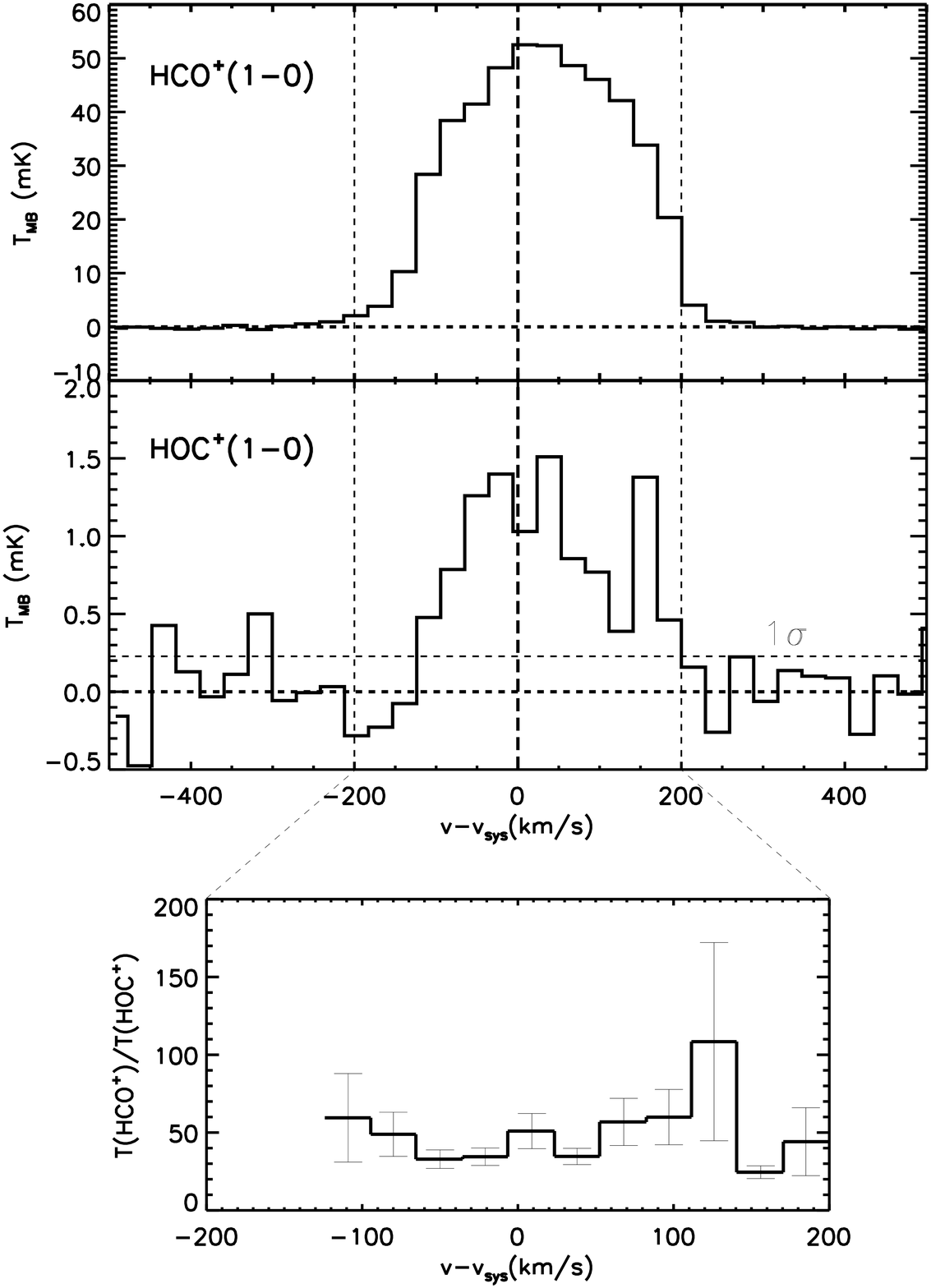}
    \end{minipage}
    \begin{minipage}{10cm}
        \includegraphics[width=10cm]{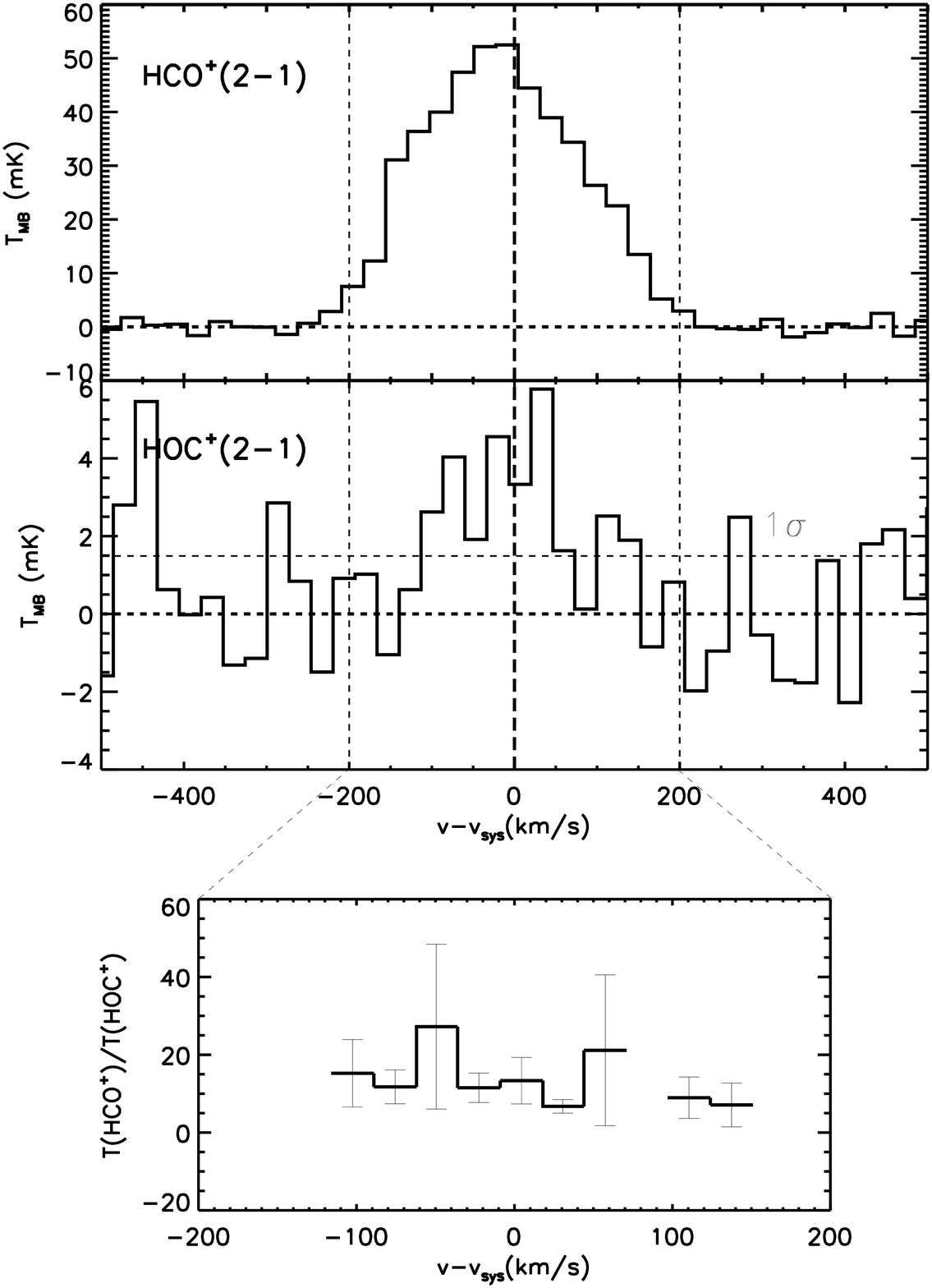}
    \end{minipage}
\caption{
  Top and middle: HCO$^{+}$ and HOC$^{+}$ spectra.
  Bottom: HCO$^{+}$-to-HOC$^{+}$ temperature ratio profile derived for channels fulfilling $T(\rm HOC^{+}) > 1 \sigma$ (left: $J = 1 - 0$ transitions; right: $J = 2 - 1$ transitions).
}
\label{f_hoc}
\end{figure}

\begin{figure}
  \centering
       \includegraphics[width=18cm]{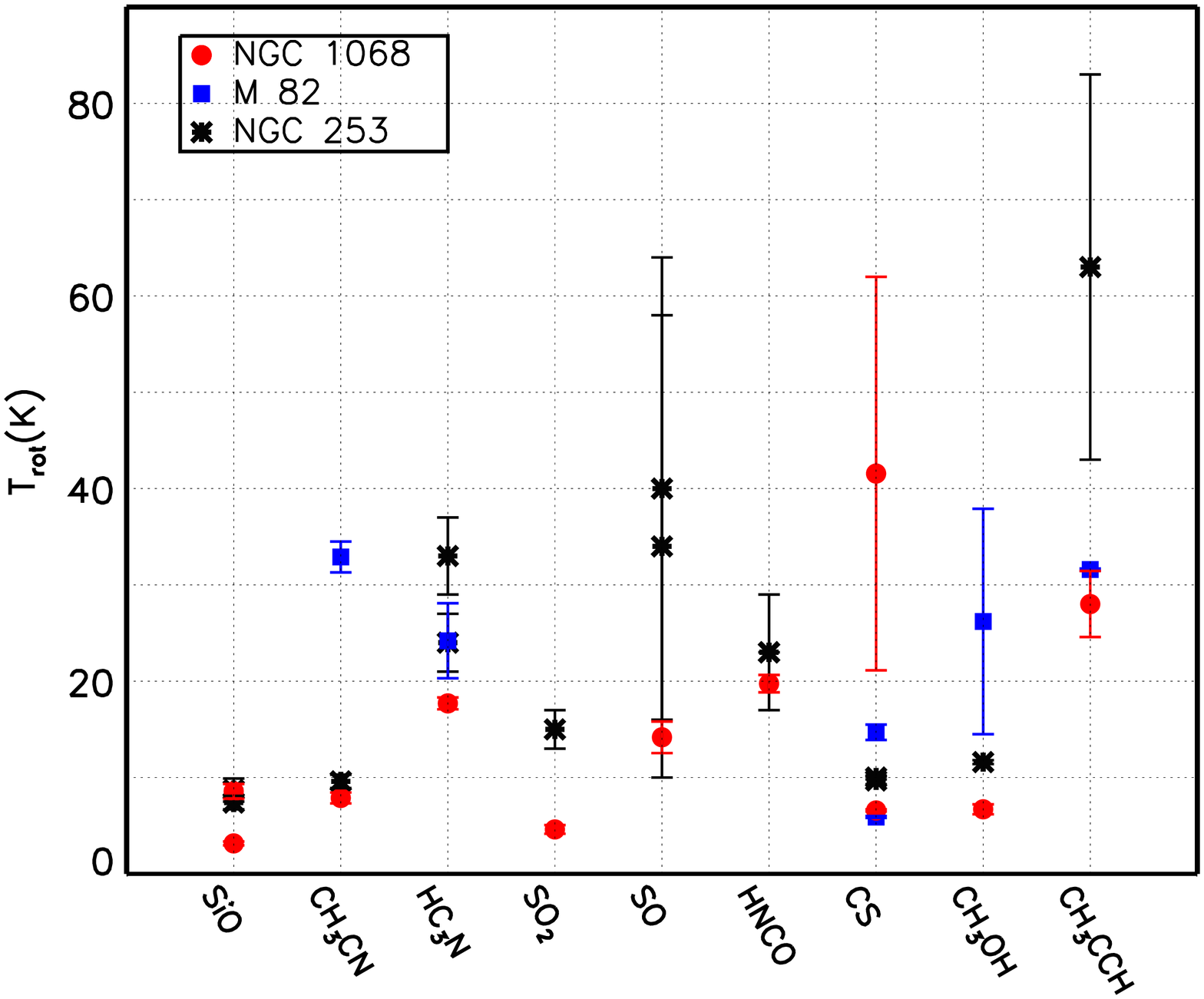}
\caption{
Comparison of the rotation temperatures in NGC\,1068, M\,82, and NGC\,253.
The molecular rotation temperatures of M\,82 and NGC\,253 are taken form \cite{Aladro11} and \cite{Martin06}.
}
\label{tvsname}
\end{figure}

\begin{figure}
  \centering
    \begin{minipage}{14cm}
        \includegraphics[width=14cm]{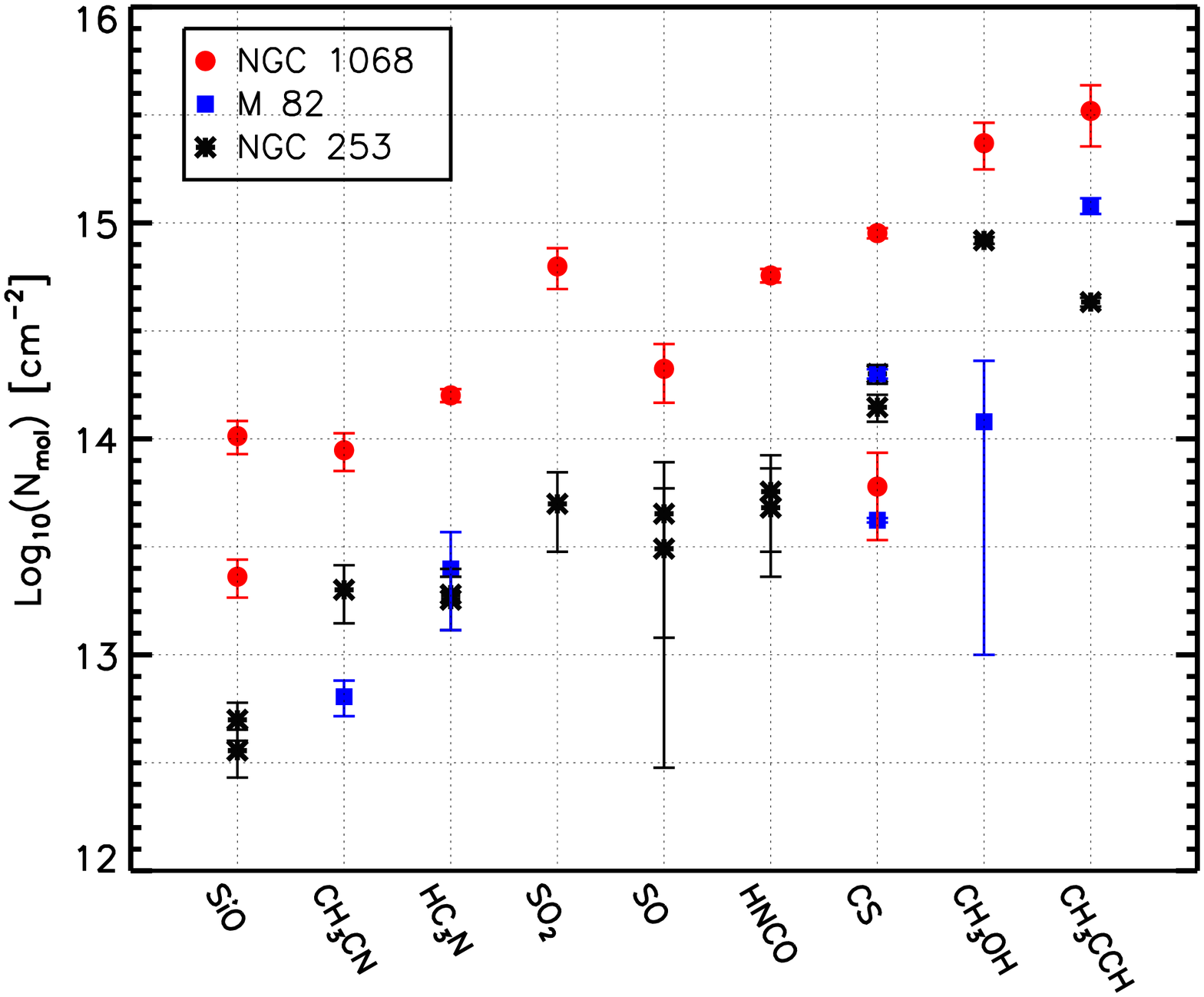}
\end{minipage}
    \begin{minipage}{14cm}
        \includegraphics[width=14cm]{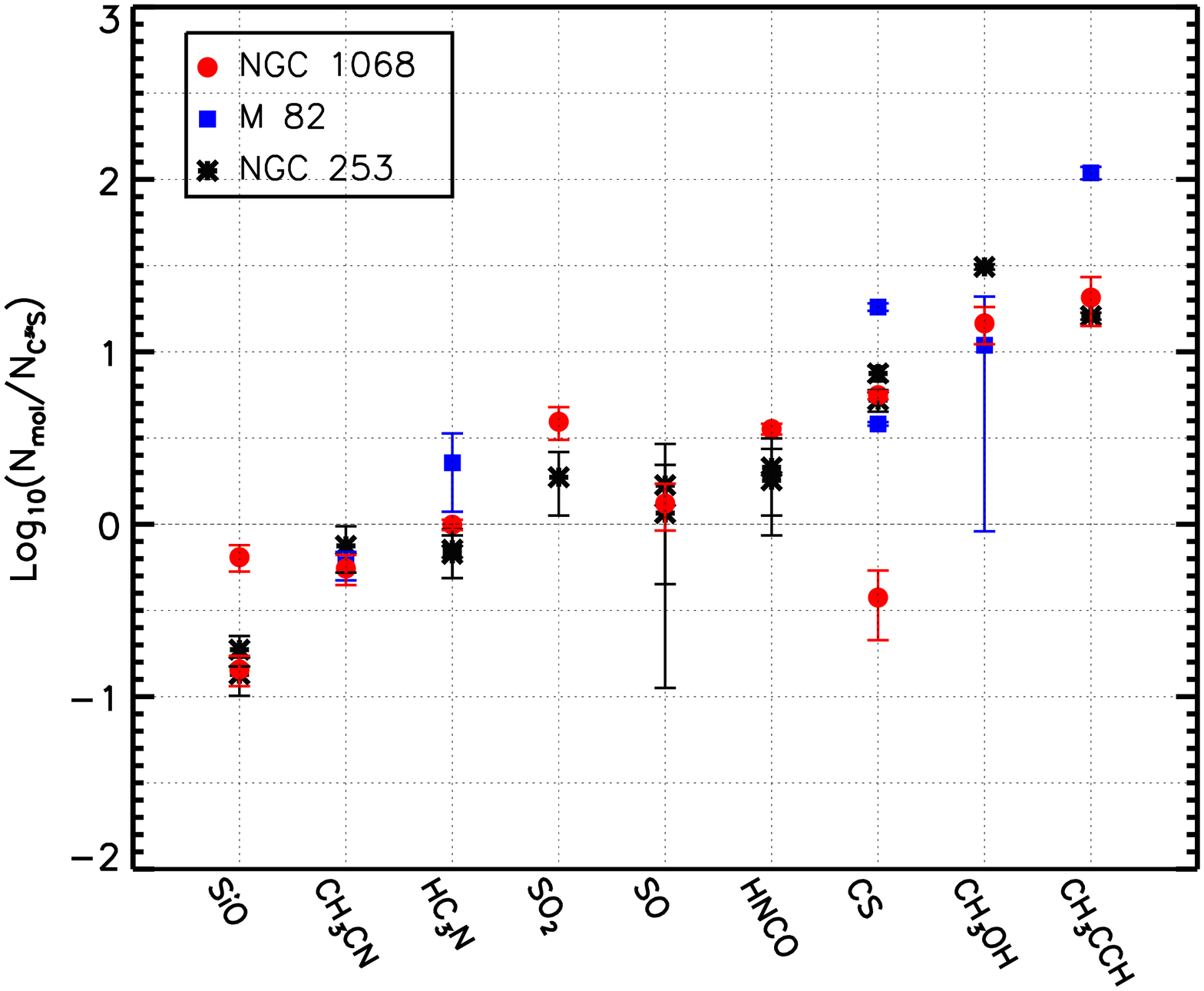}
  \end{minipage}
  \caption{
  Upper panel: Comparison of the column densities in NGC\,1068, M\,82, and NGC\,253. 
  Lower panel: Comparison of the fraction abundances with respect to C$^{34}$S. 
  The column densities of the molecules in M\,82 and NGC\,253 are taken from \cite{Aladro11} and \cite{Martin06}, respectively. 
  The column density of C$^{34}$S in NGC\,1068 is taken from \cite{Aladro13}.
  }
\label{tvsn}
\end{figure}

\restoregeometry

\clearpage

\begin{table*}
\caption{Band parameters.}
\centering
\begin{tabular}{cccccc}
\hline\hline
band range                    & $F_{\rm eff}$  & $B_{\rm eff}$    & On-time{\boldmath$^{a}$}       & HPBW{\boldmath$^{b}$} & spatial resolution \\
GHz                               & \%               &  \%                  & hour                 &$''$                               & Kpc                  \\  
\hline
75.691-83.502               &  95                &   81                  &    0.7             &  32.5 - 29.5   & 2.3 - 2.1              \\          
91.440-99.224               &  95                &   81                  &    0.7             &  26.9 - 24.8   & 1.9 - 1.7              \\  
160.756-168.567           &  93                 &   73                  &    18.0             &  15.3 - 14.6   & 1.1 -1.0               \\ 
176.498-184.309           &  93                 &   68                  &    18.0             &  13.9 - 13.3   &  0.9                     \\ 
\hline
\end{tabular}
\tablefoot{ 
\tablefoottext{a}{on-source time for each band;} 
\tablefoottext{b}{${\rm \frac{HPBW}{arcsec}}$ = ${\rm \frac{2460}{Freq}}$ $\times$ ${\rm \frac{1}{GHz}}$.} \\
 $F_{\rm eff}$ and $B_{\rm eff}$ are taken from the website of the IRAM 30m (http://www.iram.es/IRAMES/mainWiki/Iram30mEfficiencies). \\
       }
\label{table_parameter}
\end{table*}

\clearpage

\newgeometry{left=0.1cm,bottom=1cm}
\begin{longtable}{cccccccccl}
\caption{The transitions detected with single dish telescope in NGC\,1068.} \\
\hline
\hline
Line                                                               & Frequency      & Flux                                     & Velocity                                    & Line Width                          & $T_{\rm mb}$   & RMS    &  Telescope & HPBW                   & Ref. \\ 
                                                                     & MHz             & mK\,km\,s$^{-1}$                   & km\,s$^{-1}$                            & km\,s$^{-1}$                       & mK                  & mK      &                   &  $^{\prime\prime}$ &        \\
\hline
         HC$_{3}$N\,$(9-8)$ $^{a}$                     & 81881.5          & $517.3\pm160.7$                   & $1167.6$                                  & $213.2$                              & 2.28                 & 1.37     & IRAM 30m  & 30.0    & 1, P \\ 
         CH$_3$OH\,$(5_{k}-4_{k})$                     & 84521.2         & $127.2\pm28.0$                    & $1035.7\pm8.9$                        & $78.0\pm19.1$                    & 1.53                  & 0.41    & IRAM 30m  & 29.1     & 2  \\      
        HC$^{18}$O$^+$\,$(1-0)$                        & 85162.2         & $125.6\pm40.3$                    & $1124.7\pm58.3$                       & $317.6\pm107.9$                 & 0.37                 & 0.27    & IRAM 30m  & 28.9    & 2  \\
                                                                     &                     & $133.1\pm36.8$                    & $1124.5\pm49.5$                       & $300.8\pm83.3$                 & ...                     & ...        & IRAM 30m  & 28.9    & 5   \\
        c-C$_3$H$_2$\,$(2_{1,2}-1_{0,1})$            & 85338.9        & $1190.3\pm59.9$                   & $1128.6\pm6.3$                         & $249.6\pm12.4$                 & 4.48                 & 0.36    & IRAM 30m  & 28.8    & 2  \\
\end{longtable}
\restoregeometry

\begin{table*}
\caption{Flux ratios  (row/column).}
\centering
\begin{tabular}{cccccc}
\hline
\hline
\multicolumn{1}{c}{}  &\multicolumn{1}{c}{$^{13}$CCH\,$(2-1)$ } &\multicolumn{1}{c}{HCN\,$(2-1)$} &\multicolumn{1}{c}{HCO$^{+}$\,$(2-1)$}                     &\multicolumn{1}{c}{HOC$^{+}$\,$(2-1)$}          &\multicolumn{1}{c}{HNC\,$(2-1)$}    \\ 
\hline
 $^{13}$CCH\,$(2-1)$                             & 1.00 $\pm$ 0.28       &       0.013 $\pm$ 0.003   &   0.024 $\pm$ 0.005     &  0.38 $\pm$ 0.10      & 0.085 $\pm$ 0.022        \\          
 HCN\,$(2-1)$                                         & 75.3 $\pm$ 15.1  &       1.00   $\pm$ 0.01     &   1.82   $\pm$ 0.03       &  28.7 $\pm$ 5.28      & 6.42   $\pm$ 1.07          \\      
 HCO$^{+}$\,$(2-1)$                              & 41.4 $\pm$ 8.34               &       0.55   $\pm$ 0.008   &   1.00   $\pm$ 0.02       &  15.8 $\pm$ 2.91      & 3.53   $\pm$ 0.59          \\ 
 HOC$^{+}$\,$(2-1)$                              & 2.63 $\pm$ 0.72               &       0.035 $\pm$ 0.006   &   0.063 $\pm$ 0.012     &  1.00 $\pm$ 0.26      & 0.22   $\pm$ 0.056        \\  
 HNC\,$(2-1)$                                        & 11.7 $\pm$ 3.07            &      0.16   $\pm$ 0.03     &   0.28   $\pm$ 0.047     &  4.46 $\pm$ 1.11      & 1.00   $\pm$ 0.24          \\    
\hline
\end{tabular}
\label{Tab_line_ratio}
\end{table*}

\begin{table*}
\caption{Flux ratios of the blue and red components of the CDN lines.}
\centering
\begin{tabular}{lccc}
\hline\hline
Transition                                           & Flux(blue)                            & Flux(red)                         & R$_{ \rm E/w}$              \\
                                                         &  mK km s$^{-1}$                  & mK km s$^{-1}$               &                                     \\ 
\hline
HC$_{3}$N (18-17)                               &     405                                 & 175                               &  2.32 $\pm$ 0.28                     \\   
SiO ($5-4$)                                        &      281                                 & 135                               &  2.08 $\pm$ 0.57                     \\  
SO$_{2}$ ($14_{3,11}-14_{2,12}$)            &      2123                               & 1034                             &  2.05 $\pm$ 0.16                     \\        
HNC (2-1)                                           &      1994                               & 1175                              & 1.70 $\pm$ 0.46                      \\ 
HCN (2-1)                                           &     13376                               & 9116                             &  1.47 $\pm$ 0.02                     \\  
HCO$^{+}$ (2-1)                                  &     7203                                & 5231                             &  1.38 $\pm$ 0.03                     \\
\hline
\hline
\end{tabular}
\tablefoot{ 
Columns 2 and 3 list the fluxes of the blue and red components, respectively.
Column 4 shows the flux ratios of the blue-to-red (east-to-west) components.
       }
\label{tab_ratio}
\end{table*}

\begin{table*}
\caption{Rotational temperatures and column densities   of the observed molecules. } \centering
\begin{tabular}{llcccc}
\hline\hline
Molecule                &     lines                                            & $T_{\rm rot}$                       &   $N_{\rm mol}$                                                     &   Telescope      & Ref       \\
                            &                                                        & K                                        &   cm$^{-2}$                                                           &                       &            \\
\hline
CS                        &   (2-1) (3-2) (5-4)                             &   6.60 $\pm$ 0.12                 &  (8.78 $\pm$ 0.47) $\times$ 10$^{14}$ cm$^{-2}$      & IRAM 30m       & 1, 5, 2     \\
                            &   (5-4)(7-6)                                      & 41.56 $\pm$ 17.82                &  (6.01 $\pm$ 2.31) $\times$ 10$^{13}$ cm$^{-2}$       & IRAM, JCMT    & 2, 6         \\
HC$_{3}$N             &   (9-8)(10-9)(11-10)(12-11)(18-17)        & 17.75  $\pm$ 0.60                 &  (1.59 $\pm$ 0.10) $\times$ 10$^{14}$ cm$^{-2}$       & IRAM 30m       & 1, 2, 7, 7,1 \\
HCN                      &   (1-0)(2-1)(3-2)(3-2)(4-3)                 &   17.32 $\pm$ 0.49                &  (6.78 $\pm$ 0.15) $\times$ 10$^{13}$ cm$^{-2}$      & IRAM, JCMT    & 2, 1, 8, 9, 4 \\
HCO$^{+}$             &   (1-0)(2-1)                                       &    3.03 $\pm$ 0.02                 &  (5.81 $\pm$ 0.05) $\times$ 10$^{14}$ cm$^{-2}$     & IRAM 30m       & 2, 1            \\
\hline 
\hline 
\end{tabular}
\tablefoot{ 
\tablefoottext{a}{Owing to the emission of CH$_{3}$OH $(5_{-1,5} - 4_{0,4}) E$ comes from a mega-maser (Wang et al. 2014), the rotation temperature of its warmer component is an upper limit.} \tablefoottext{b}{The rotation temperature is assumed to be 10 K. \\}
{\bf Refrence}: (1) this work; (2) \cite{Qiu18}; (3) \cite{Aladro13}; (4) \cite{Perez09}; (5) \cite{Mauersberger89}; 
(6) \cite{Bayet09}; (7) \cite{Aladro15}; (8) \cite{Krips08}; (9) \cite{Perez07}; (10) \cite{Tan18}; 
(11) \cite{Usero04}; (12) \cite{Martin09b}.
 }
\label{table_RT}
\end{table*}

\clearpage

\newgeometry{left=0.3cm,bottom=1cm}
\begin{table*}
\tiny
\caption{Integrated intensity ratios shown in Fig. \ref{figure10}.}
\centering
\begin{tabular}{lccccccccccc}
\hline\hline
Transition                                                             & Frequency               & $I_{1}$                             & $I_{2}$                                  & $R$                               & $\theta_{b1}$ & $\theta_{b2}$  & $\theta_s$ & Ref(1)  & Ref(2)  & Marks \\
                                                                           & GHz                       &  mK km s$^{-1}$              &  mK km s$^{-1}$                     &                                      &     $''$           &   $''$              &  $''$          &           &            &           \\    
\hline                                                                                                           
             $^{13}$CO\,$(1-0)$                                  &   110.2                     &       8900 $\pm$ 200          &      13100 $\pm$ 240                &     0.68 $\pm$ 0.02      &      16.5    &    22.3   &  ...                        &   3    &    5   & NRO45m / IRAM30m  \\ 
             CH$_{3}$OH\,$(2_k-1_k)$                         &    96.7                     &        1100 $\pm$ 200          &        1613 $\pm$ 271                &     0.68 $\pm$ 0.17       &      19.0    &    25.4   &  ...                        &   3    &    1   & NRO45m / IRAM30m  \\     
             HNC\,$(1-0)$                                          &    90.7                     &       5600 $\pm$ 400         &       7836 $\pm$ 35                  &     0.71 $\pm$ 0.05       &      19.1     &    27.1    &  ...                        &   3    &    2   & NRO45m / IRAM30m  \\ 
             $^{13}$CO\,$(2-1)$                                  &  220.4                     &      13314 $\pm$ 127           &      17100 $\pm$ 2000              &     0.78 $\pm$ 0.09      &       11.2     &    21.0   &  ...                        &   2    &    8   & IRAM30m / JCMT15m \\ 
             HCN\,$(3-2)$                                         &   265.9                     &     19000 $\pm$ 600          &     22700 $\pm$ 2500              &     0.84 $\pm$ 0.10       &       9.8     &    18.0    &  ...                        &   7    &    10   & IRAM30m / JCMT15m \\    
             C$^{18}$O\,$(1-0)$                                  &   109.8                     &      3300 $\pm$ 200          &      3900 $\pm$ 90                   &     0.85 $\pm$ 0.05      &      16.5     &    22.4   &  ...                        &   3    &    5   & NRO45m / IRAM30m  \\    
             CO\,$(1-0)$                                            &    115.3                     &   155000 $\pm$ 1000         &   172200 $\pm$ 4500               &     0.90 $\pm$ 0.02      &      15.2     &    21.3    &  ...                        &   3    &    4   & NRO45m / IRAM30m  \\        
             SiO\,$(2-1)$                                           &     86.8                     &        800 $\pm$ 200          &        879 $\pm$ 65                   &     0.91 $\pm$ 0.24      &      18.7     &    28.3   &  ...                         &   3    &    2   & NRO45m / IRAM30m  \\       
             HCO$^{+}$\,$(1-0)$                                 &    89.2                     &     13300 $\pm$ 700          &     14519 $\pm$ 56                    &     0.92 $\pm$ 0.05     &      19.1      &    27.6   &  ...                         &   3    &    2   & NRO45m / IRAM30m  \\ 
             CS\,$(2-1)$                                            &     98.0                     &      5400 $\pm$ 200          &       5701 $\pm$ 239                 &     0.95 $\pm$ 0.05     &      19.0     &    25.1    &  ...                         &   3    &    1   & NRO45m / IRAM30m  \\       
             c-C$_3$H$_2$\,$(21_2-10_1)$                  &    85.3                     &       1200 $\pm$ 100          &       1190 $\pm$ 60                    &     1.01 $\pm$ 0.10       &      19.0     &    28.8   &  239 $\pm$ 1448    &   3    &    2   & NRO45m / IRAM30m  \\     
             CN\,$(1_{3/2,5/2}-0_{1/2,3/2})$                &    113.5                     &    27600 $\pm$ 400          &    25790 $\pm$ 190                  &     1.07 $\pm$ 0.02      &      15.2     &    21.7    &  56 $\pm$ 8           &   3    &    5   & NRO45m / IRAM30m  \\      
             H$^{13}$CN\,$(1-0)$                                &     86.3                     &      1400 $\pm$ 200          &      1250 $\pm$ 37                    &     1.12 $\pm$ 0.16       &      18.7     &    28.5   &  59 $\pm$ 44         &   3    &    2   & NRO45m / IRAM30m  \\        
             CO\,$(2-1)$                                            &   230.5                     &   266000 $\pm$ 32000      &  220000 $\pm$ 26000              &     1.21 $\pm$ 0.20      &      10.6     &    21.0    &  38 $\pm$ 20         &   8    &    8   & IRAM30m / JCMT15m \\     
             C$_2$H\,$(1-0)$                                     &     87.3                     &      10000 $\pm$ 300         &      7940 $\pm$ 53                   &     1.26 $\pm$ 0.04      &      19.0     &    28.2   &  36 $\pm$ 3           &   3    &    2   & NRO45m / IRAM30m  \\  
             HCN\,$(1-0)$                                          &     88.6                     &     26400 $\pm$ 600         &     20754 $\pm$ 29                  &     1.27 $\pm$ 0.03      &      19.1      &    28.0   &  34 $\pm$ 2           &   3    &    2   & NRO45m / IRAM30m  \\      
             CN\,$(1_{1/2,3/2}-0_{1/2,3/2})$                &    113.2                     &      18500 $\pm$ 300         &      13150 $\pm$ 100                 &     1.41 $\pm$ 0.03       &      15.2     &    21.7    &  19 $\pm$ 1            &   3    &    5   & NRO45m / IRAM30m  \\ 
             CH$_3$CN\,$(6_k-5_k)$                          &    110.4                     &        1000 $\pm$ 200         &        700 $\pm$ 490                 &     1.43 $\pm$ 1.04      &      16.5      &    22.3   &  16 $\pm$ 40         &   3    &    5   & NRO45m / IRAM30m  \\   
             N$_2$H$^{+}$\,$(1-0)$                            &     93.2                     &        2100 $\pm$ 400         &       1407 $\pm$ 229                 &     1.49 $\pm$ 0.37      &     19.1       &    26.4   &  18 $\pm$ 15          &   3    &    1   & NRO45m / IRAM30m  \\  
             HC$_3$N\,$(12-11)$                                &    109.2                      &       1300 $\pm$ 100          &        850 $\pm$ 40                  &     1.53 $\pm$ 0.14       &     16.5      &    22.5    &  13 $\pm$ 4           &   3    &    5   & NRO45m / IRAM30m  \\       
             SO\,$(2_3-1_2)$                                      &    99.3                      &         800 $\pm$ 200         &        470 $\pm$ 60                  &     1.70 $\pm$ 0.48      &      16.5      &    24.8   &  15 $\pm$ 11           &   3    &    5   & NRO45m / IRAM30m  \\       
             CN\,$(2_{5/2,7/2}-1_{3/2,5/2})$                &  226.9                      &      16479 $\pm$ 342         &      8464 $\pm$ 672                 &     1.95 $\pm$ 0.16      &      10.9      &    21.0    &  15 $\pm$ 2           &   2    &    9   & IRAM30m / JCMT15m \\   
             $^{13}$CN\,$(1_{1/2}-0_{1/2})$                  &   108.7                      &         400 $\pm$ 100          &        200 $\pm$ 100                 &     2.00 $\pm$ 1.12      &      16.5      &    22.5    &  ...                         &   3    &   4   & NRO45m / IRAM30m  \\  
             HC$_3$N\,$(11-10)$                                 &   100.1                      &        1600 $\pm$ 300          &        800 $\pm$ 40                 &     2.00 $\pm$ 0.34      &     16.5      &    24.6    &  8 $\pm$ 8            &   3    &   5   & NRO45m / IRAM30m  \\       
             HNCO\,$(4_{0,4}-3_{0,3})$                        &    87.9                      &        1600 $\pm$ 200         &        675 $\pm$ 45                  &     2.37 $\pm$ 0.34     &      19.1      &    28.0    &   ...                        &   3    &   2   & NRO45m / IRAM30m  \\
             HNCO\,$(5_{0,5}-4_{0,4})$                        &   109.9                      &         900 $\pm$ 100         &         300 $\pm$ 150                &     3.00 $\pm$ 1.54      &      16.5     &    22.4    &   ...                        &   3    &   4   & NRO45m / IRAM30m  \\    
             CN\,$(2_{5/2,7/2}-1_{3/2,5/2})$                &  226.7                       &      10231 $\pm$ 373         &      2808 $\pm$ 831                 &     3.64 $\pm$ 1.09      &      10.9     &    21.0    &   2 $\pm$ 14           &   2    &   9   & IRAM30m / JCMT15m \\   
             HC$_3$N\,$(10-9)$                                  &    91.0                       &       4700 $\pm$ 1000        &        929 $\pm$ 36                  &     5.06 $\pm$ 1.09      &      19.1     &    27.0    &   ...                         &   3    &    1   & NRO45m / IRAM30m  \\             
             H$^{13}$CO$^{+}$\,$(1-0)$                       &    86.8                       &             <        150            &        758 $\pm$ 39                  &     < 0.20                    &      18.7     &    28.4   &   ...                         &   3    &    2   & NRO45m / IRAM30m  \\     
             CH$_3$CCH\,$(5_k-4_k)$                        &    85.5                        &            <         140           &        348 $\pm$ 107                 &     < 0.40                    &      19.0     &    28.8   &   ...                         &   3    &    2   & NRO45m / IRAM30m  \\     
             SO\,$(3_2-2_1)$                                      &   109.3                       &             <        250           &         200 $\pm$ 100                &     < 1.25                    &      16.5     &    22.5   &  < 26                      &   6    &    4   & NRO45m / IRAM30m  \\     
             SO\,$(5_4-4_4)$                                     &   100.0                        &             <        500          &        200 $\pm$ 100                 &     < 2.00                    &      16.5     &    24.6   &  ...                          &   6    &    4   & NRO45m / IRAM30m  \\     
\hline
\hline
\end{tabular}
\tablefoot{
Columns 1 and 2 list the transitions and frequencies, respectively. Columns 3 and 4 list the detected integrated intensities with the NRO 45m (or IRAM 30m) and  IRAM 30m (ro JCMT 15m) telescopes, respectively.
Column 5 lists the  ratios between the intensities given in Cols.~3 and  4. 
Columns 6 and 7 list the beam sizes of the NRO 45m (or IRAM 30m) and   the IRAM 30m (or JCMT 15m) telescopes, respectively.
Column 8 lists the estimated source sizes.
 Columns 9 and 10 list the references for the data listed in Cols. 1 and 2, respectively. \\
 {\bf Reference:} (1) this work; (2) \citet{Qiu18}; (3) \citet{Takano19}; (4) \citet{Aladro13}; (5) \citet{Aladro15};  \\
 (6) \citet{Nakajima18}; (7) \citet{Krips08}; (8) \citet{Israel09}; (9) \citet{Perez09}; (10) \citet{Perez07}. \\
       }
\label{flux_ratio}
\end{table*}
\restoregeometry
\normalsize

\restoregeometry

\end{document}